\documentclass[preprint,aps,superscriptaddress,preprintnumbers,nofootinbib]{revtex4-1}

\usepackage{graphicx}
\usepackage{amssymb}
\usepackage{amsmath}
\usepackage{color}
\usepackage{multirow}
\usepackage{textcomp}
\usepackage{rotfloat}

\begin{document}
\newcommand{\beq}{\begin{equation}}
\newcommand{\eeq}{\end{equation}}
\newcommand{\bea}{\begin{eqnarray}}
\newcommand{\eea}{\end{eqnarray}}
\newcommand{\nn}{\nonumber}

\newcommand{\uL}{u_{\tmbx{L}}}
\newcommand{\dL}{d_{\tmbx{L}}}
\newcommand{\nL}{\nu_{\tmbx{L}}}
\newcommand{\eL}{e_{\tmbx{L}}}
\newcommand{\uR}{u_{\tmbx{R}}}
\newcommand{\dR}{d_{\tmbx{R}}}
\newcommand{\nR}{\nu_{\tmbx{R}}}
\newcommand{\eR}{e_{\tmbx{R}}}
\newcommand{\tto}{G(221)}
\newcommand{\tbox}[1]{\mbox{\tiny #1}}
\newcommand{\tmbx}[1]{\mbox{\tiny{$#1$}}}
\newcommand{\VEV}[1]{\langle  #1 \rangle}
\newcommand{\mfrac}[2]{\frac{ \mbox{$#1$} }{ \mbox{$#2$} }}
\newcommand{\tbeta}{\tilde{\beta}}

\newcommand{\tphi}{\tilde{\phi}}
\newcommand{\tth}{\tilde{\theta}}
\newcommand{\tbet}{\tilde{\beta}}
\newcommand{\tal}{\tilde{\alpha}}

\allowdisplaybreaks[4]

\newcommand{\tabincell}[2]{\begin{tabular}{@{}#1@{}}#2\end{tabular}}

%\title{Simple Non-Abelian Extensions and Diboson Excesses at LHC}
\title{Simple non-Abelian extensions of the standard model gauge group and the diboson excesses at the LHC} 

\author{Qing-Hong Cao}
\email{qinghongcao@pku.edu.cn}
\affiliation{Department of Physics and State Key Laboratory of Nuclear Physics
and Technology,
\\Peking University, Beijing 100871, China}
\affiliation{Collaborative Innovation Center of Quantum Matter, Beijing, China}
\affiliation{Center for High Energy Physics, Peking University, Beijing 100871, China}

\author{Bin Yan}
\email{binyan@pku.edu.cn}
\affiliation{Department of Physics and State Key Laboratory of Nuclear Physics
and Technology,
\\Peking University, Beijing 100871, China}

\author{Dong-Ming Zhang}
\email{zhangdongming@pku.edu.cn}
\affiliation{Department of Physics and State Key Laboratory of Nuclear Physics
and Technology,
\\Peking University, Beijing 100871, China}

\begin{abstract}

The ATLAS collaboration reported excesses at around 2 TeV in the di-boson production decaying into hadronic final states. We consider the possibility of explaining the excesses with extra gauge bosons in two simple non-Abelian extensions of the Standard Model. One is the so-called $G(221)$ models with a symmetry structure of $SU(2)_1\otimes SU(2)_2\otimes U(1)_X$ and the other is the $G(331)$ models with an extended symmetry of $SU(3)_C\otimes SU(3)_L\otimes U(1)_X$. The $W'$ and $Z'$ bosons emerge after the electroweak symmetry is spontaneously broken. Two patterns of symmetry breaking in the $G(221)$ models are considered in this work: one is $SU(2)_L\otimes SU(2)_2 \otimes U(1)_X \to SU(2)_L\otimes U(1)_Y$, the other is $SU(2)_1\otimes SU(2)_2 \otimes U(1)_Y \to SU(2)_L\otimes U(1)_Y$. The symmetry breaking of the $G(331)$ model is $SU(3)_L\otimes U(1)_X \to SU(2)_L \otimes U(1)_Y$.
We perform a global analysis of $W^\prime$ and $Z^\prime$ phenomenology in ten new physics models, including all the channels of $W^\prime/Z^\prime$ decay. Our study shows that the leptonic mode and the dijet mode of $W^\prime/Z^\prime$ decays impose a very stringent bound on the parameter space in several new physics models. Such tight bounds provide a useful guide for building new physics models to address on the diboson anomalies. We also note that the Left-Right and Lepton-Phobic models can explain the $3.4\sigma$ $WZ$  excess if the $2.6\sigma$ deviation in the $W^+W^-$ pair around 2~TeV were confirmed to be a fluctuation of the SM backgrounds.  

\end{abstract}

\maketitle

\section{Introduciton}

Searches for new physics (NP) effects in the final state of vector boson pairs have been carried out recently by both ATLAS~\cite{Aad:2015owa} and CMS~\cite{Khachatryan:2014hpa,Khachatryan:2014gha} Collaborations using the technique of jet substructure. It was reported recently by the ATLAS collaboration~\cite{Aad:2015owa} that, using a data sample with $20~{\rm fb}^{-1}$ integrated luminosity, a $3.6\sigma$ deviation is observed in the invariant mass distribution of the $WZ$ pair, which requires a NP contribution to the cross section of the $WZ$ production as $\sigma(WZ) \sim  4-8~{\rm fb}$. Also a $2.6\sigma$ and $2.9\sigma$ deviation is observed in the invariant mass distribution of $WW$ and $ZZ$ pair production, respectively. The NP contributions of $\sigma(WW) \sim 3-7~{\rm fb}$ and $\sigma(ZZ)\sim 3-9~{\rm fb}$ are needed to explain the excesses. All the three excesses occur around $2~{\rm TeV}$ in the invariant mass distribution of vector boson pair~\footnote{The CMS collaboration also performed similar searches in the diboson channel~\cite{Khachatryan:2014hpa,Khachatryan:2014gha} but no excess was observed. In this study we focus on the ATLAS results and explore the NP explanation of those diboson excesses. }. 
The vector boson pair production is highly correlated with the associated production of a vector boson and Higgs boson. The CMS collaboration has obtained a bound on the cross section of $WH$ and $ZH$ productions~\cite{Khachatryan:2015bma}, $\sigma(WH)\leq 7.1~{\rm fb}$ and $\sigma(ZH)\leq 6.8~{\rm fb}$, respectively. 

As the final state involves two gauge bosons, it is natural to consider the excesses are induced by a spin-one resonances in new physics (NP) beyond the SM. Those heavy gauge bosons might arise from an extension of the SM with additional non-Abelian gauge symmetry. It is interesting to ask whether or not the deviation can be addressed by heavy gauge bosons after one takes into account other precision data. There has been recent excitement among theorists for this measurement at the LHC~\cite{Fukano:2015hga, Hisano:2015gna, Franzosi:2015zra, Cheung:2015nha,Dobrescu:2015qna,Aguilar-Saavedra:2015rna,Gao:2015irw,Thamm:2015csa,Alves:2015mua}.

In this work we consider two kinds of non-Abelian gauge extension to the SM: one is the so-called $G(221)$ models with a symmetry of $SU(2)_1\otimes SU(2)_2 \otimes U(1)_X$~\cite{Hsieh:2010zr,Berger:2011hn,Cao:2012ng} and the other is the $G(331)$ model with a symmetry of $SU(3)_C\otimes SU(3)_L\otimes U(1)_X$~\cite{Frampton:1992wt,Pisano:1991ee}. 
Both charged extra boson $W^\prime$ and new neutral boson $Z^\prime$ arise after the symmetry breaking.  Several $G(221)$ and $G(331)$ models are examined in this work. 
We demonstrate that the leptonic decay and dijet decay modes of $W^\prime/Z^\prime$ impose a very stringent bound on the parameter space of those NP models. 
In order to explain the $WW/WZ$ excess under the two simple extensions, the leptonic and dijet decay modes of those extra gauge bosons need to be largely reduced in a more complete NP theory. 

There are a few bounds from the $W^\prime/Z^\prime$ searches in their fermionic decays at the LHC,  e.g. for a 2~TeV $W^\prime/Z^\prime$, $\sigma(pp\to Z^\prime/W^\prime \to jj) \leq 102~{\rm fb}$~\cite{Aad:2014aqa,Khachatryan:2015sja}, $\sigma(pp\to Z^\prime \to t\bar{t}) \leq 11~{\rm fb}$~\cite{Aad:2014xea}, $\sigma(pp \to W^\prime_R \to t\bar{b}) \leq 124~{\rm fb}$, $\sigma(pp \to W^\prime_L \to t\bar{b}) \leq 162~{\rm fb}$~\cite{Khachatryan:2015sma}, $\sigma(pp\to Z^\prime \to e^+e^-/\mu^+\mu^-) \leq 0.2~{\rm fb}$~\cite{Aad:2014cka,Khachatryan:2014fba} and $\sigma(pp\to W^\prime\to e\nu/\mu\nu)\leq 0.7~{\rm fb}$~\cite{ATLAS:2014wra,Khachatryan:2014tva}. We also take all the above bounds into account and perform a global analysis on each individual NP model. 

It is hard to explain the $ZZ$ excess in the simple non-Abelian gauge extension of the SM. The difficulty has been discussed extensively in Refs.~\cite{Franzosi:2015zra,Cheung:2015nha,Gao:2015irw}. For example, having an  extra neutral gauge boson decaying to the $ZZ$ mode would require the violation in $P$ or $CP$ symmetry~\cite{Franzosi:2015zra}. An alternative way is to introduce an extra scalar which predominately decays into $ZZ$ and $WW$ pairs. Unfortunately, the cross section of the scalar production is usually too tiny to explain the $ZZ$ excess~\cite{Cheung:2015nha}. Therefore, we focus our attention on the $WW$ and $WZ$ excesses in this work.

The paper is organized as follows. In Sec.~\ref{221} we briefly review the $G(221)$ models. In Sec.~\ref{sec:wpzp} we present the NLO cross section of $W^\prime/Z^\prime$ production at the LHC Run-1 and the PDF uncertainties. In Sec.~\ref{221a} we focus our attention on the first breaking pattern of $G(221)$ and discuss the Left-Right, Lepto-Phobic, Hadro-Phobic and Fermio-Phobic models. In Sec.~\ref{221b} we study the second breaking pattern of $G(221)$ and explore the phenomenology of the un-unified and non-universal models. In Sec.~\ref{331} we study the $G(331)$ model. Finally we conclude in Sec.~\ref{summary}.

\section{G(221) Models}\label{221}

The $G(221)$ model is the minimal extension of the SM, which consists of both $W^\prime$ and $Z^\prime$, exhibits a gauge structure of $SU(2)_1 \otimes SU(2)_2 \otimes U(1)_X$, named as $G(221)$ model~\cite{Mohapatra:1974gc, Mohapatra:1974hk, Mohapatra:1980yp,Barger:1980ix,Barger:1980ti,Georgi:1989ic,Georgi:1989xz,Li:1981nk,Malkawi:1996fs,
He:1999vp,Chivukula:2003wj,Chivukula:2006cg,Hsieh:2010zr,Berger:2011xk,Du:2012vh,Abe:2012fb,Wang:2013jwa,Cao:2015doa}. The model can be viewed as the low energy effective theory of many NP models with extended gauge structure when all the heavy particles other than the $W^\prime$ and $Z^\prime$ bosons decouple. In particular, we consider several $G(221)$ models categorized as follows: left-right (LR)~\cite{Mohapatra:1974gc, Mohapatra:1974hk, Mohapatra:1980yp,Patra:2015bga}, lepto-phobibc (LP), hadron-phobic (HP), fermio-phobic (FP)~\cite{Barger:1980ix,Barger:1980ti,Chivukula:2006cg}, un-unified (UU)~\cite{Georgi:1989ic,Georgi:1989xz} and non-universal (NU)~\cite{Li:1981nk,Malkawi:1996fs,He:1999vp,Berger:2011xk}. The charge assignments of the SM fermion in those models are listed in Table~\ref{tb:models}. 

\begin{table}
\begin{center}
\caption{\it 
The charge assignments of the SM fermions under
the $G(221)$ gauge groups.
Unless otherwise specified, the charge assignments
apply to all three generations.
}
\label{tb:models}
\vspace{0.125in}
\begin{tabular}{|c|c|c|c|}
\hline Model & $SU(2)_1$ & $SU(2)_2$ & $U(1)_{X}$ \\
\hline
Left-right (LR) &
$\begin{pmatrix} \uL \\ \dL \end{pmatrix}, \begin{pmatrix} \nL \\ \eL \end{pmatrix}$ &
$\begin{pmatrix} \uR \\ \dR \end{pmatrix}, \begin{pmatrix} \nR \\ \eR \end{pmatrix}$ &
$\begin{matrix} \tfrac{1}{6}\ \mbox{for quarks,} \\ -\tfrac{1}{2}\ \mbox{for leptons.} \end{matrix}$
\\
\hline
Lepto-phobic (LP) &
$\begin{pmatrix} \uL \\ \dL \end{pmatrix}, \begin{pmatrix} \nL \\ \eL \end{pmatrix}$ &
$\begin{pmatrix} \uR \\ \dR \end{pmatrix}$ &
$\begin{matrix} \tfrac{1}{6}\ \mbox{for quarks,} \\ Y_{\tbox{SM}}\ \mbox{for leptons.} \end{matrix}$
\\
\hline
Hadro-phobic (HP) &
$\begin{pmatrix} \uL \\ \dL \end{pmatrix}, \begin{pmatrix} \nL \\ \eL \end{pmatrix}$ &
$\begin{pmatrix} \nR \\ \eR \end{pmatrix}$ &
$\begin{matrix} Y_{\tbox{SM}}\ \mbox{for quarks,} \\ -\tfrac{1}{2}\ \mbox{for leptons.} \end{matrix}$
\\
\hline
Fermio-phobic (FP) &
$\begin{pmatrix} \uL \\ \dL \end{pmatrix}, \begin{pmatrix} \nL \\ \eL \end{pmatrix}$ &
 &
$ \begin{matrix} Y_{\tbox{SM}}\ \mbox{for all fermions.} \end{matrix}$
\\
\hline
Un-unified (UU) &
$\begin{pmatrix} \uL \\ \dL \end{pmatrix}$ &
$\begin{pmatrix} \nL \\ \eL \end{pmatrix}$ &
$ \begin{matrix} Y_{\tbox{SM}}\ \mbox{for all fermions.} \end{matrix}$
\\
\hline
Non-universal (NU) &
$\begin{pmatrix} \uL \\ \dL \end{pmatrix}_{1^{\tbox{st}},2^{\tbox{nd}}},
 \begin{pmatrix} \nL \\ \eL \end{pmatrix}_{1^{\tbox{st}},2^{\tbox{nd}}}$ &
$\begin{pmatrix} \uL \\ \dL \end{pmatrix}_{3^{\tbox{rd}}},
 \begin{pmatrix} \nL \\ \eL \end{pmatrix}_{3^{\tbox{rd}}}$ &
$ \begin{matrix} Y_{\tbox{SM}}\ \mbox{for all fermions.} \end{matrix}$
\\
\hline
\end{tabular}
\end{center}
\end{table}

We classify the $G(221)$ models based on the pattern of symmetry breaking and quantum number assignment of the SM fermions.  The symmetry breaking is assumed to be induced by fundamental scalar fields whose quantum number under the $G(221)$ gauge group depends on the breaking pattern. The NP models mentioned above fall into the following two patterns of symmetry breaking:
\begin{itemize}
\item[(a)] breaking pattern I (BP-I): \\
$SU(2)_1$ is identified as the $SU(2)_L$ of the SM. The first stage of symmetry breaking 
$SU(2)_2 \times U(1)_{X}\to U(1)_{Y}$ occurs at the TeV scale, while the second stage of 
symmetry breaking $SU(2)_{L}\times U(1)_Y \to U(1)_{\rm em}$ takes place at the electroweak scale;
\item[(b)] breaking pattern II (BP-II):\\
$U(1)_X$ is identified as the $U(1)_Y$ of the SM. The first stage of symmetry breaking 
$SU(2)_{1}\times SU(2)_{2}\to SU(2)_{L}$ occurs at the TeV scale, while the second stage of
symmetry breaking $SU(2)_{L}\times U(1)_Y \to U(1)_{\rm em}$ happens at the electroweak scale.
\end{itemize}
The $W^\prime$ and $Z^\prime$ arise after the symmetry breaking at the TeV scale. The most general interaction of the $Z^\prime$ and $W^\prime$ to SM fermions is 
\bea
\mathcal{L}_{f}=Z^\prime_{\mu}\,\bar{f}\,\gamma^{\mu}(g_L P_L+g_R P_R)f
+ W^\prime_{\mu}\,\bar{f}\,\gamma^{\mu}(g_L^\prime P_L + g_R^\prime P_R) f^\prime +h.c.\,,
\label{eq:effcoup}
\eea
where $P_{L,R}=(1\mp\gamma_{5})/2$ are the usual chirality projectors.
For simplicity, we use $g_L$ and $g_R$ for both $Z^\prime$ and $W^\prime$ bosons
from now on. Note that throughout this work only SM fermions are considered, despite in certain models new heavy fermions are necessary to cancel gauge anomalies.

\section{The $W^\prime/Z^\prime$ Production Cross section}\label{sec:wpzp}

The $W^\prime$ and $Z^\prime$ are produced singly through the Drell-Yan process. Following the experimental searches, we adapt the narrow width approximation (NWA) to factorize the process of $pp \to W^\prime/Z^\prime \to V_1V_2$ as follows: 
\beq 
\sigma(pp \to V^\prime \to XY) \simeq  \sigma(pp \to V^\prime)\otimes {\rm BR}(V^\prime \to XY) \equiv \sigma(V^\prime)\times {\rm BR}(V^\prime \to XY),
\eeq
where $X$ and $Y$ denote the decay products of the $V^\prime$ boson.
Next we consider a few $G(221)$ models and discuss their implications on the $VV^\prime$ and $VH$ productions.

\begin{figure}[b]
\includegraphics[width=0.32\textwidth]{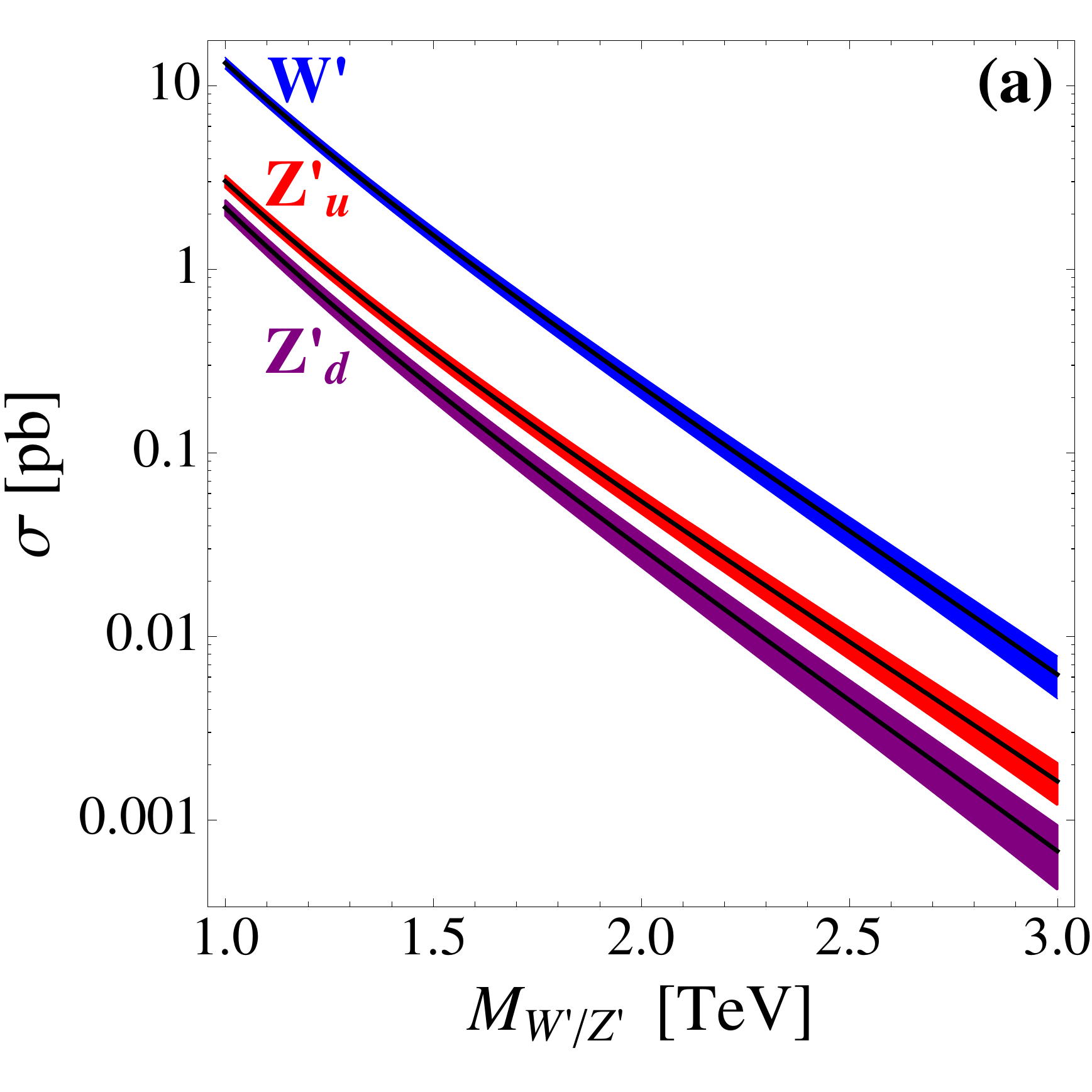}
\includegraphics[width=0.32\textwidth]{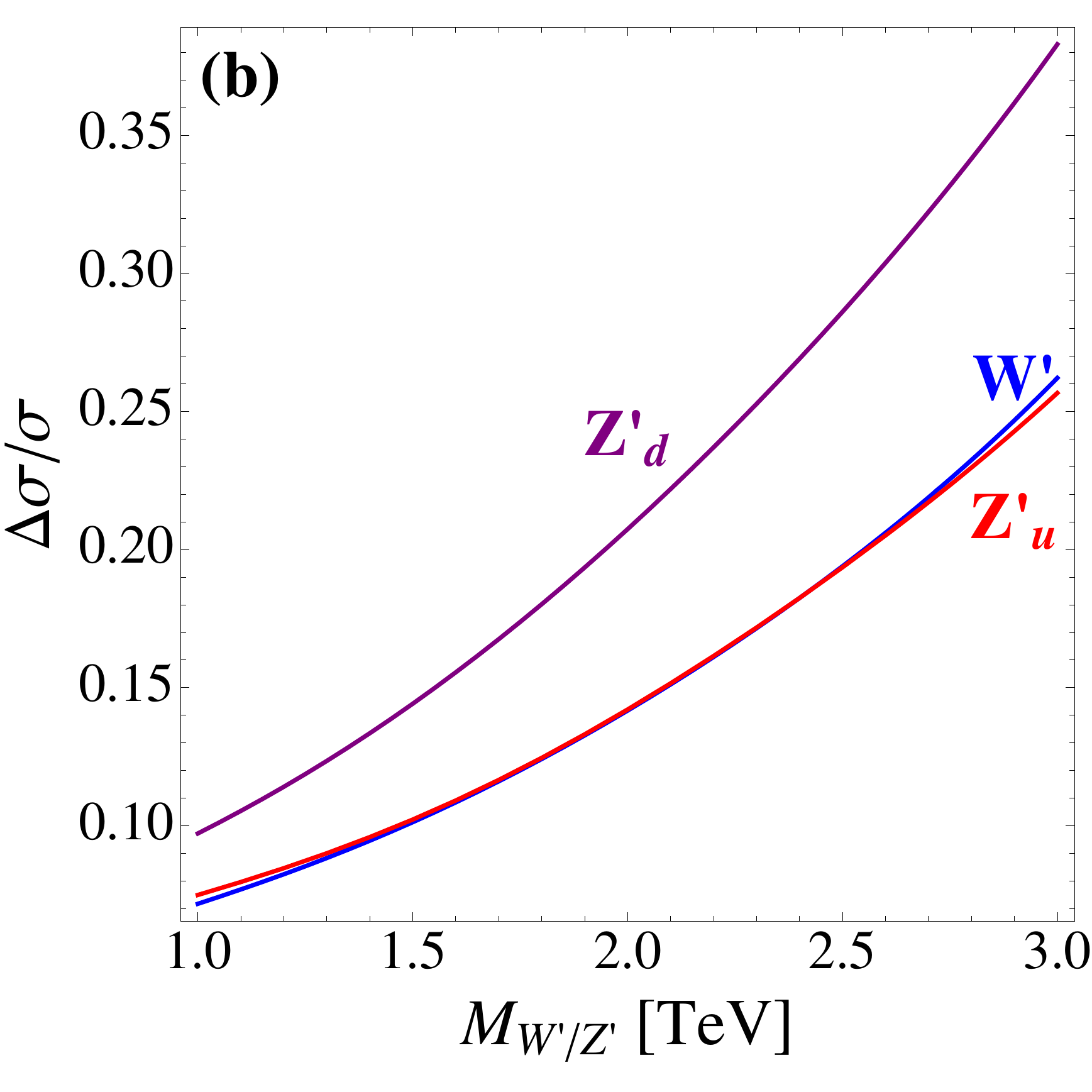}
\caption{\it The NLO cross section of $pp\to W^\prime/Z^\prime$ with a sequential coupling as a function of $M_{W^\prime/Z^\prime}$ calculated with the CT14 NNLO PDFs at LHC Run-1. (a) The PDF uncertainty bands and (b) the relative PDF uncertainties $\Delta \sigma/\sigma$ of $\sigma_{W^\prime}$ and $\sigma_{Z^\prime}^u$ and $\sigma_{Z^\prime}^d$, where $\sigma_{Z^\prime}^u$ and $\sigma_{Z^\prime}^d$ represent the cross sections induced by up-type and down-type quark initial states, respectively. 
}\label{unpdf}
\end{figure}

An accurate theory prediction of the cross section of $W^\prime$ and $Z^\prime$ productions is crucial for disentangling the NP signal from the SM backgrounds. 
We calculate the quantum chromodynamics (QCD) corrections to cross section of a sequential $W^{\prime}/Z^\prime$ boson production at the next-to-leading-order (NLO). For simplicity we set the renormalization scale ($\mu_R$) and the factorization scale ($\mu_F$) to be equal. The cross section exhibits two theoretical uncertainties: one is from the Parton Distribution Function (PDF), the other is from the choice of $\mu=\mu_R=\mu_F$. 
In this work we adapt the CT14 NNLO PDFs~\cite{Dulat:2015mca} to calculate the NLO QCD corrections to the cross section of a sequential $W^\prime/Z^\prime$ boson production $\sigma(W^\prime/Z^\prime)$.  The 57 sets of the CT14 NNLO PDFs are used to evaluate the PDF uncertainties.  Figure~\ref{unpdf} displays $\sigma(W^\prime/Z^\prime)$ as  a function of  $M_{W^\prime/Z^\prime}$. The default renormalization and factorization scales are chosen as the mass of extra gauge bosons $\mu_R=\mu_F=M_{W^\prime/Z^\prime}$. As a rule of thumb, we vary the scale $\mu$ by a factor of 2 to estimate the higher order corrections. The scale uncertainties are about 5\% in the $W^\prime$ and $Z^\prime$ production, which are found to be much smaller than the PDF uncertainties. We thus focus on the PDF uncertainties of $\sigma(W^\prime/Z^\prime)$. Figure~\ref{unpdf}(a) shows the NLO cross section of $pp \to W^\prime/Z^\prime$ and the corresponding PDF uncertainties denoted by the shaded band as a function of $M_{W^\prime/Z^\prime}$ at the LHC Run-1. In order to model the NP effects, we treat the up-type quark and down-type quark initial states separately in the $Z^\prime$ production; see the $Z^\prime_u$ and $Z^\prime_d$ bands. The relative uncertainties of PDFs are plotted in Fig.~\ref{unpdf}(b), which shows the uncertainties are about 10\% for $M_{W^\prime/Z^\prime}\sim {\rm TeV}$ and 30\% for $M_{W^\prime/Z^\prime} \sim 3~{\rm TeV}$.  Following Ref.~\cite{Berger:2009qy}, we fit the theory prediction of the cross section by  a simple three parameter analytic expression,
\bea
\log\left[\dfrac{\sigma(M_{V^\prime})}{\rm pb}\right]=A\left(\dfrac{M_{V^\prime}}{\rm TeV}\right)^{-1}+B+C\left(\dfrac{M_{V^\prime}}{\rm TeV}\right),
\eea
where $V^\prime=W^\prime/Z^\prime$. The cross sections are normalized to picobarn (pb) while $M_{W^\prime/Z^\prime}$ to TeV. The fitting functions of the production cross sections of $W^\prime$ and $Z^\prime$ are
\bea
W^{\prime}&\quad:\quad& 4.59925+1.34518x^{-1}-3.37137 x\nn\\
Z^{\prime}_u&:&2.82225+1.51681x^{-1}-3.24437 x \nn\\
Z^{\prime}_d&:&2.88763+1.42266x^{-1}-3.54818 x,
\eea
where $x=M_{W^\prime/Z^\prime}/{\rm TeV}$.

To explain the diboson excess of the ATLAS collaboration results, we consider a 2~TeV $W^\prime/Z^\prime$ boson in this work. The production cross sections of a sequential $W^\prime/Z^\prime$ boson at the LHC Run-1 are
\bea
\sigma_{W^\prime}^{SQ}  &=&229.67\pm 32.54 ~({\rm PDF})^{+12.54}_{-12.49}~ {\rm fb} ~({\rm scale}), \nn\\
\sigma_{Z^\prime u}^{SQ} &=&~54.50\pm ~~7.74 ~({\rm PDF})^{+2.87}_{-2.86} ~{\rm fb} ~({\rm scale}), \nn\\
\sigma_{Z^\prime d}^{SQ} &=&~30.25\pm ~~6.27 ~({\rm PDF})^{+1.71}_{-1.71}~ {\rm fb} ~({\rm scale}).\label{eq:sqwz}
\eea
The PDF uncertainties are $\sim 14\%$ for both $\sigma(W^\prime)$ and $\sigma(Z^\prime_u)$ while it is $\sim 21\%$ for $\sigma(Z^\prime_d)$. Using CT10 NLO PDFs~\cite{Lai:2010vv} slightly increases the PDF uncertainties. For example, the uncertainty of $\sigma(W^\prime)$ and $\sigma(Z^\prime_u)$ are $\sim 17\%$ and that of $\sigma(Z^\prime_d)$ is about $24\%$. In this work we choose the benchmark points shown in Eq.~\eqref{eq:sqwz} as a reference to calculate the production cross sections of $W^\prime$ and $Z^\prime$ in several NP models.  

\begin{figure}
\includegraphics[width=0.32\textwidth]{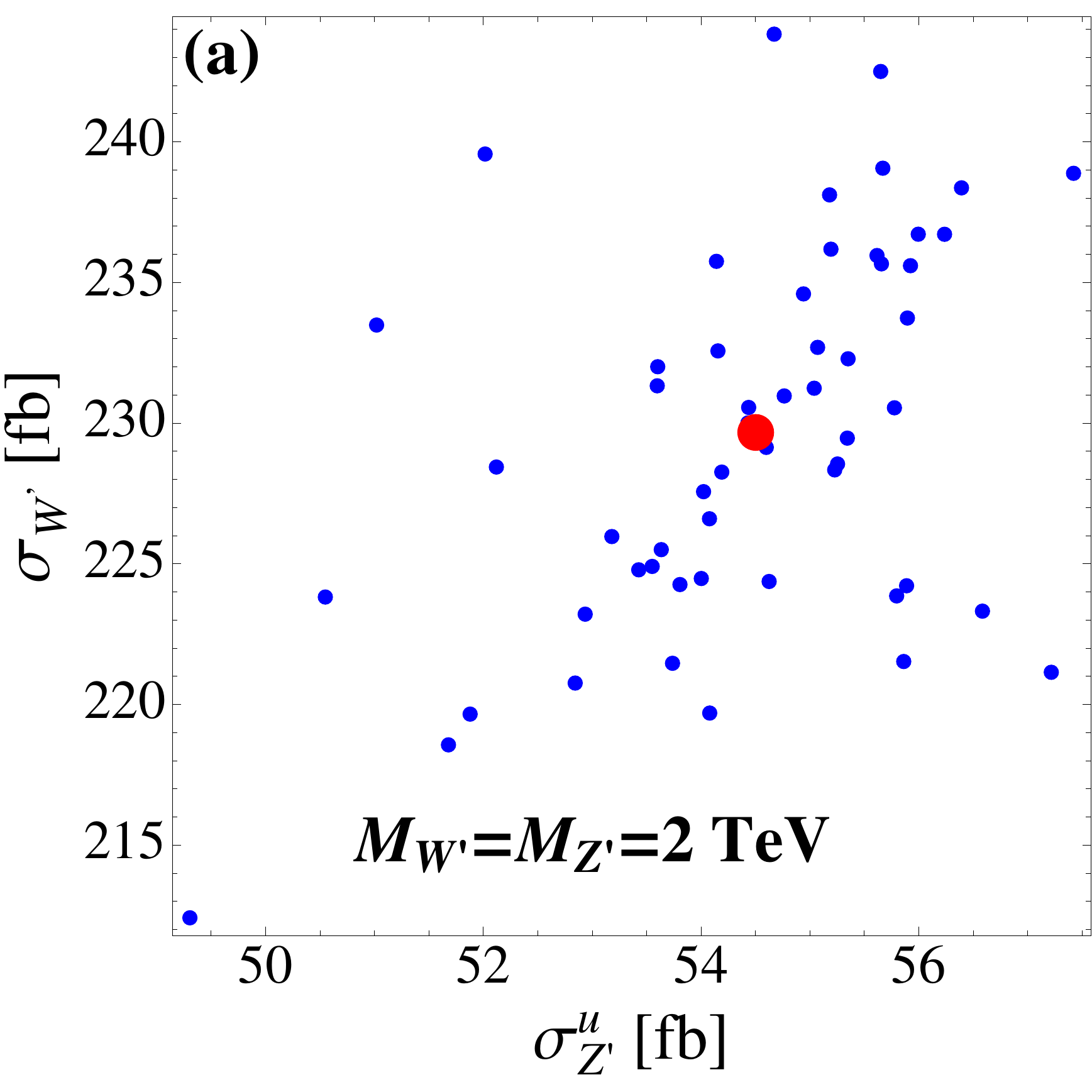}
\includegraphics[width=0.32\textwidth]{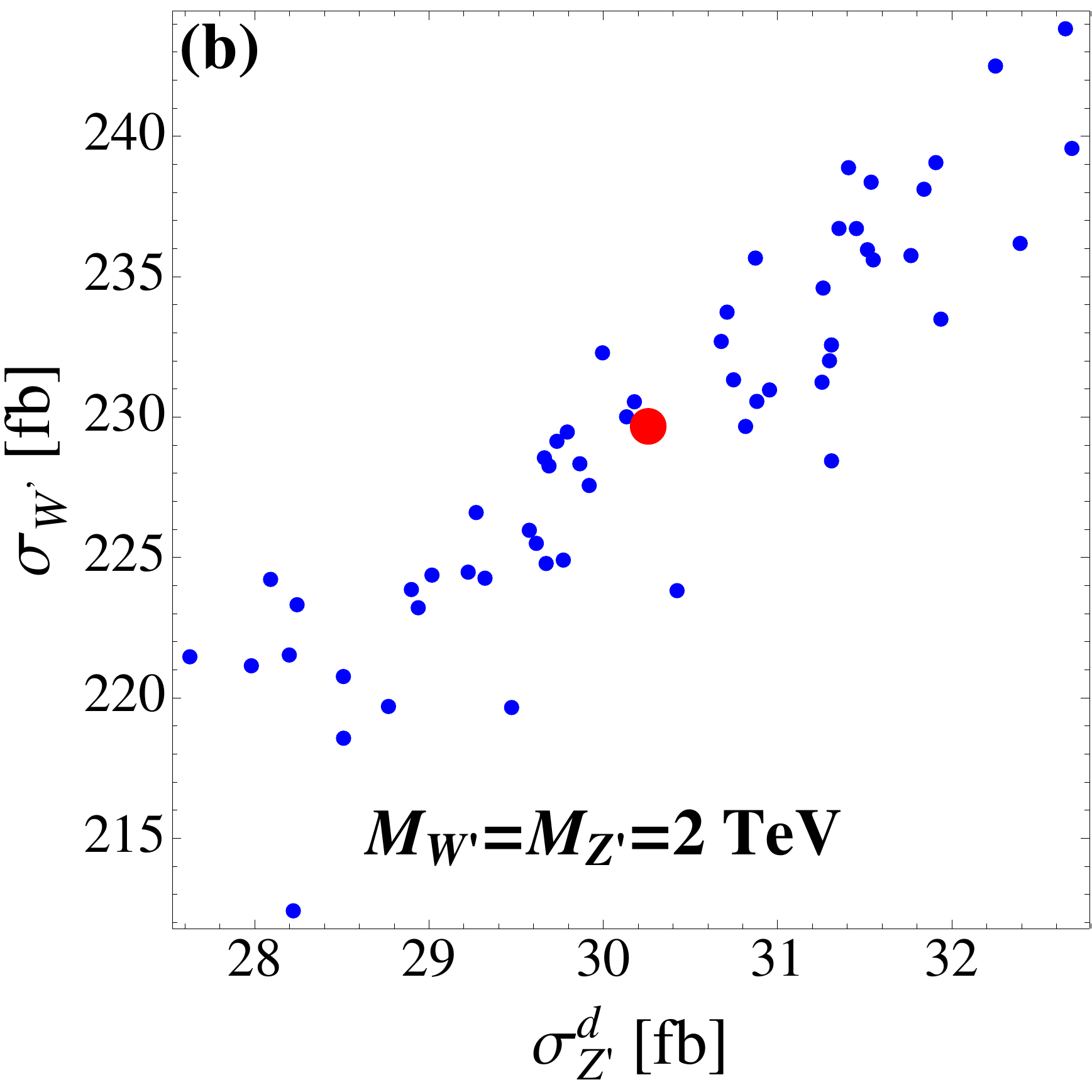}
\includegraphics[width=0.32\textwidth]{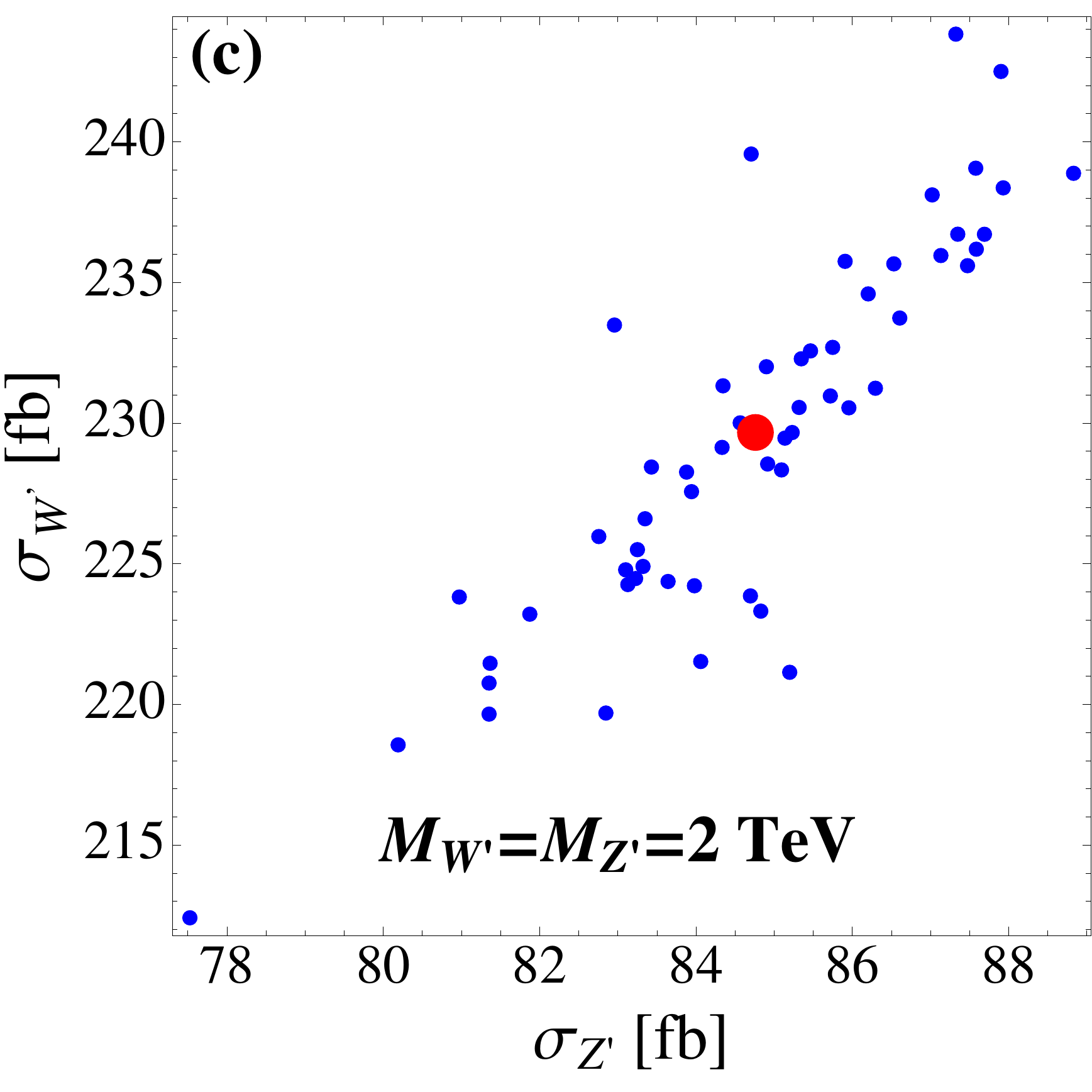}
\caption{\it $\sigma_{W^\prime}$ versus $\sigma_{Z^\prime_u}$(a), $\sigma_{Z^\prime_d}$ (b) and $\sigma_{Z^\prime}$ (c) for $M_{W^\prime}=M_{Z^\prime}=2~{\rm TeV}$. The blue point represents the cross sections calculated with 56 sets of PDFs while the red spot label the cross section evaluated with the central PDF. 
}\label{pdfcorr}
\end{figure}

As the $W^\prime$ and $Z^\prime$ are correlated in NP models with non-Abelian extension gauge structures, we explore the correlation between $\sigma(W^\prime)$ and $\sigma(Z^\prime_{u,d})$ for the 56 sets of CT14 NNLO PDFs.
Figure~\ref{pdfcorr} displays $\sigma(W^\prime)$ versus $\sigma(Z^\prime_u)$ (a) and $\sigma(Z^\prime_d)$ (b) at the LHC Run-1. The red point represents the cross section from the PDF set which the global fitting variables with central values, while the blue points denote the cross section from other PDF sets. The 56 PDF sets yield a correlation between $\sigma(W^\prime)$ and $\sigma(Z^\prime_d)$. On the other hand, the correlation is diluted in $\sigma(W^\prime)$ versus $\sigma(Z^\prime_u)$. In Fig.~\ref{pdfcorr}(c) we plot the production cross sections of the sequential $W^\prime$ and $Z^\prime$ boson, which exhibit a linear correlation. 

\section{$G(211)$ models: Breaking Pattern I}\label{221a}

We first consider several NP models exhibiting the first type symmetry breaking pattern. In the BP-I, $SU(2)_1$ is identified as the $SU(2)_L$ of the SM. The first stage of symmetry breaking 
$SU(2)_2 \times U(1)_{X}\to U(1)_{Y}$ occurs at the TeV scale, which 
could be induced by a scalar doublet field $\Phi\sim(1,2,1/2)$,
or a triplet scalar field $\Sigma\sim (1,3,1)$ with a vacuum expectation value
(VEV) $u$. The explicit form of the doublet and triplet as well as their vacuum expectation values are given as follows:
\begin{align}
&\Phi = \begin{pmatrix} \phi^+ \\ \phi^0\end{pmatrix},
&\VEV{\Phi} &=
\mfrac{1}{\sqrt{2}}
\begin{pmatrix} 0 \\ u \end{pmatrix}, \nn \\
&\Sigma=\frac{1}{\sqrt{2}}
\begin{pmatrix}\phi^{+} &\sqrt{2}\phi^{++} \\ \sqrt{2}\phi^{0} & -\phi^{+} \end{pmatrix},
&\VEV{\Sigma} &= \mfrac{1}{\sqrt{2}}
\begin{pmatrix}0 & 0 \\ u & 0 \end{pmatrix}.
\end{align}
The second stage of symmetry breaking $SU(2)_{L}\times U(1)_Y \to U(1)_{em}$ takes place at the electroweak scale. It is via another scalar field $H\sim(2,\bar{2},0)$ with two VEVs $v_1$ and $v_2$, which can be 
redefined as a VEV $v = \sqrt{v_1^2 + v_2^2}$ and a mixing
angle $\beta \equiv \arctan(v_1/v_2)$. The detailed form of $H$ and its VEV are 
\beq
H=\begin{pmatrix}h_1^{0} & h_1^{+} \\ h_2^{-} & h_2^{0} \end{pmatrix},\qquad
\VEV{H} = \mfrac{1}{\sqrt{2}}
\begin{pmatrix} v_1 & 0 \\ 0 & v_2 \end{pmatrix}.
\eeq 

We denote $g_1$, $g_2$ and $g_X$ as the coupling of $SU(2)_1$, $SU(2)_2$
and $U(1)_X$, respectively. In the BP-I, the three couplings are
\begin{eqnarray}
g_{1}=\frac{e}{s_{W}},\quad g_{2}=\frac{e}{c_{W}s_{\phi}},\quad g_{X}=\frac{e}{c_{W}c_{\phi}},
\end{eqnarray}
where $s_{W}$ and $c_W$ are sine and cosine of the SM weak mixing angle, 
while $s_{\phi}$ and $c_{\phi}$ are sine and cosine of the new mixing angle 
$\phi\equiv\arctan(g_X/g_2)$ appearing after the TeV symmetry breaking.
After symmetry breaking both $W^\prime$ and $Z^\prime$ bosons obtain masses
and mix with the SM gauge bosons. Different electroweak symmetry breaking (EWSB) patterns will induce different $W^\prime$ and $Z^\prime$ mass relations.
When the first stage breaking of BP-I is realized by the doublet $\Phi$,  the masses of the $W^\prime$ and $Z^\prime$ are 
\bea
M_{{W^{\prime}}^{\pm}}^{2} = \frac{e^{2}v^{2}}{4c_{W}^{2}s_{\phi}^{2}}\left(x+1\right)~,~
M_{Z^{\prime}}^{2} & = & \frac{e^{2}v^{2}}{4c_{W}^{2}s_{\phi}^{2}c_{\phi}^{2}}\left(x+c_{\phi}^{4}\right),
\label{mzp_bp1}
\eea
where $x=u^2/v^2$. Note that the precision data constraints (including those from CERN LEP and SLAC SLC experiment data) pushed the TeV symmetry breaking
higher than 1~TeV. Therefore, we assume $x$ is much larger than 1 and approximate the predictions of physical observables by taking Taylor expansion in $1/x$. As a result, the masses of $W^\prime$ and $Z^\prime$ are almost degenerated in the region of $c_\phi\sim 1$. 

If the symmetry breaking is realized  by the triplet $\Sigma$, the $Z^\prime$ mass is much larger than the $W^\prime$ mass
\bea
M_{{W^{\prime}}^{\pm}}^{2} = \frac{e^{2}v^{2}}{4c_{W}^{2}s_{\phi}^{2}}\left(2x+1\right)~,~
M_{Z^{\prime}}^{2} & = & \frac{e^{2}v^{2}}{4c_{W}^{2}s_{\phi}^{2}c_{\phi}^{2}}\left(4x+c_{\phi}^{4}\right).
\label{mzp_bp1t}
\eea
The recent discovered excesses occur around $M_{W^\prime} \simeq M_{Z^\prime} \sim 2~{\rm TeV}$~\cite{Aad:2015owa}. That leads us to focus on the doublet model throughout this work. The triplet model is studied in Ref.~\cite{Gao:2015irw}

After the second stage of symmetry breaking at the electroweak scale, a non-abelian coupling of the $W^\prime$ and $Z^\prime$ to the SM bosons are generated as follows:
\bea
H~W_{\nu}~W_{\rho}^{\prime}  &\quad:\quad&  -\frac{1}{2}\frac{e^{2}s_{2\beta}}{c_{W}s_{W}s_{\phi}}vg_{\nu\rho}\biggl[1+\frac{\left(c_{W}^{2}s_{\phi}^{2}-s_{W}^{2}\right)}{xs_{W}^{2}}\biggr], \nn\\
H~Z_{\nu}~Z_{\rho}^{\prime} &:& -\frac{1}{2}\frac{e^{2}c_{\phi}}{c_{W}^{2}s_{W}s_{\phi}}vg_{\nu\rho}\biggl[1-\frac{c_{\phi}^{2}\left(c_{\phi}^{2}s_{W}^{2}-s_{\phi}^{2}\right)}{xs_{W}^{2}}\biggr], \nn\\
W_{\mu}^{+}~W_{\nu}^{\prime-}~Z_{\rho}&:& \frac{es_{2\beta}s_{\phi}}{xs_{W}^{2}}, \nn\\
W_{\mu}^{+}~W_{\nu}^{-}~Z_{\rho}^{\prime} &:& \frac{es_{\phi}c_{W}c_{\phi}^{3}}{xs_{W}^{2}},
\eea
where the Lorentz index $\left[g^{\mu\nu}(k_{1}-k_{2})^{\rho}+g^{\nu\rho}(k_{2}-k_{3})^{\mu}+g^{\rho\mu}(k_{3}-k_{1})^{\nu}\right]$
in the three gauge boson couplings is implied.

The detailed expressions of the partial decay widths of $W^\prime/Z^\prime$  are listed in the Appendix. The equivalence theorem tells us that one can treat the final state vector bosons as Nambu-Goldstone bosons in the high energy limit. We compare the bosonic decay of $W^\prime/Z^\prime$ in the limit of $x\gg 1$ and $M_{W^\prime/Z^\prime}\gg m_{W/Z/H}$ and verify in the BP-I that
\beq
\frac{{\rm BR}(W^\prime \to WZ)}{{\rm BR}(W^\prime \to WH)} \sim 1\quad,\quad \frac{{\rm BR}(Z^\prime \to WW)}{{\rm BR}(Z^\prime \to ZH)} \sim 1. 
\eeq
It is worth mentioning that the $WH$ mode might be suppressed in an UV completion model which exhibits a rather complicated scalar potential.  

The couplings of the $W^\prime$ bosons to the SM fermions in the notation in Eq.~\eqref{eq:effcoup} are 
\bea
g_L^{W^\prime\bar{f}f^\prime} &=& -\frac{e}{\sqrt{2}s_{W}^{2}}\gamma_{\mu}\frac{c_{W}s_{2\beta}s_{\phi}}{x},\nn\\  
g_R^{W^\prime\bar{f}f^\prime} &=& \frac{e}{\sqrt{2}c_{W}s_{\phi}}\gamma_{\mu},
\eea
while those of the $Z^\prime$ boson are 
\bea 
g_L^{Z^{\prime}\bar{f}f} &=&  
\frac{e}{c_{W}c_{\phi}s_{\phi}}\gamma_{\mu}\left[\left(T_{3}^1-Q\right)s_{\phi}^{2}-\frac{c_{\phi}^{4}s_{\phi}^{2}\left(T_{3}^1-Qs_{W}^{2}\right)}{xs_{W}^{2}}\right],\nn\\
g_R^{Z^{\prime}\bar{f}f} &=&\frac{e}{c_{W}c_{\phi}s_{\phi}}\gamma_{\mu}\left[\left(T_{3}^2-Qs_{\phi}^{2}\right)+Q\frac{c_{\phi}^{4}s_{\phi}^{2}}{x}\right],\label{zpf}
\eea
where $T_3^1$ and $T_3^2$ are the third components of the generator of gauge groups $SU(2)_1$ and $SU(2)_2$, and $Q$ is the  electric charge of fermion $f$.

Next we consider specific NP models and discuss their implications in the production of $W^\prime/Z^\prime$ and their decay modes of the $WZ/WW$ pair at the LHC. 

\subsection{Left-Right doublet model}

\subsubsection{The $W^\prime$ constraints}

\begin{figure}
\includegraphics[width=0.32\textwidth]{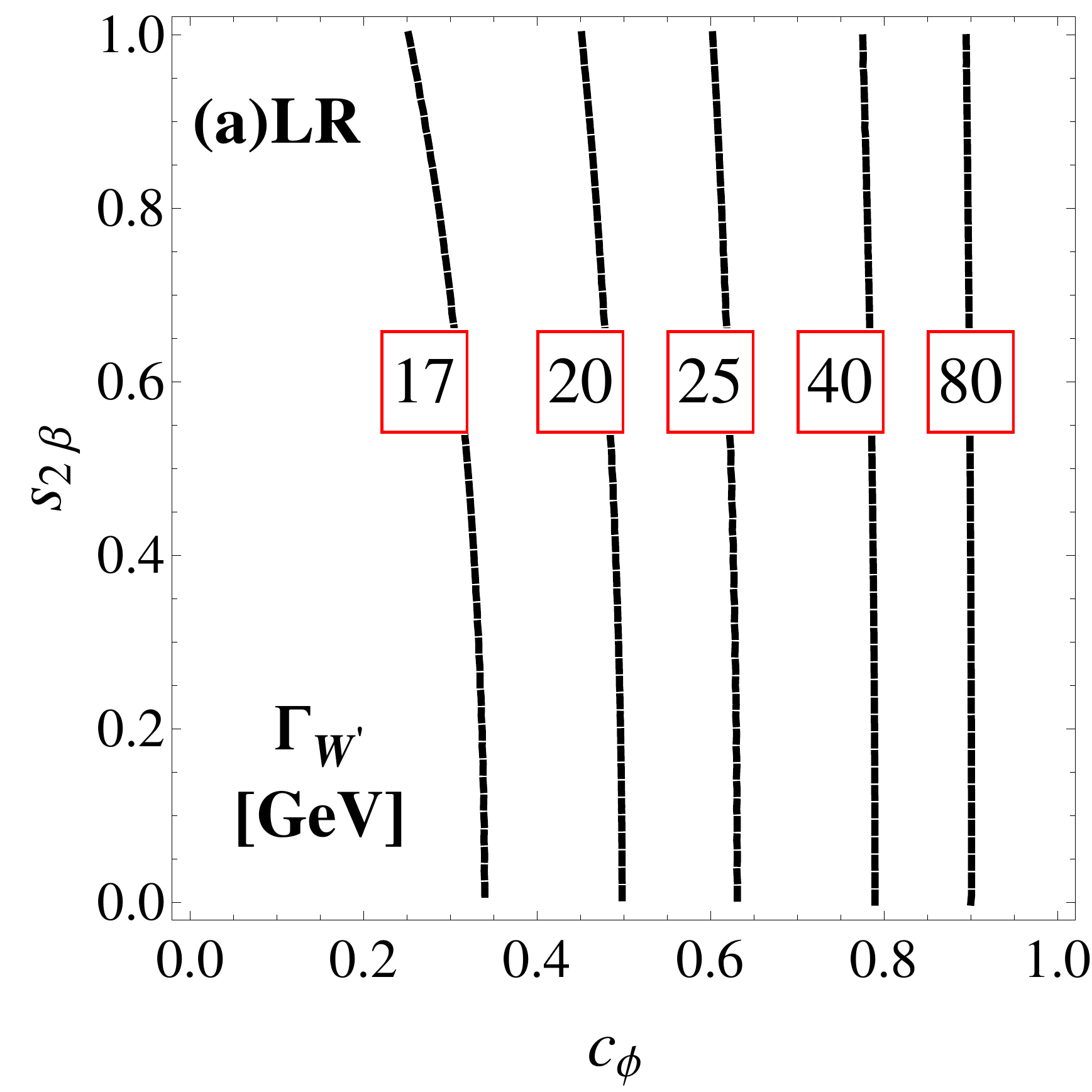}
\includegraphics[width=0.32\textwidth]{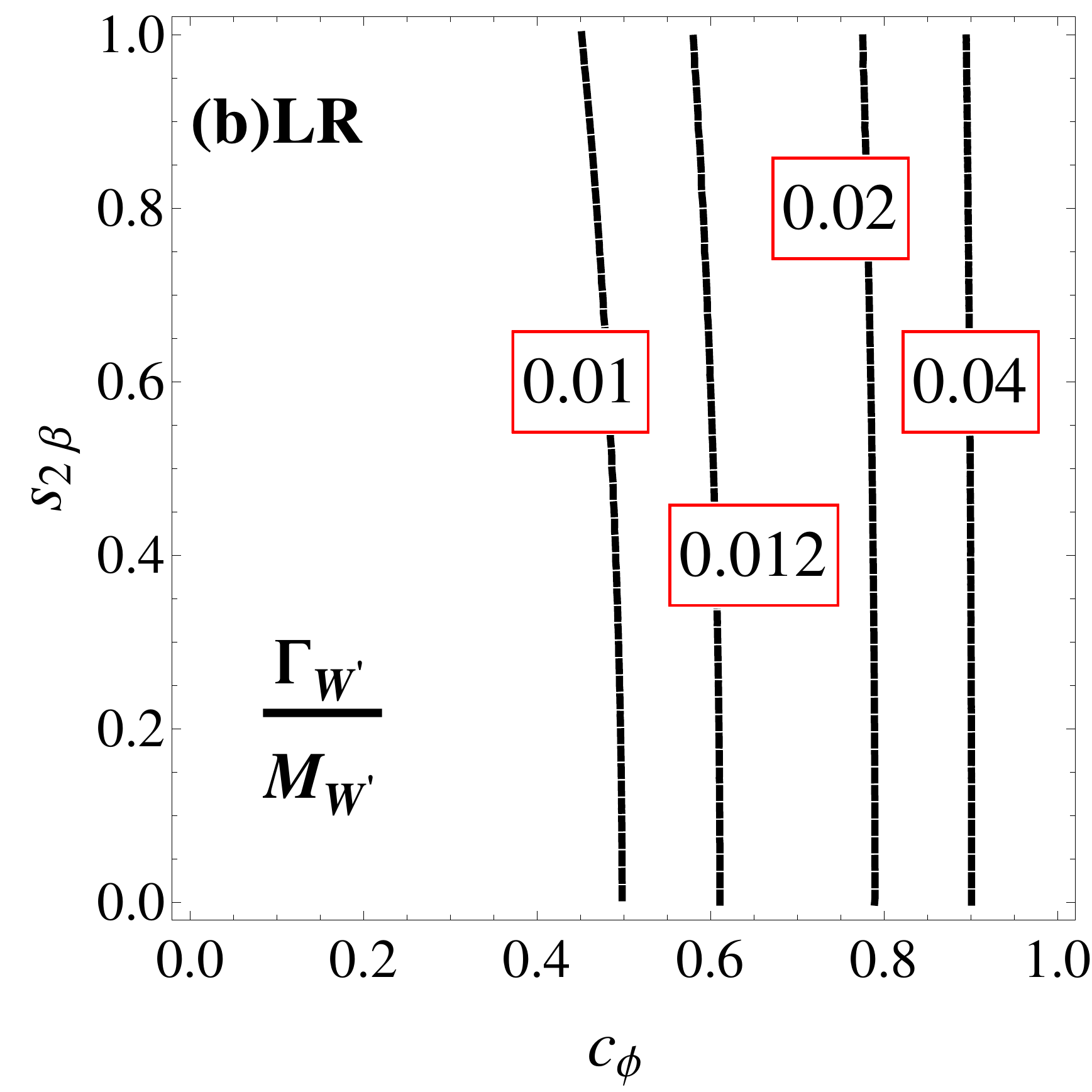}
\caption{\it The contours of the total width of $W^{\prime}$ (a) and the ratio of total width and mass of $W^{\prime}$ (b) in the plane of $c_\phi$ and $s_{2\beta}$ in the Left-Right model. 
}\label{LRw1}
\end{figure}

We begin with the Left-Right model in which the left-handed and right-handed fermion doublets are gauged under $SU(2)_1$ and $SU(2)_2$, respectively. 
Figure~\ref{LRw1} displays the contour of the total width $\Gamma_{W^\prime}$ and the ratio $\Gamma_{W^\prime}/M_{W^\prime}$ in the plane of $c_\phi$ and $s_{2\beta}$.  It is clear that $\Gamma_{W^\prime}\ll M_{W^\prime}$ in all of the parameter space such that it is reasonable to factorize the $\sigma(pp\rightarrow V^\prime\rightarrow V_1 V_2)\equiv\sigma(V^\prime)\times BR(V^\prime\rightarrow V_1 V_2)$. The ratio $\Gamma_{W^\prime}/M_{W^\prime}$ depends on $c_\phi$ mildly but it is not sensitive to $s_{2\beta}$.  Note that $s_{2\beta}$ appears only in the left-handed couplings of $W^\prime$ to the SM fermions which is suppressed by $x$. On the other hand, the right-handed coupling of $W^\prime$ depends only on $c_\phi$.  

\begin{figure}
\includegraphics[width=0.32\textwidth]{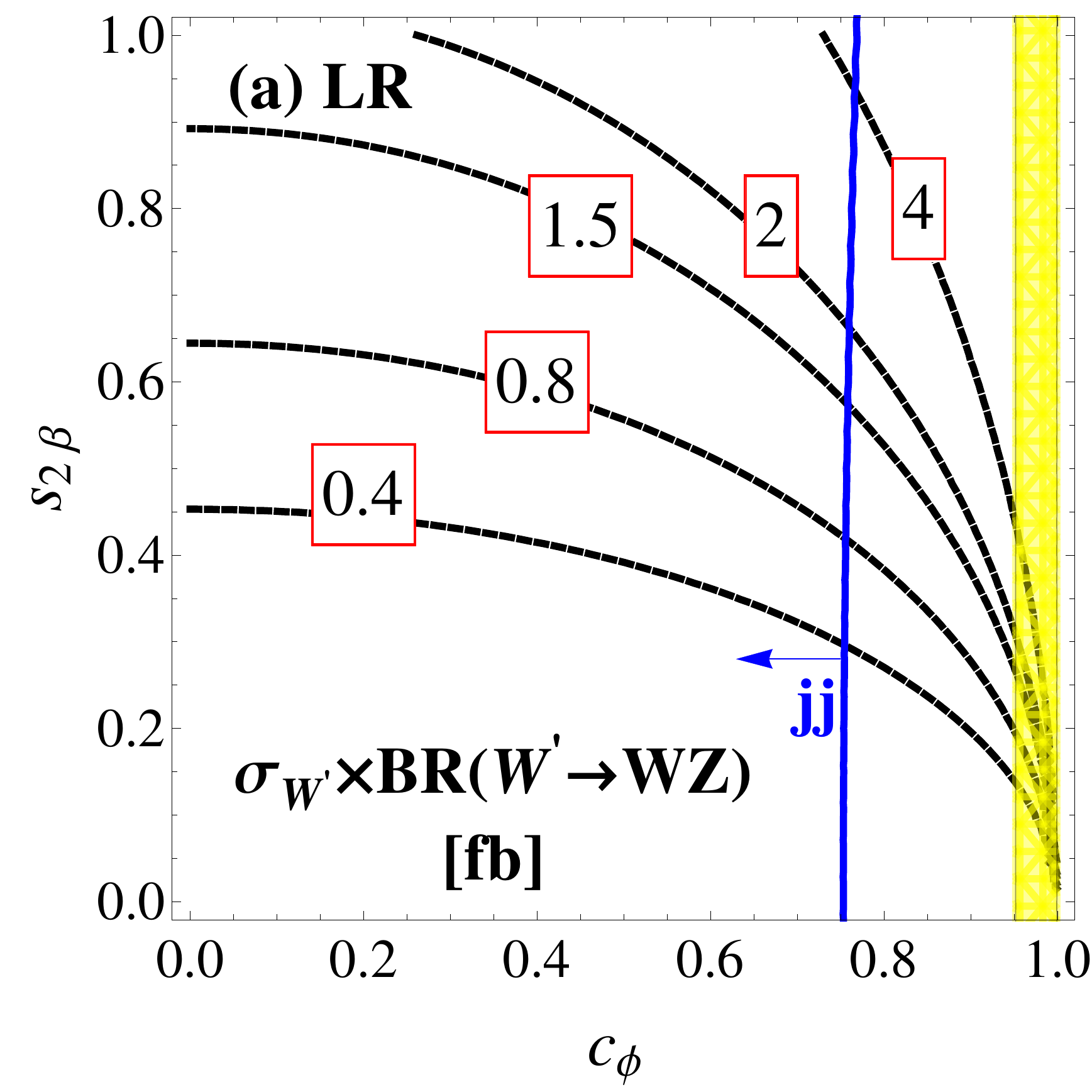}
\includegraphics[width=0.32\textwidth]{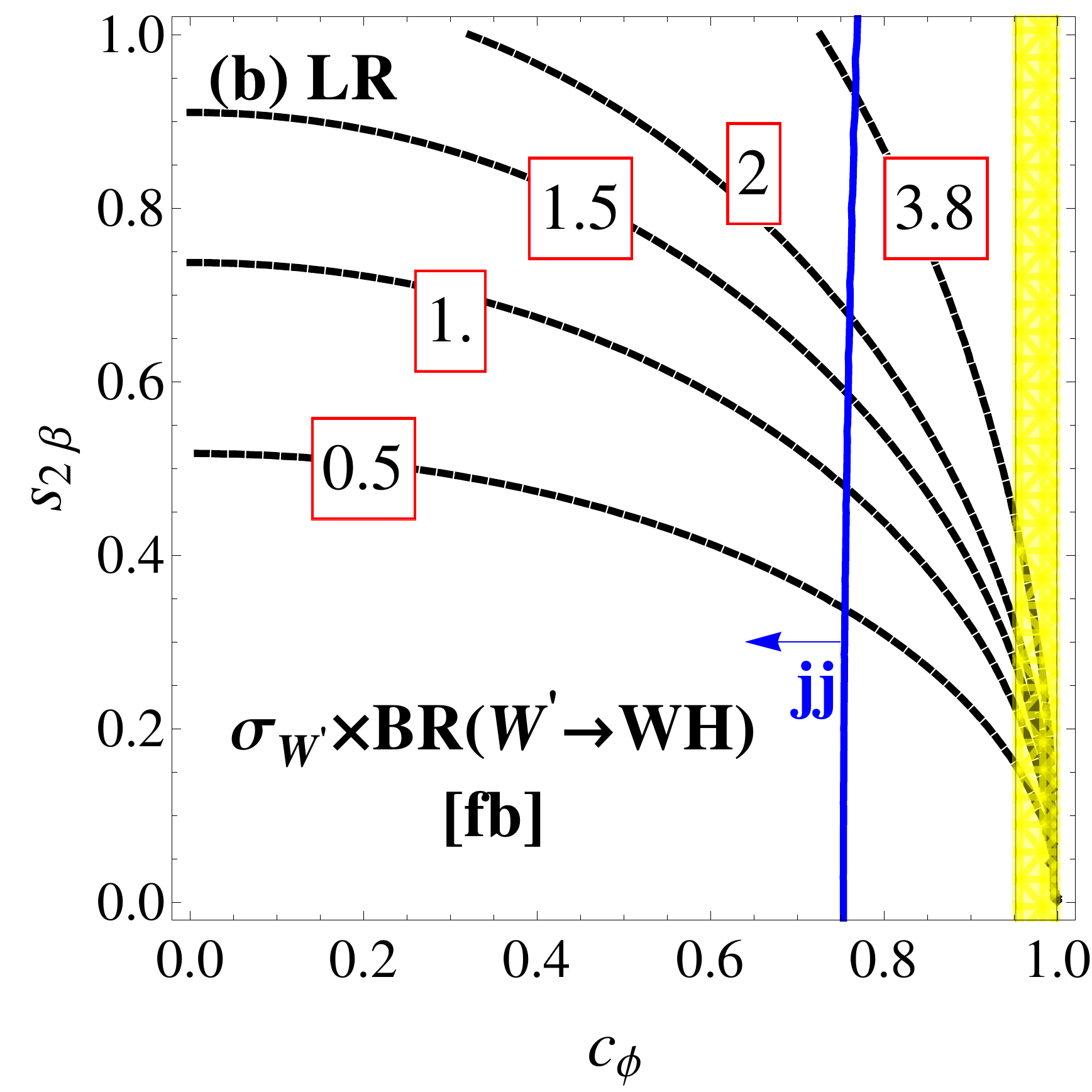}\\
\includegraphics[width=0.32\textwidth]{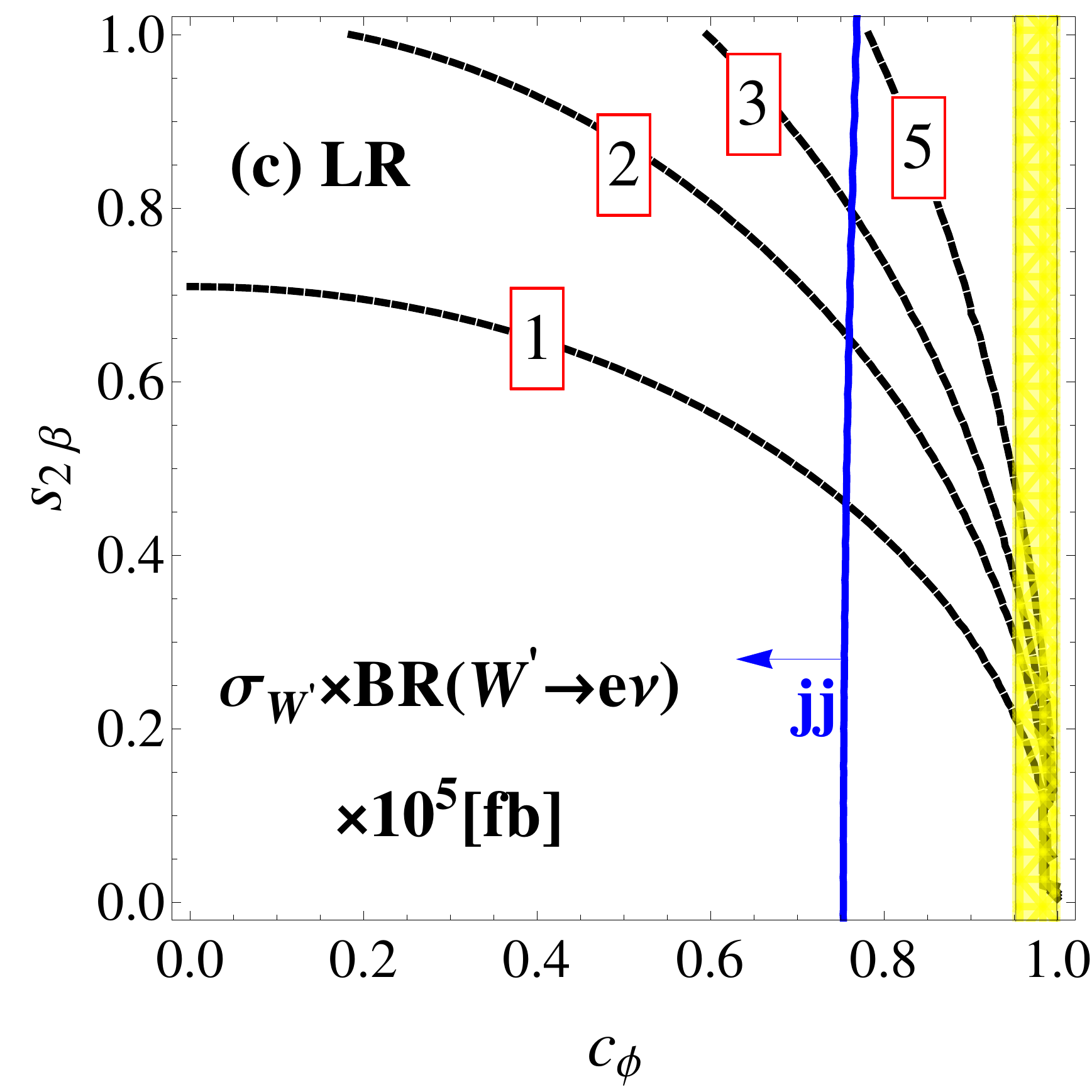}
\includegraphics[width=0.32\textwidth]{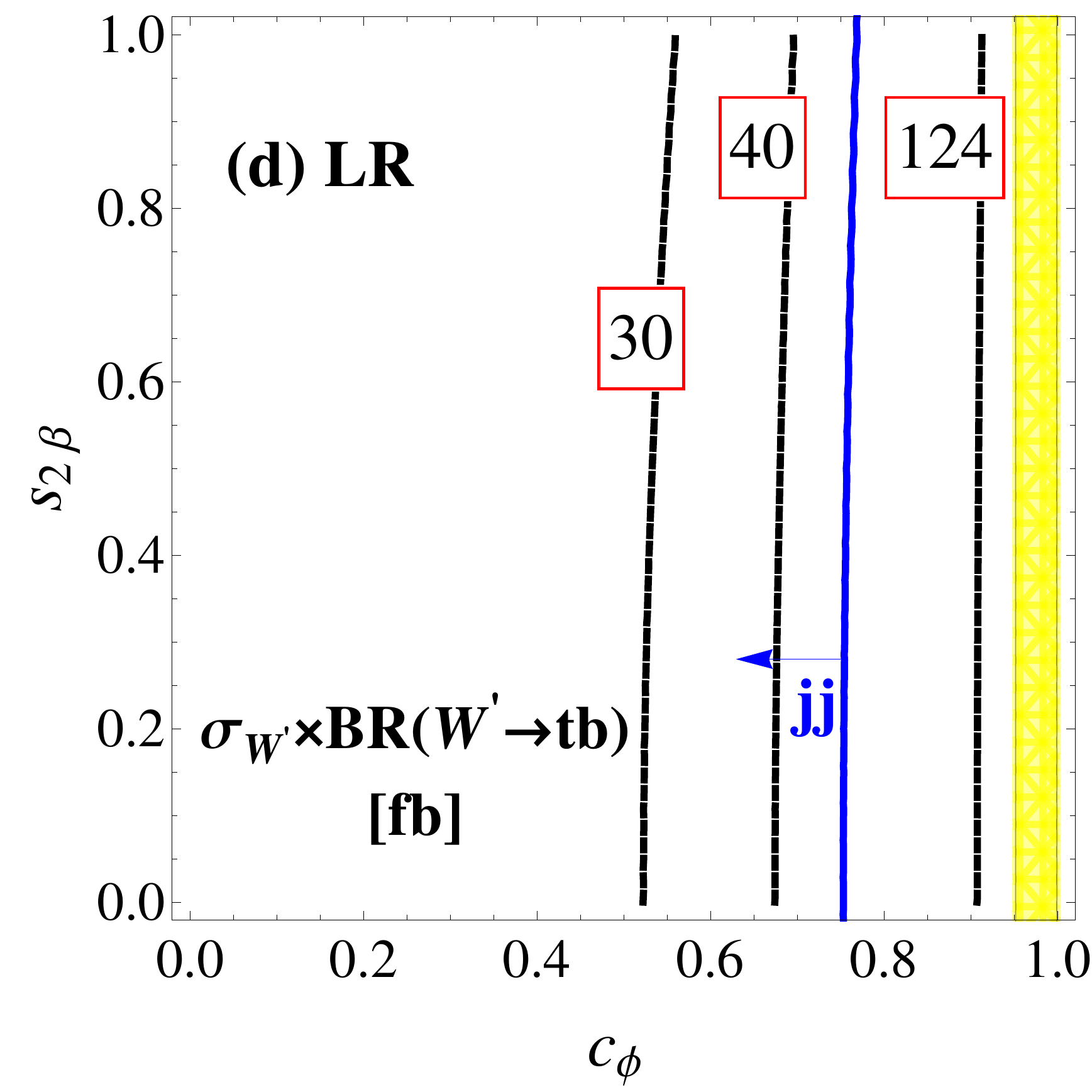}
\caption{\it The contours of the cross section (a) $\sigma(W^\prime) \times {\rm BR}(W^\prime \to WZ)$, (b) $\sigma(W^\prime)\times {\rm BR}(W^\prime \to WH)$, (c) $\sigma(W^\prime)\times {\rm BR}(W^\prime \to e\nu)$ and (d) $\sigma(W^\prime)\times {\rm BR}(W^\prime \to tb)$ in the plane of $c_\phi$ and $s_{2\beta}$. The vertical line ($jj$) denotes the constraint from the di-jet measurements. The yellow band represents the degenerated mass region of $W^\prime$ and $Z^\prime$.
}\label{LRw2}
\end{figure}

Figure~\ref{LRw2}(a) displays the contour of the cross section of $\sigma(W^\prime)\times {\rm BR}(W^\prime \to WZ)$ in the plane of $c_\phi$ and $s_{2\beta}$. The yellow bands represent the degenerated region of $M_{W^\prime}$ and $M_{Z^\prime}$. In order to produce $\sigma(WZ) \sim 4-8~{\rm fb}$ and $\sigma(W^\prime)\times {\rm BR}(W^\prime \to jj)\leq 102~{\rm fb}$~\cite{Khachatryan:2015sja}, one needs $0.73<c_\phi<0.75$ and $s_{2\beta}\gtrsim0.9$.

In accord to the equivalence theorem, the vector-boson pair production is highly correlated with the associated production of the vector boson and Higgs boson. We also plot in Fig.~\ref{LRw2}(b) the contour of the cross section of $\sigma({W^\prime})\times {\rm BR}(W^\prime \to WH)$ in the plane of $c_\phi$ and $s_{2\beta}$. In the vicinity of $c_\phi \sim 0.73$ and $s_{2\beta}\sim 0.9$, $\sigma(W^\prime) \times {\rm BR}(W^\prime \to WH) \sim 3~{\rm fb}$ which is below the current experimental limit of $\sigma(W^\prime) \times {\rm BR}(W^\prime \to WH)<7.1~{\rm fb}$~\cite{Khachatryan:2015bma}. 

The cross section of $\sigma(W^\prime)\times {\rm BR}(W^\prime \to e\nu)$ is shown in Fig.~\ref{LRw2}(c), which satisfies the current experimental upper limit $\sigma(pp\to W^\prime\to e\nu/\mu\nu)\leq 0.7~{\rm fb}$ in the whole parameter space. The current bound on the $tb$ mode demands $c_\phi<0.91$;  see Fig.~\ref{LRw2}(d).

\begin{figure}
\includegraphics[width=0.32\textwidth]{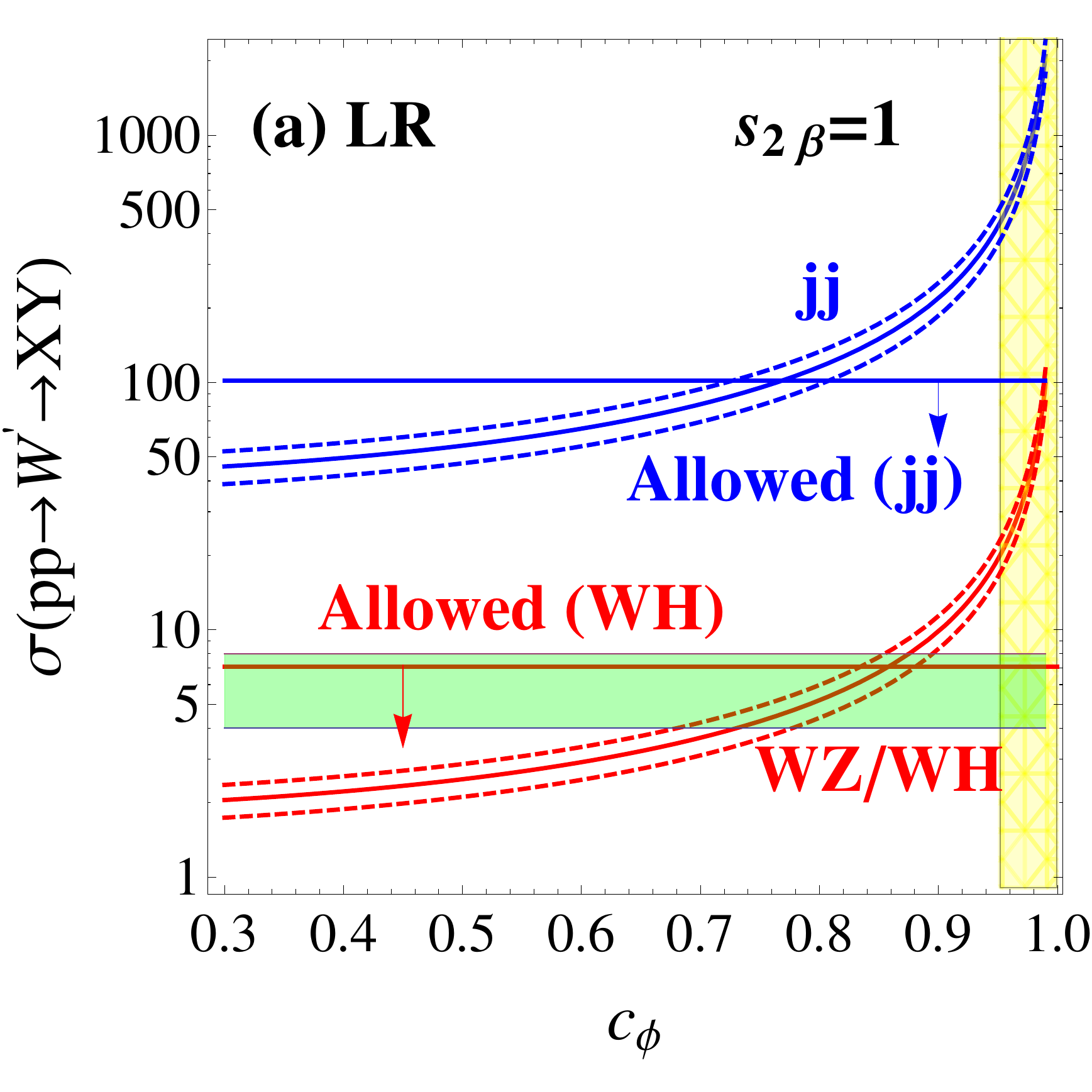}
\caption{\it The cross section of $pp \to W^\prime \to WZ/WH$  (red curves) and $pp \to W^\prime \to jj$ (blue curves) as a function of $c_\phi$ with $s_{2\beta}=1$. The dashed curves represent the PDF uncertainties. The green shaded region represents the parameter space compatible with the $WZ$ excess.  The yellow shaded region is required for $M_{W^\prime} \simeq M_{Z^\prime}$. The current experimental limits of $\sigma(pp \to W^\prime \to jj) < 102~{\rm fb}$ and $\sigma(pp \to W^\prime \to WH)<7.1~{\rm fb}$ are also plotted.  }
\label{LRwPDF}
\end{figure} 

In Fig.~\ref{LRwPDF} we present the cross section $\sigma(W^\prime) \times {\rm BR}(W^\prime \to XY)$ as a function of $c_\phi$, where $X$ and $Y$ denote the SM particles in the $W^\prime$ decay.
To see the maximally allowed region of $c_\phi$, we consider the PDF uncertainties of the production cross section of $W^\prime$ and choose $s_{2\beta}=1$. The outer dashed-curves represent the PDF uncertainties. The green shaded region represents the parameter space compatible with the $WZ$ excess.  The yellow shaded region is required for $M_{W^\prime} \simeq M_{Z^\prime}$. The current experimental limits of $\sigma(pp \to W^\prime \to jj) < 102~{\rm fb}$
 and $\sigma(pp \to W^\prime \to WH)<7.1~{\rm fb}$ are also plotted. The parameter space of $0.68<c_\phi<0.81$ can explain $WZ$ excess and the current experimental upper limits of $WH$ and $jj$. However, it predicts $2.47~{\rm TeV}<M_{Z^\prime}<2.94~{\rm TeV}$ which is in contradiction with the $WW$ excess around 2~TeV. 
If further experiments confirm that the $WW$ excess is owing to a fluctuation of the SM backgrounds, then the $W^\prime$ in the Left-Right model could explain the $WZ$ excess.

\subsubsection{The $Z^\prime$ constraints}

\begin{figure}
\includegraphics[width=0.32\textwidth]{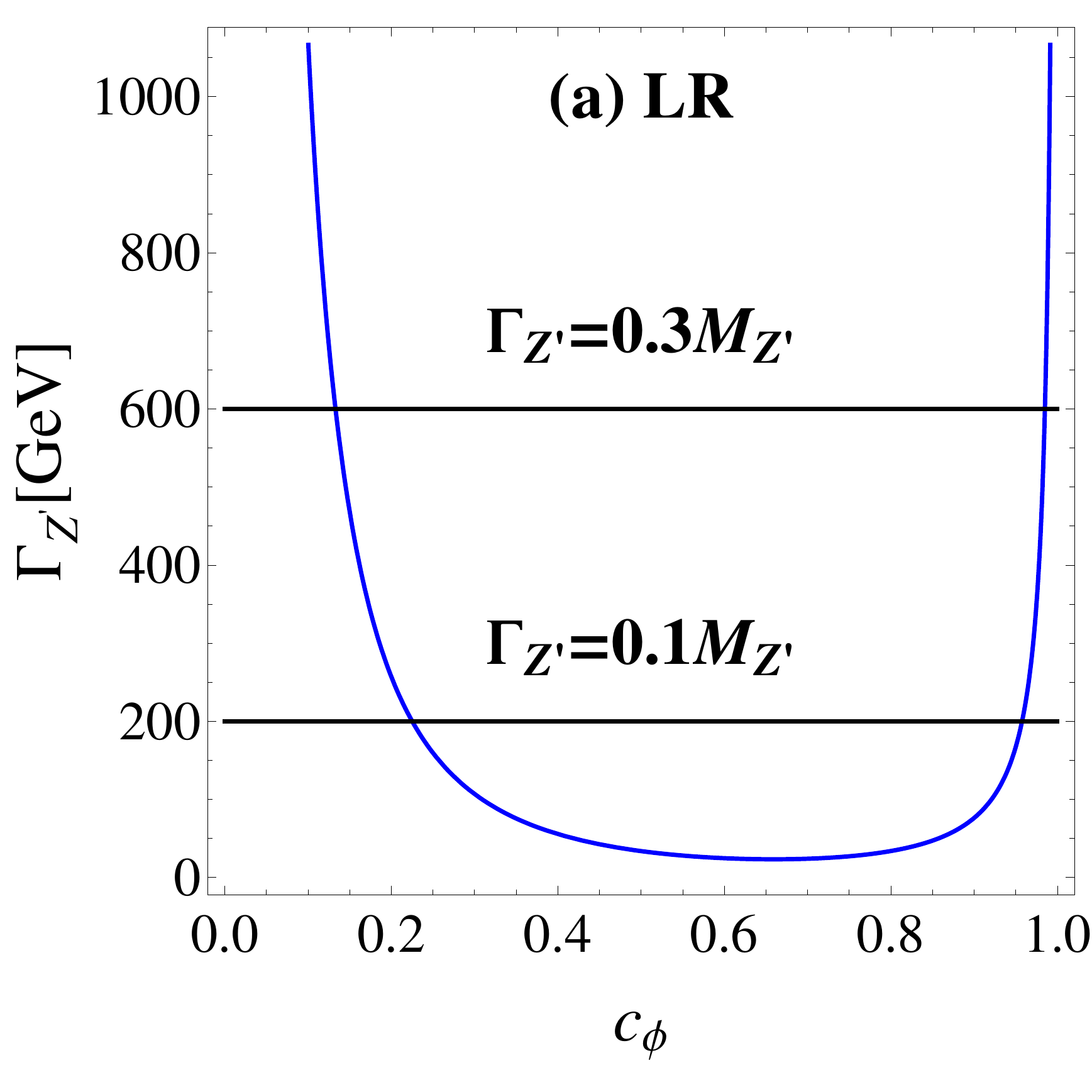}
\includegraphics[width=0.32\textwidth]{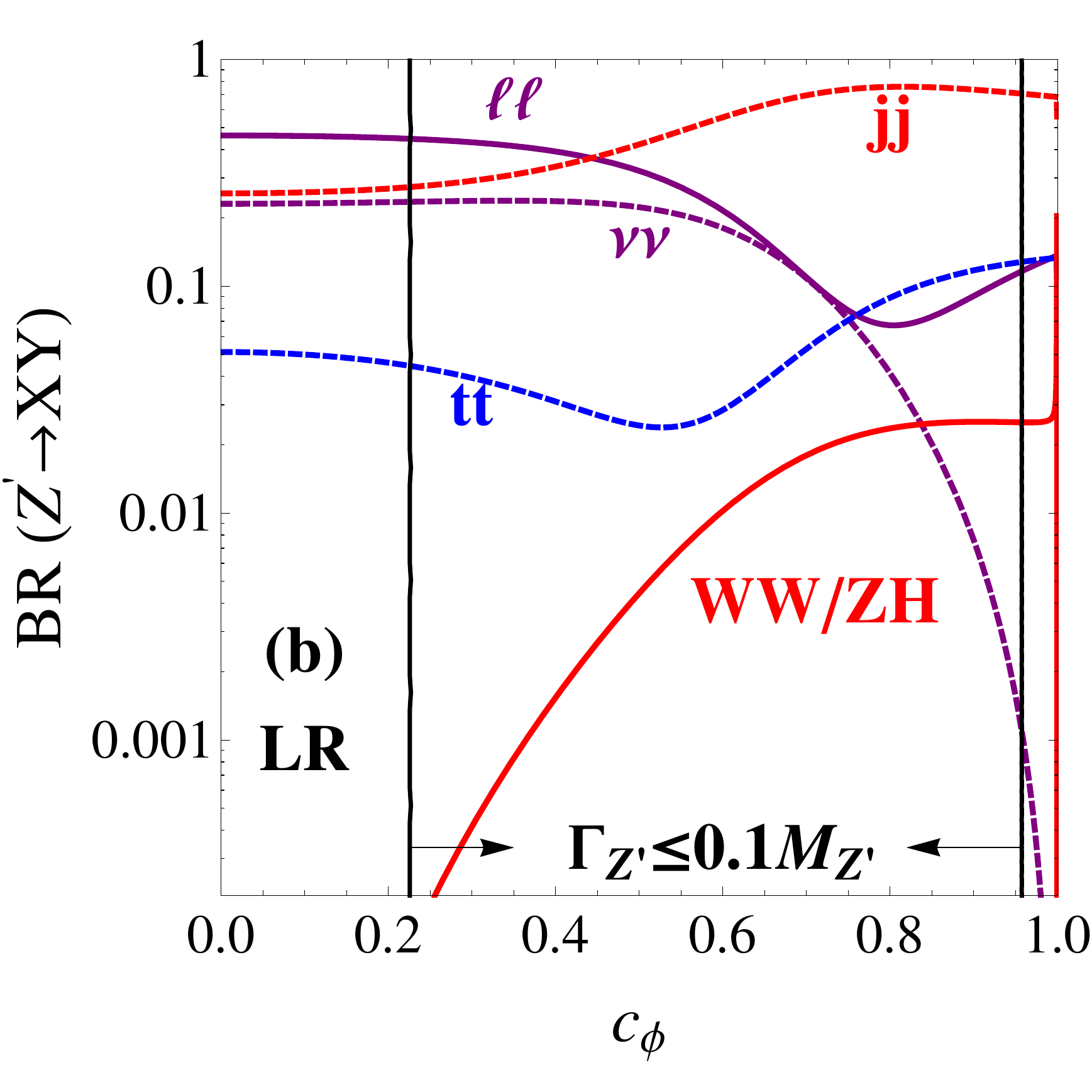}
\caption{\it The total width (a)  and the branching ratios of all the decay modes (b) of $Z^\prime$ as a function of $c_{\phi}$.  The $jj$ mode includes all the light quark flavors ($u,d,c,s,b$),  the $tt$ mode denotes the top-quark pair final state, the $\ell\ell$ mode sums over the charged leptons while the $\nu\nu$ mode sums over all the three neutrino final states. }
\label{LRzp1}
\end{figure}

The coupling of $Z^\prime$ to the SM fermions is very sensitive to the mixing angle $\phi=\arctan(g_X/g_2)$. In the limit of $x\gg 1$, $g_{L/R}^{Z^\prime \bar{f}f}  \sim 1 /s_\phi c_\phi$. 
The couplings tend to be non-perturbative in the region of $c_\phi\sim 0$ or $c_\phi \sim 1$, yielding a large decay width of $Z^\prime$; see the Fig.~\ref{LRzp1}(a). We demand $\Gamma(Z^\prime)\leq 0.1 M_{Z^\prime}$ in this work, which requires $0.23 \leq c_\phi \leq 0.96$. Figure~\ref{LRzp1}(b) displays the branching ratios of all the decay modes of $Z^\prime$. The $jj$ mode includes all the light quark flavors ($u,d,c,s,b$), the $\ell\ell$ mode sums over the charged leptons while the $\nu\nu$ mode sums over all the three neutrino final states. We single out the top-quark pair mode ($tt$) to compare to the latest experimental data.  The $WW$ and $ZH$ modes are much smaller than other modes; see the red-solid curve. 

\begin{figure}
\includegraphics[width=0.32\textwidth]{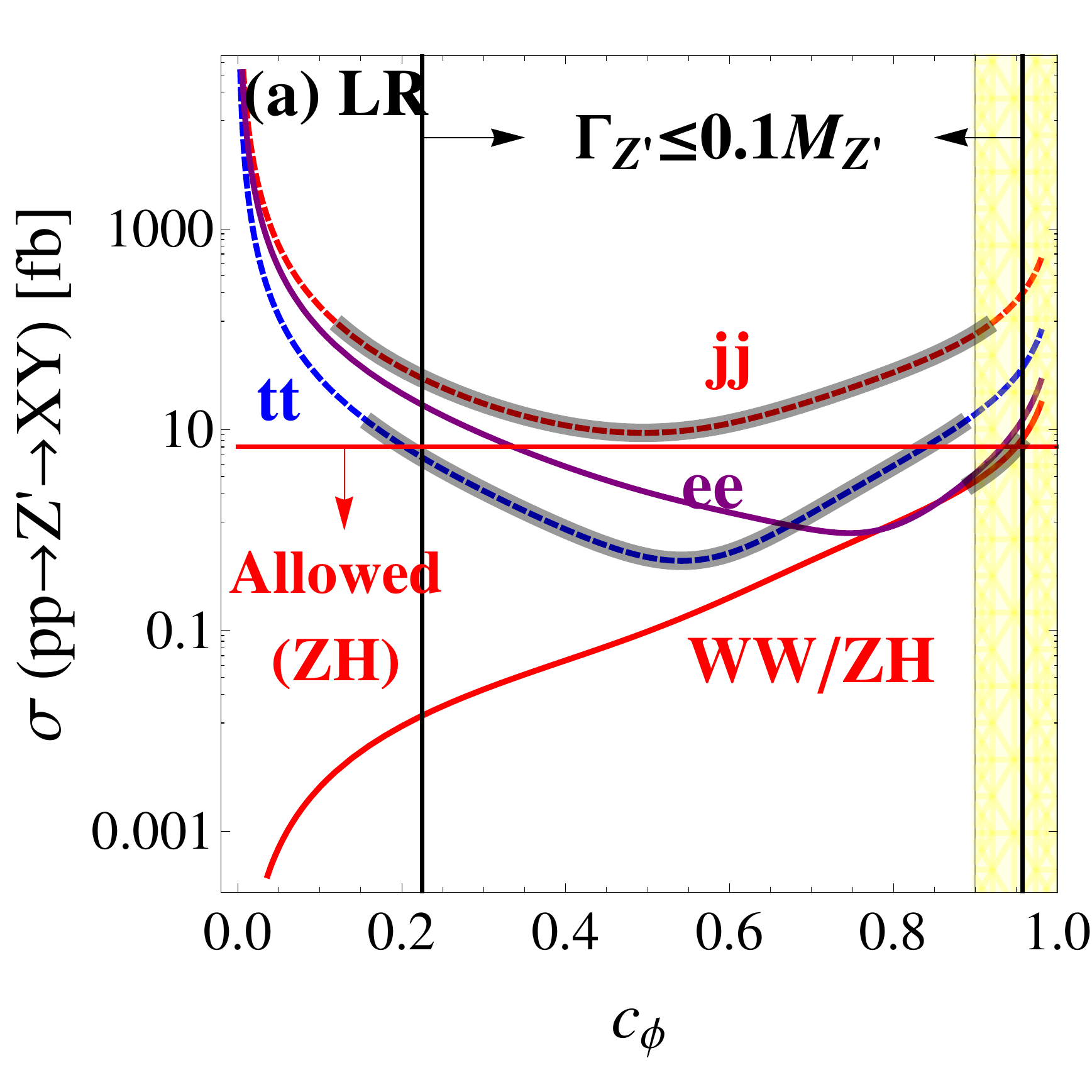}
\includegraphics[width=0.30\textwidth]{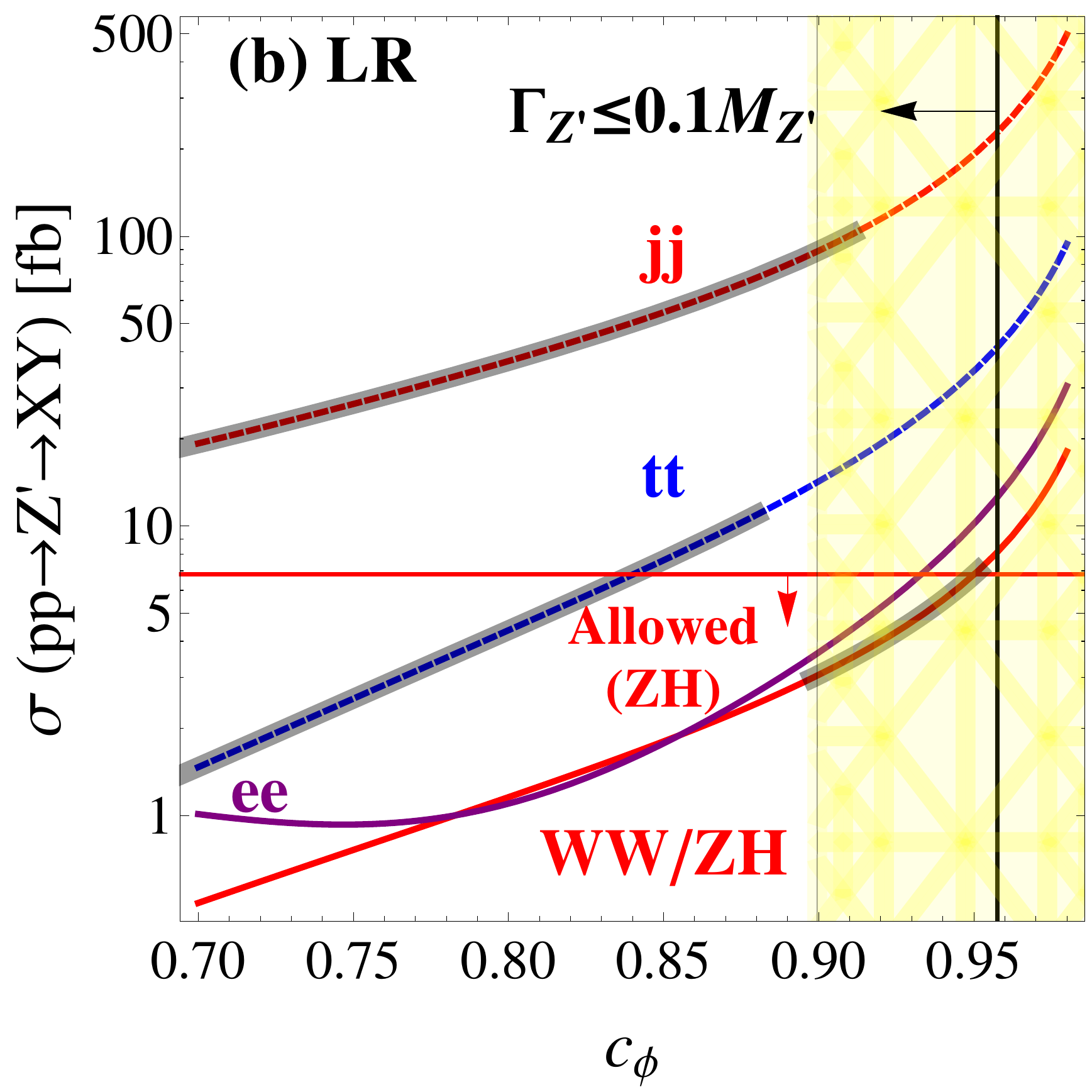}
\caption{\it The contours of the cross section $\sigma(Z^\prime) \times {\rm BR}(Z^\prime \to XY)$, where $X$ and $Y$ denote the SM particles in the $Z^\prime$ decay as a function of $c_\phi$. The shaded bands along each curve represent the region compatible with the current experimental data. The yellow shaded region is required for $M_{W^\prime} \simeq M_{Z^\prime}$. 
}\label{LRzp2}
\end{figure}

In Fig.~\ref{LRzp2} we present the cross section $\sigma(Z^\prime) \times {\rm BR}(Z^\prime \to XY)$ as a function of $c_\phi$, where $X$ and $Y$ denote the SM particles in the $Z^\prime$ decay. The curves show the theoretical predictions while the shaded bands along each curve represent the parameter space compatible with current experimental data. The current bound on $\sigma(Z^\prime)\times {\rm BR}(Z^\prime \to t\bar{t})$ mode demands $0.16\leq c_\phi \leq 0.88$; see the blue-dotted curve with the $tt$ label. The di-jet ($jj$) constraint is slightly weaker than the $tt$ constraint. 
The shaded band along the $WW/ZH$ curve (red-solid) represents the required $c_\phi$ to explain the $WW$ excess. 
However, all the parameter space of interest to us is excluded by the leptonic decay mode, which imposes much tighter constraint of $\sigma(Z^\prime)\times {\rm BR}(Z^\prime \to e^+e^-)\leq 0.2~{\rm fb}$~\cite{Aad:2014cka,Khachatryan:2014fba}.  As shown in Fig.~\ref{LRzp2}(b), $\sigma(Z^\prime)\times {\rm BR}(Z^\prime \to e^+ e^-) \sim 1~{\rm fb}$ for a 2~TeV $Z^\prime$  boson; see the purple curve. 
We thus conclude that, if the $WW$ excess is induced by the $Z^\prime$ boson in the Left-Right model, one needs to extend the model to suppress the leptonic decays of the $Z^\prime$ boson.

\subsection{Lepto-Phobic doublet model}

\subsubsection{The $W^\prime$ constraints}

\begin{figure}
\includegraphics[width=0.32\textwidth]{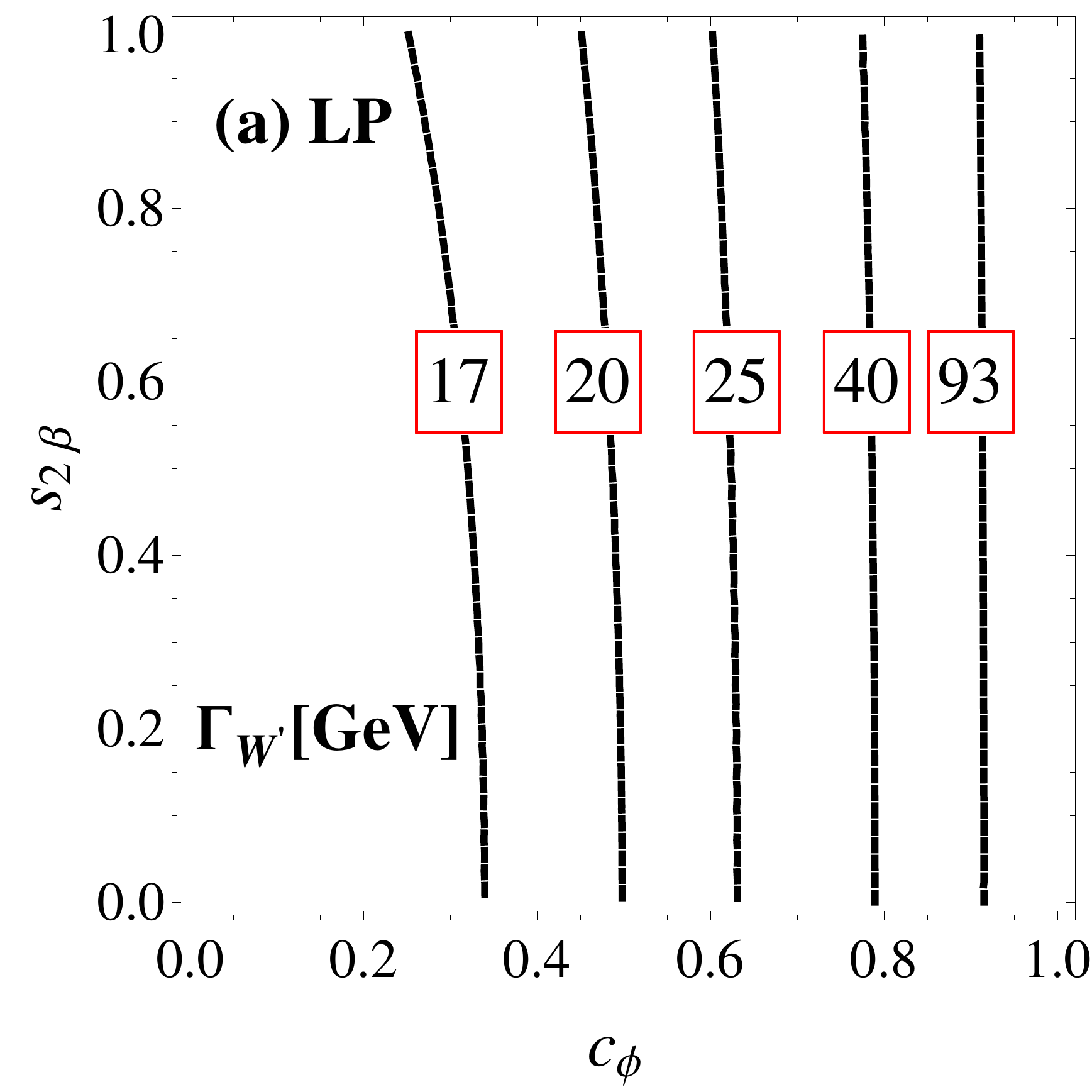}
\includegraphics[width=0.32\textwidth]{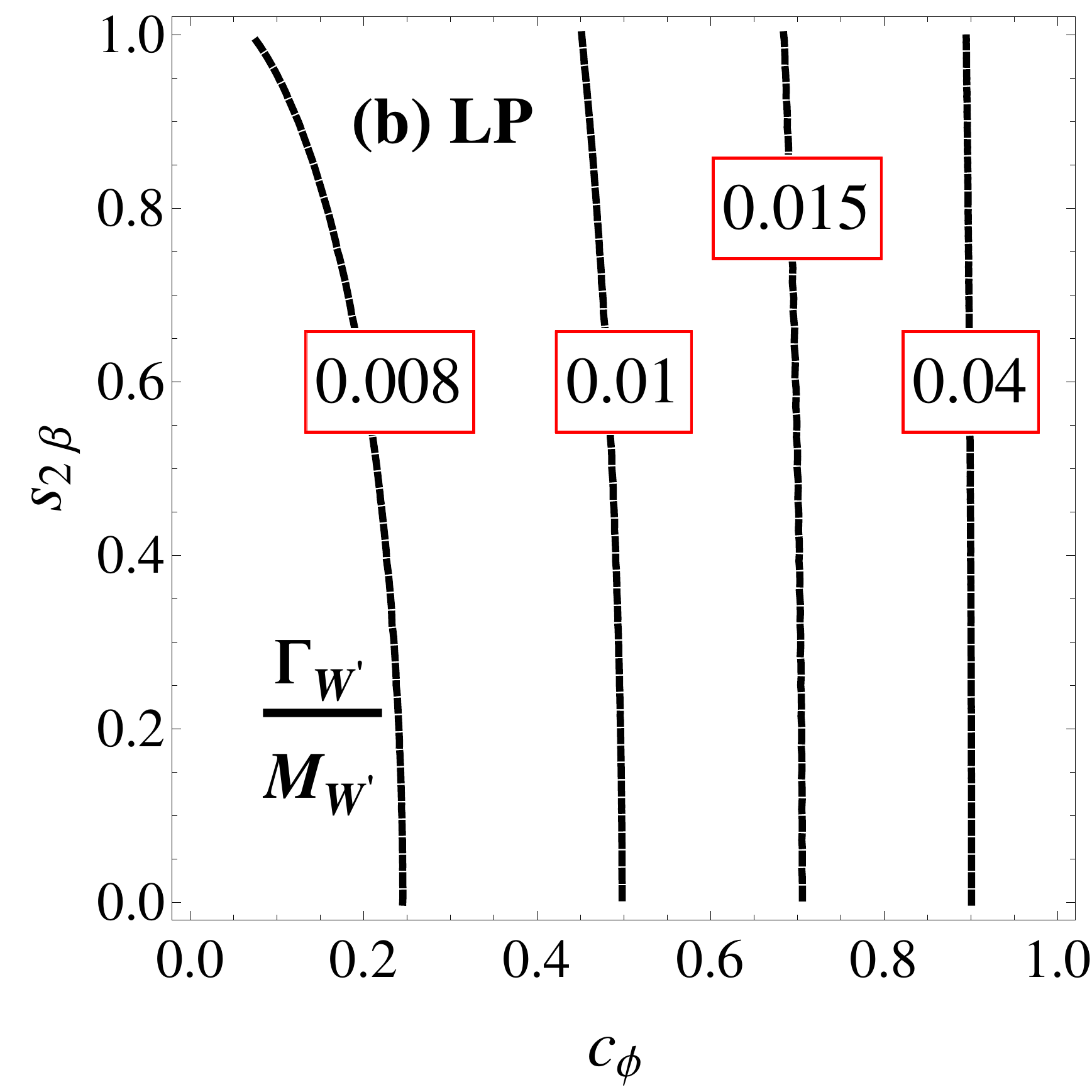}
\caption{\it The total width $\Gamma_{W^\prime}$ (a)  and $\Gamma_{W^\prime}/M_{W^\prime}$ (b) in the plane of $c_\phi$ and $s_{2\beta}$ in the Lepto-Phobic doublet model. 
}\label{LPw1}
\end{figure}

The Lepto-Phobic doublet model is similar to the Left-Right model but the leptonic doublet is gauged only under $SU(1)_1$; see Table~\ref{tb:models}.
Figure~\ref{LPw1} displays the contour of the total width $\Gamma_{W^\prime}$ and $\Gamma_{W^\prime}/M_{W^\prime}$ in the plane of $c_\phi$ and $s_{2\beta}$. It shows the NWA is also a good approximation to describe the production and decay of $W^\prime$.

\begin{figure}
\includegraphics[width=0.3\textwidth]{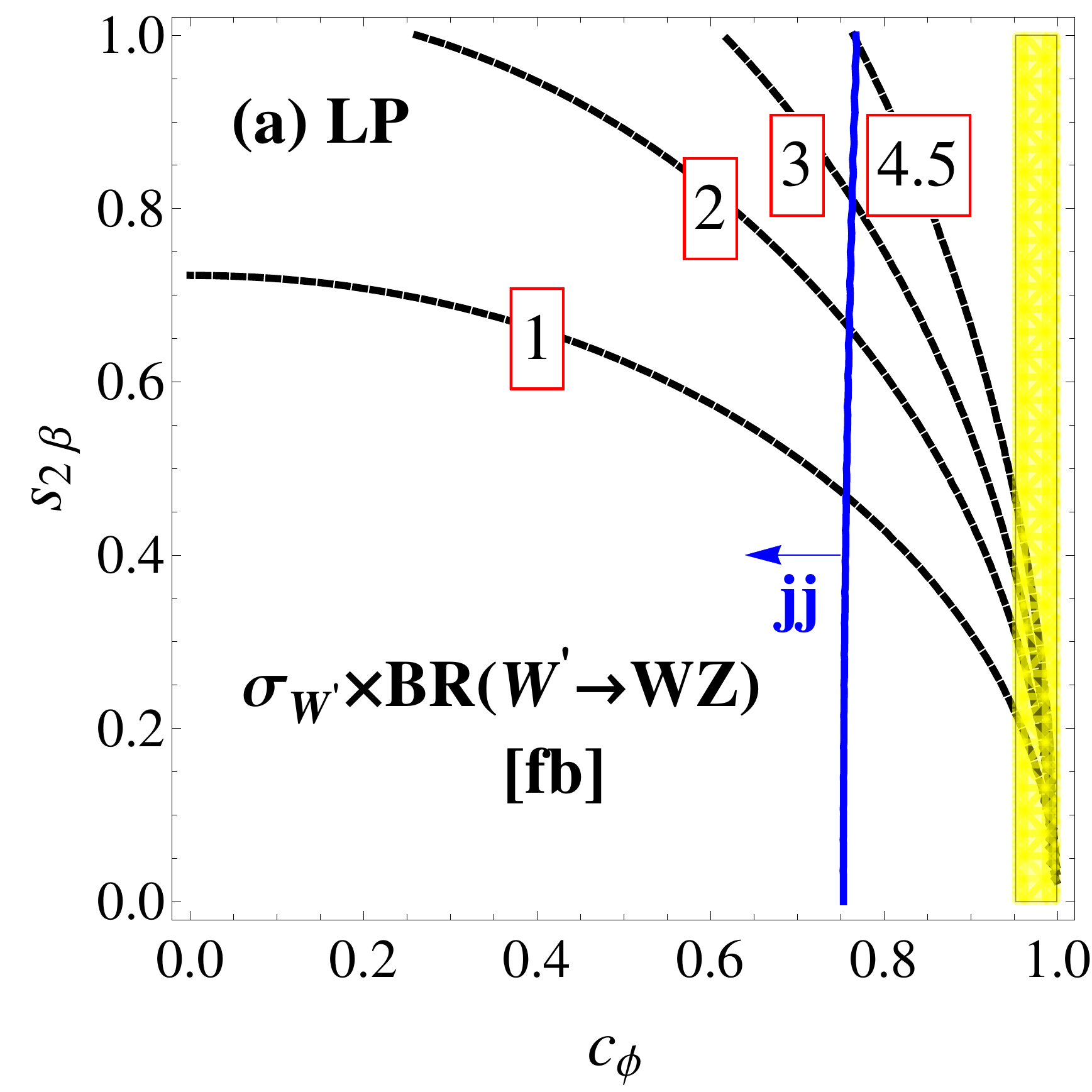}
\includegraphics[width=0.3\textwidth]{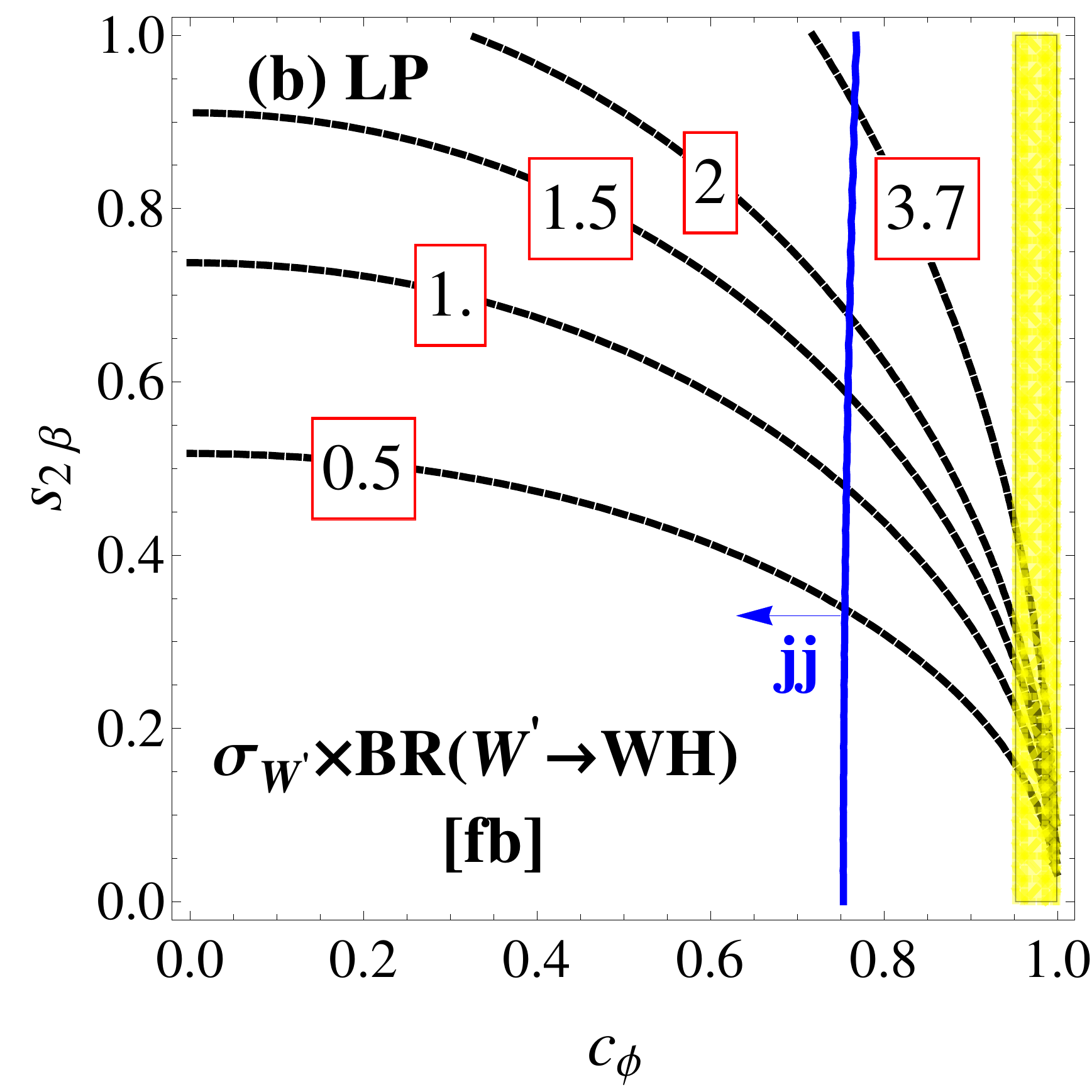}\\
\includegraphics[width=0.3\textwidth]{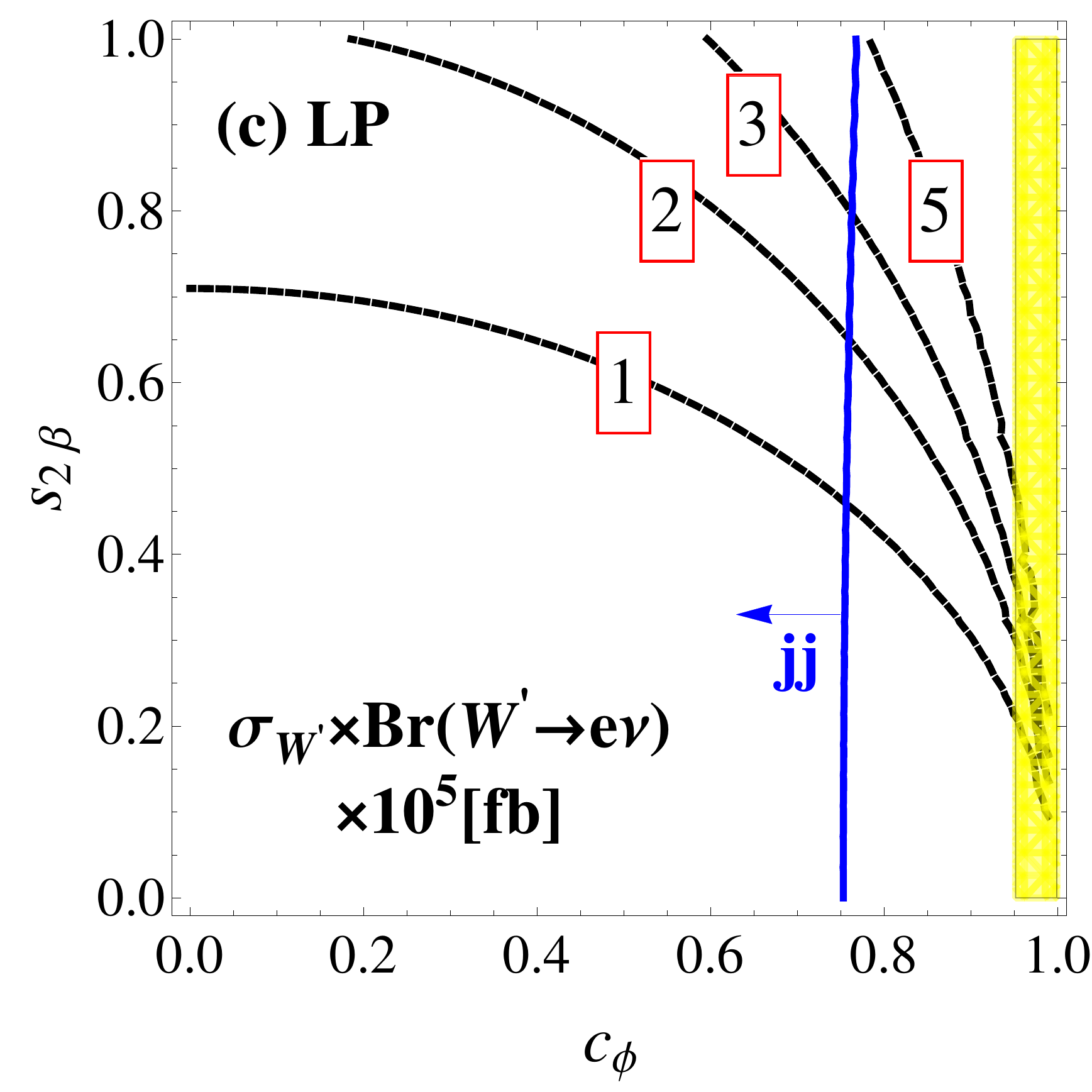}
\includegraphics[width=0.3\textwidth]{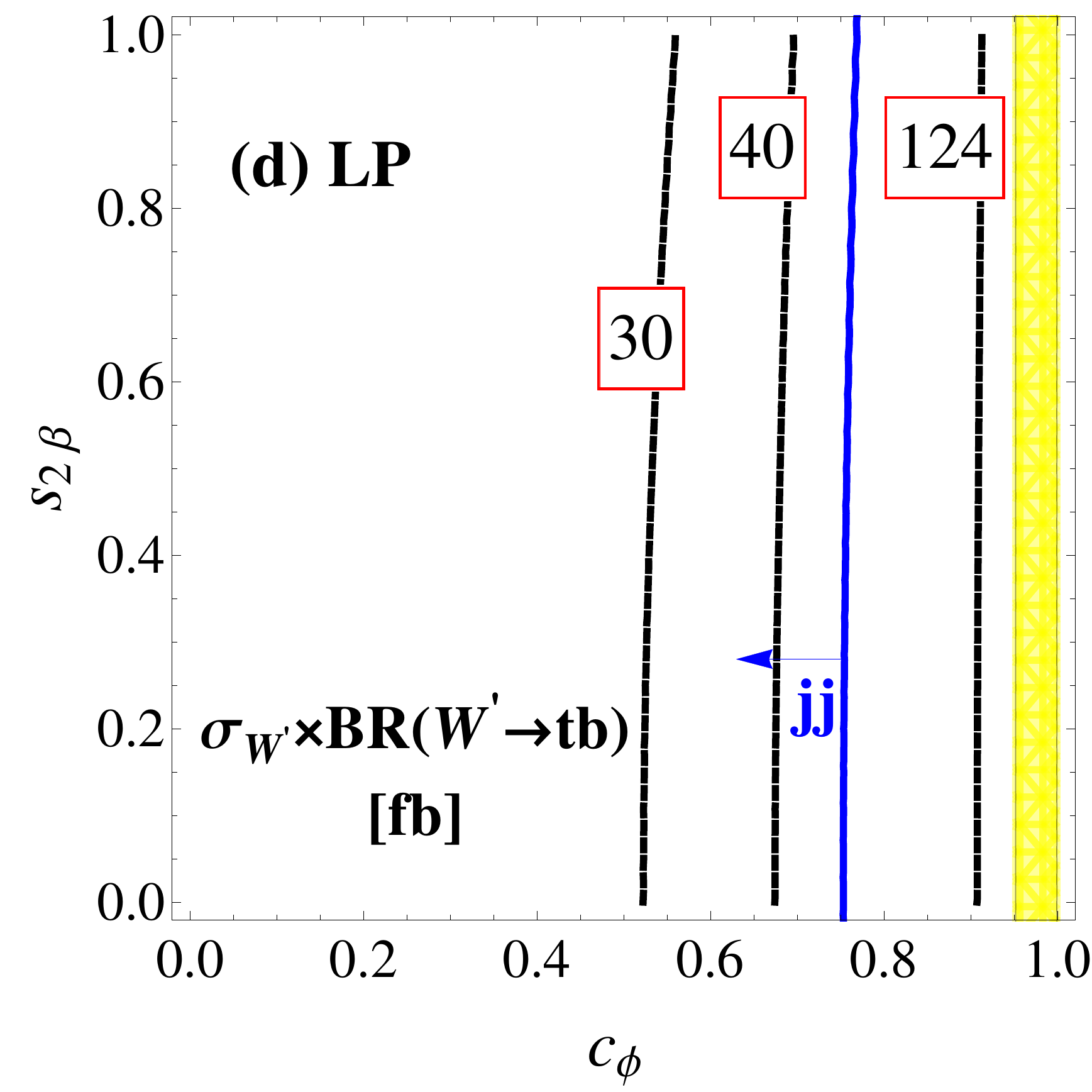}
\caption{\it The contours of the cross section (a) $\sigma(W^\prime) \times {\rm BR}(W^\prime \to WZ)$, (b) $\sigma(W^\prime)\times {\rm BR}(W^\prime \to WH)$, (c) $\sigma(W^\prime)\times {\rm BR}(W^\prime \to e\nu)$ and (d) $\sigma(W^\prime)\times {\rm BR}(W^\prime \to tb)$ in the plane of $c_\phi$ and $s_{2\beta}$. The vertical line ($jj$) denotes the constraint from the di-jet measurements. The yellow band represents the degenerated mass region of $W^\prime$ and $Z^\prime$.
}\label{LPw2}
\end{figure}

\begin{figure}
\includegraphics[width=0.3\textwidth]{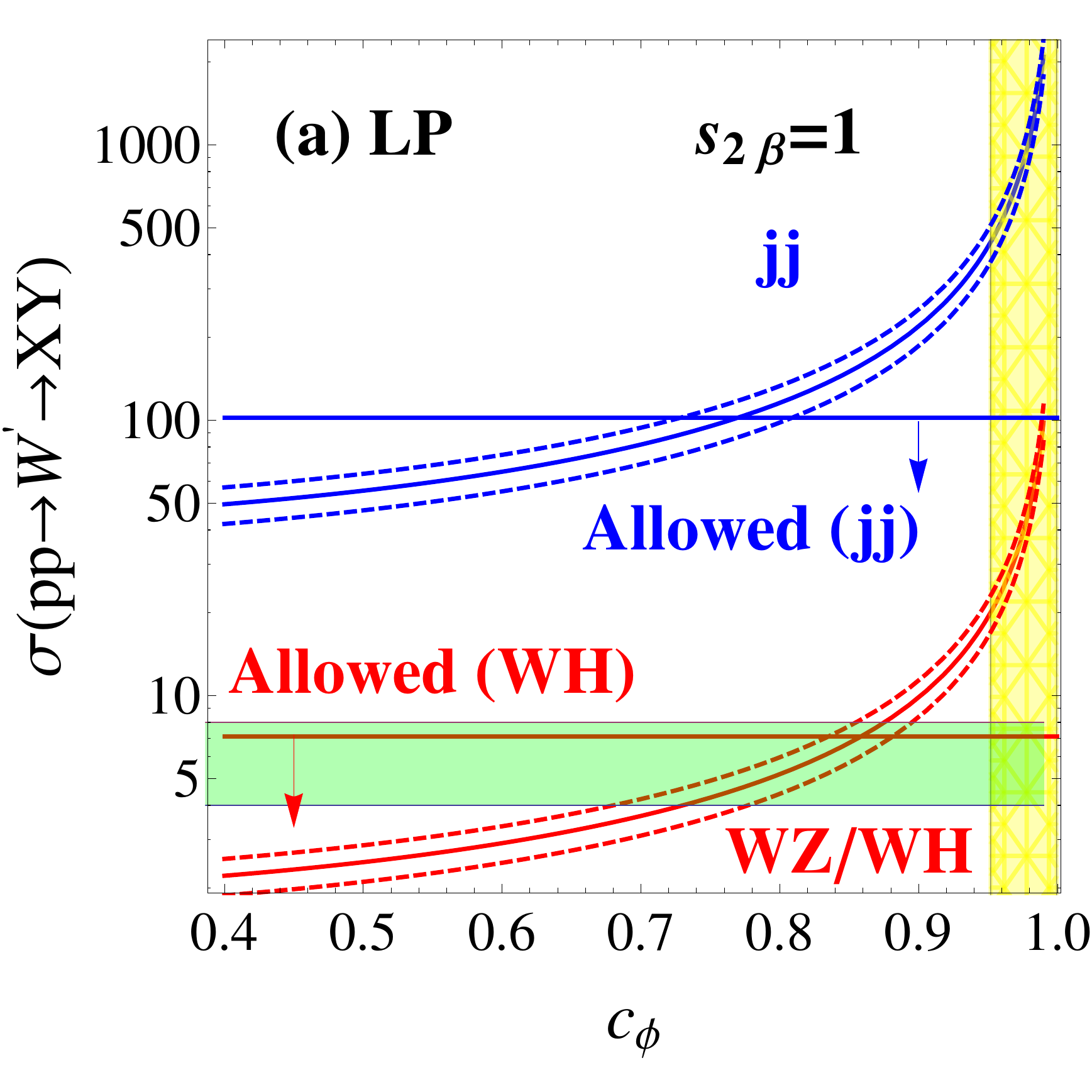}
\caption{\it The cross section of $pp \to W^\prime \to WZ/WH$  (red curves) and $pp \to W^\prime \to jj$ (blue curves) as a function of $c_\phi$ with $s_{2\beta}=1$. The dashed curves represent the PDF uncertainties. The green shaded region represents the parameter space compatible with the $WZ$ excess.  The yellow shaded region is required for $M_{W^\prime} \simeq M_{Z^\prime}$. The current experimental limits of $\sigma(pp \to W^\prime \to jj) < 102~{\rm fb}$
 and $\sigma(pp \to W^\prime \to WH)<7.1~{\rm fb}$ are also plotted. 
}\label{LPPDF}
\end{figure}

Figure~\ref{LPw2}(a) displays the contour of the cross section of $\sigma(W^\prime)\times {\rm BR}(W^\prime \to WZ)$ in the plane of $c_\phi$ and $s_{2\beta}$. The yellow bands represent the degenerated region of $M_{W^\prime}$ and $M_{Z^\prime}$. In order to produce $\sigma(WZ) \sim 4-8~{\rm fb}$ and $\sigma(W^\prime)\times {\rm BR}(W^\prime \to jj)\leq 102~{\rm fb}$~\cite{Khachatryan:2015sja}, one needs $0.73<c_\phi<0.75$ and $s_{2\beta}\gtrsim0.9$. However,  the $Z^\prime$ mass in those parameter space is much larger than the $W^\prime$ mass, e.g. $2.67~{\rm TeV}\leq M_{Z^\prime}<2.74~{\rm TeV}$ for $M_{W^\prime}=2~{\rm TeV}$. 
As analogous to the Left-Right model, the Lepto-Phobic model can explain the $WZ$ excess if the $WW$ excess is a result of the fluctuation of SM backgrounds.

Figure~\ref{LPw2}(b-d) shows the cross sections of $\sigma(W^\prime)\times {\rm BR}(W^\prime \to WH/e\nu/tb)$, respectively. In the region of $0.73<c_\phi<0.75$,
all of those three modes satisfy the current experimental upper limits.

Similar to the Left-Right model, we choose $s_{2\beta}=1$ and plot the cross section of $pp \to W^\prime \to WZ/WH$  (red curves) and $pp \to W^\prime \to jj$ (blue curves) as a function of $c_\phi$ in Fig.~\ref{LPPDF}. The outer dashed-curves represent the PDF uncertainties. The green shaded region represents the parameter space compatible with the $WZ$ excess. The yellow shaded region is required for $M_{W^\prime} \simeq M_{Z^\prime}$. The current experimental limits of $\sigma(pp \to W^\prime \to jj) <102~{\rm fb}$ and $\sigma(pp \to W^\prime \to WH)<7.1~{\rm fb}$ are also plotted. To explain the excess of the $WZ$ and satisfy $WH$ limit, it requires $0.68<c_{\phi}<0.88$, while the di-jet experimental limit requires $c_{\phi}<0.81$. 
Thus, we conclude that the Lepto-Phobic model could explain the $WZ$ excess in the region $0.68<c_{\phi}<0.81$ with $s_{2\beta}\sim 1$.  However, it predicts a heavier $Z^\prime$ as $2.47~{\rm TeV}\leq M_{Z^\prime}<2.94~{\rm TeV}$ for $M_{W^\prime}=2~{\rm TeV}$, which contradicts the $WW$ excess around 2~TeV. Bearing in mind that the $2.6\sigma$ $WW$ excess might be owing to the fluctuation of the SM backgrounds, we await the forthcoming LHC Run-2 data to make an affirmative conclusion. 

\subsubsection{The $Z^\prime$ constraints}

\begin{figure}[b]
\includegraphics[width=0.32\textwidth]{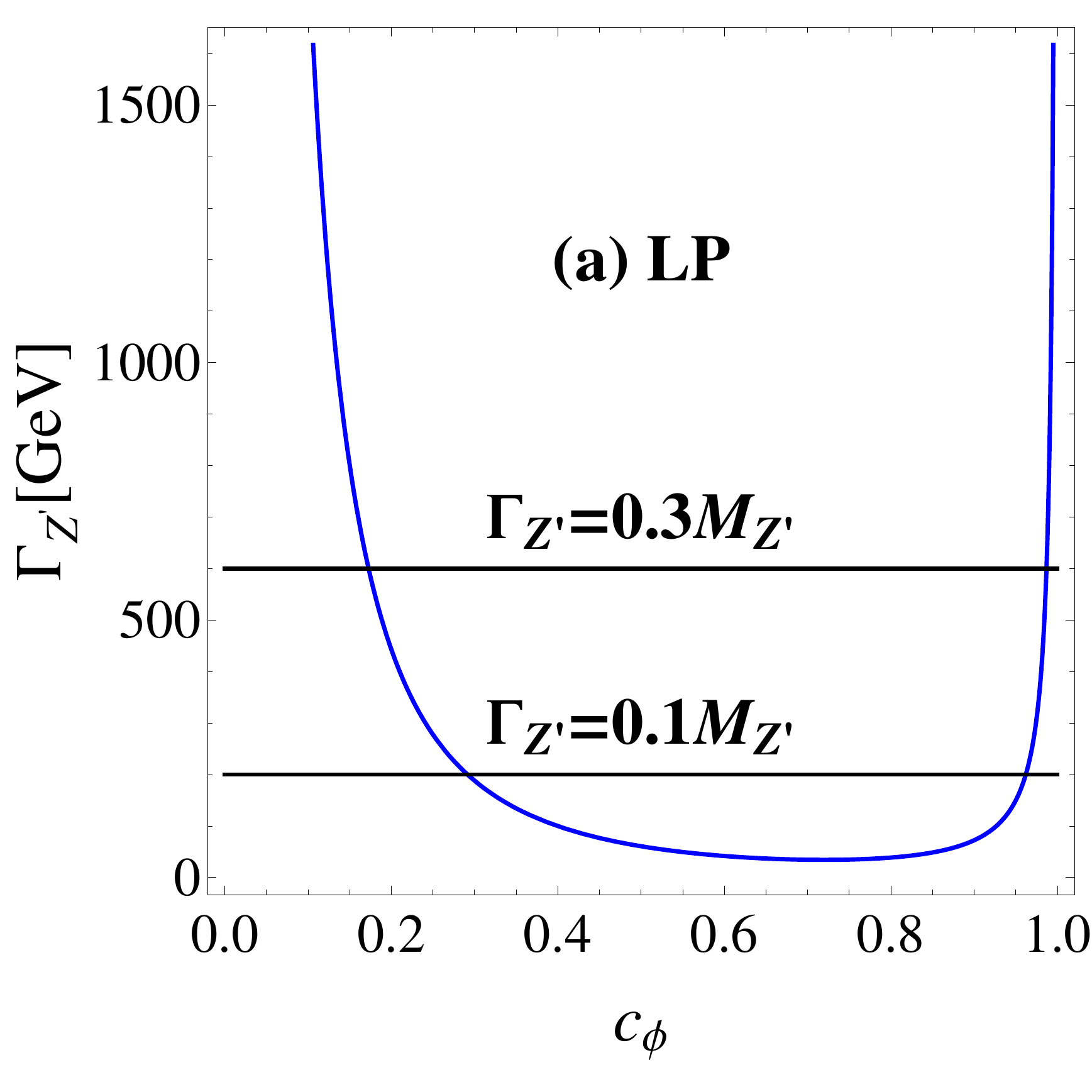}
\includegraphics[width=0.32\textwidth]{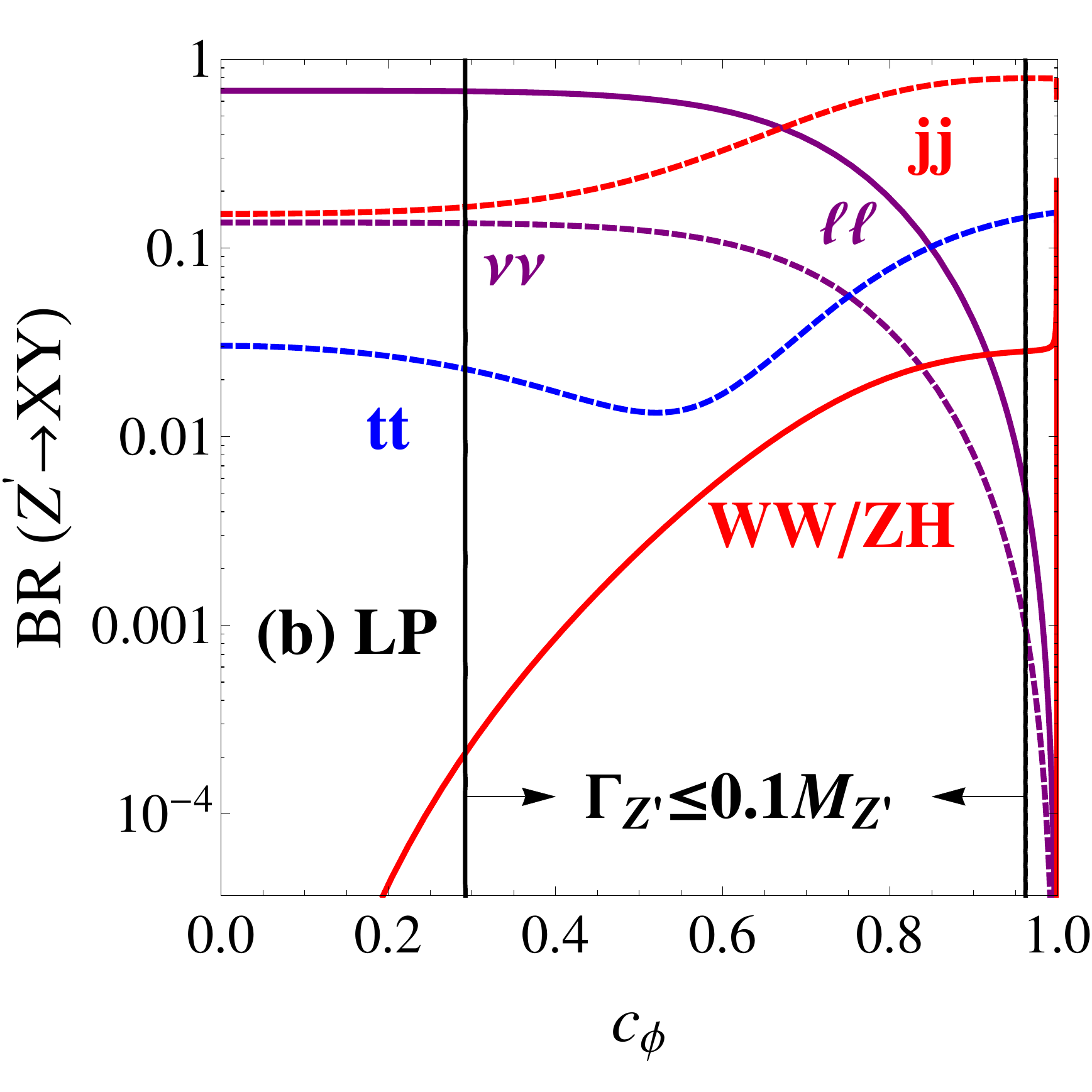}\\
\caption{\it The total width (a)  and the branching ratios of all the decay modes (b) of $Z^\prime$ as a function of $c_{\phi}$ in Lepto-Phobic model. 
}\label{LPz1}
\end{figure}

Although the couplings of $W'$ to the SM leptons are highly suppressed in the Lepto-Phobic model, the couplings of $Z'$ to the SM leptons are not.  For a small $c_\phi$ (large $g_X$), the $U(1)_X$ component in the $Z^\prime$ gives rise to a large coupling to the SM leptons.  That yields a large decay width of $Z^\prime$ in the vicinity of $c_\phi\sim 0$. 
We also require $\Gamma(Z^\prime)\leq 0.1 M_{Z^\prime}$ which leads to $0.29 \leq c_\phi \leq 0.96$; see Fig.~\ref{LPz1}(a). Figure~\ref{LPz1}(b) displays the branching ratios of the $Z^\prime$ decay. It shows the branching ratios of $Z^\prime \to \nu\nu$ and $Z^\prime \to \ell \ell $ are suppressed for a large $c_\phi$ while the $jj$ and $t\bar{t}$ decay modes tend to be dominate. Such a behavior can be understood from the fact that heavy gauge bosons are predominately coupled to the SM quarks. The $WW$ and $ZH$ modes are also much smaller than other modes; see the red-solid curve.

\begin{figure}
\includegraphics[width=0.32\textwidth]{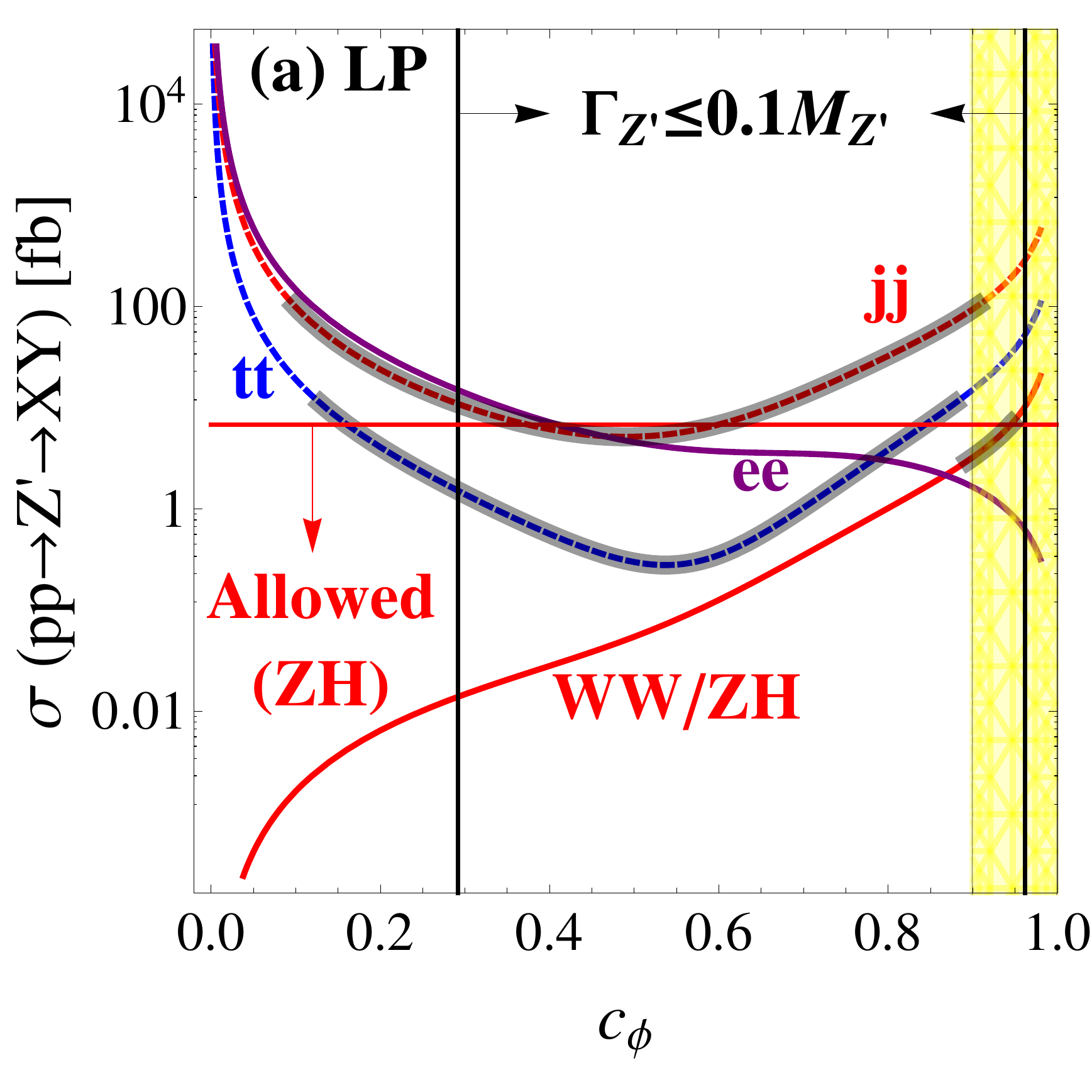}
\includegraphics[width=0.32\textwidth]{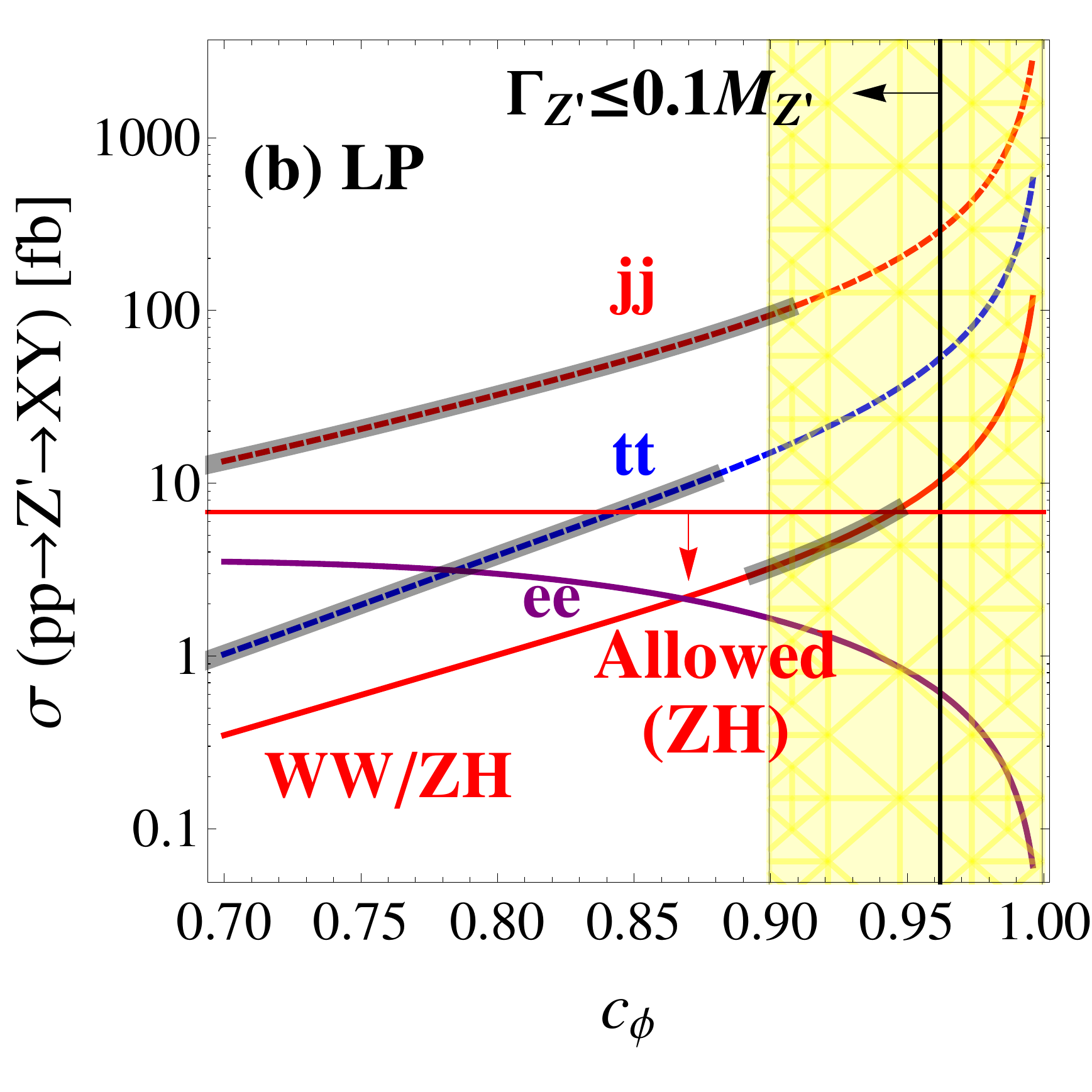}\\
\caption{\it The contours of the cross section $\sigma(Z^\prime) \times {\rm BR}(Z^\prime \to XY)$, where $X$ and $Y$ denote the SM particles in the $Z^\prime$ decay as a function of $c_\phi$ in the Lepto-Phobic model. 
}\label{LPz2}
\end{figure}

In Fig.~\ref{LPz2} we present $\sigma(Z^\prime) \times {\rm BR}(Z^\prime \to XY)$ as a function of $c_\phi$ where $X$ and $Y$ denote the SM particles in the $Z^\prime$ decay. The curves show the theoretical predictions while the shaded bands are allowed by current experimental data. 
The current bound on $\sigma(Z^\prime)\times {\rm BR}(Z^\prime \to t\bar{t})$ mode demands $0.13\leq c_\phi \leq 0.88$; see the blue-dotted curve with the $tt$ label. The di-jet ($jj$) constraint is slighter weaker than the $tt$ constraint.
The shaded band along the $WW/ZH$ curve (red-solid) represents the required $c_{\phi}$ to explain the $WW$ excess, i.e.  $0.89<c_\phi<0.95$. However, all the parameter space of interest to us is excluded by the leptonic decay mode, which imposes much tighter constraint of $\sigma(Z^\prime)\times {\rm BR}(Z^\prime \to e^+e^-)\leq 0.2~{\rm fb}$~\cite{Aad:2014cka,Khachatryan:2014fba}; see the purple-solid curve. Figure~\ref{LPz2}(b) shows the details in the vicinity of $c_{\phi}\sim 0.9$. 
The cross section of $\sigma(Z^\prime)\times {\rm BR}(Z^\prime \to e^+e^-)\sim 1~{\rm fb}$, which is much larger than the current constraint. 
Therefore, it is difficult to explain the $WW$ excess in the Lepto-Phobic model unless one can sizeably reduce the leptonic decay branching ratio of $Z^\prime$.

\subsection{Hadro-Phobic doublet model}

\subsubsection{The $W^\prime$ constraints}

In the Hadro-Phobic doublet model the right-handed leptons form a doublet gauged under the $SU(2)_2$; see Table~\ref{tb:models} for detailed quantum number assignments. The $W^\prime$ and $Z^\prime$ arise from the symmetry breaking of  $SU(2)_2\times U(1)_X \to U(1)_Y$ and therefore are coupled predominately to the SM leptons.

\begin{figure}
\includegraphics[width=0.32\textwidth]{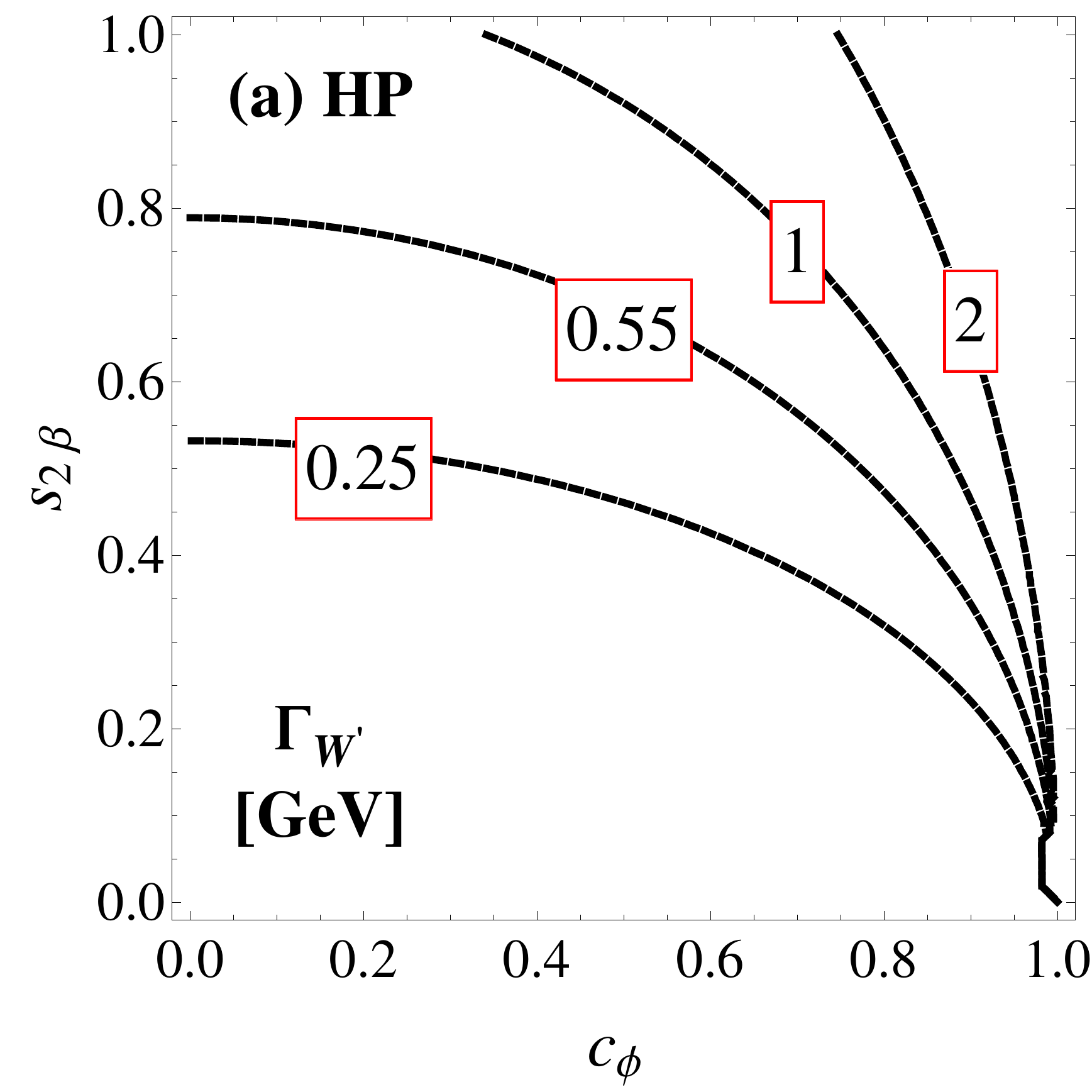}
\includegraphics[width=0.32\textwidth]{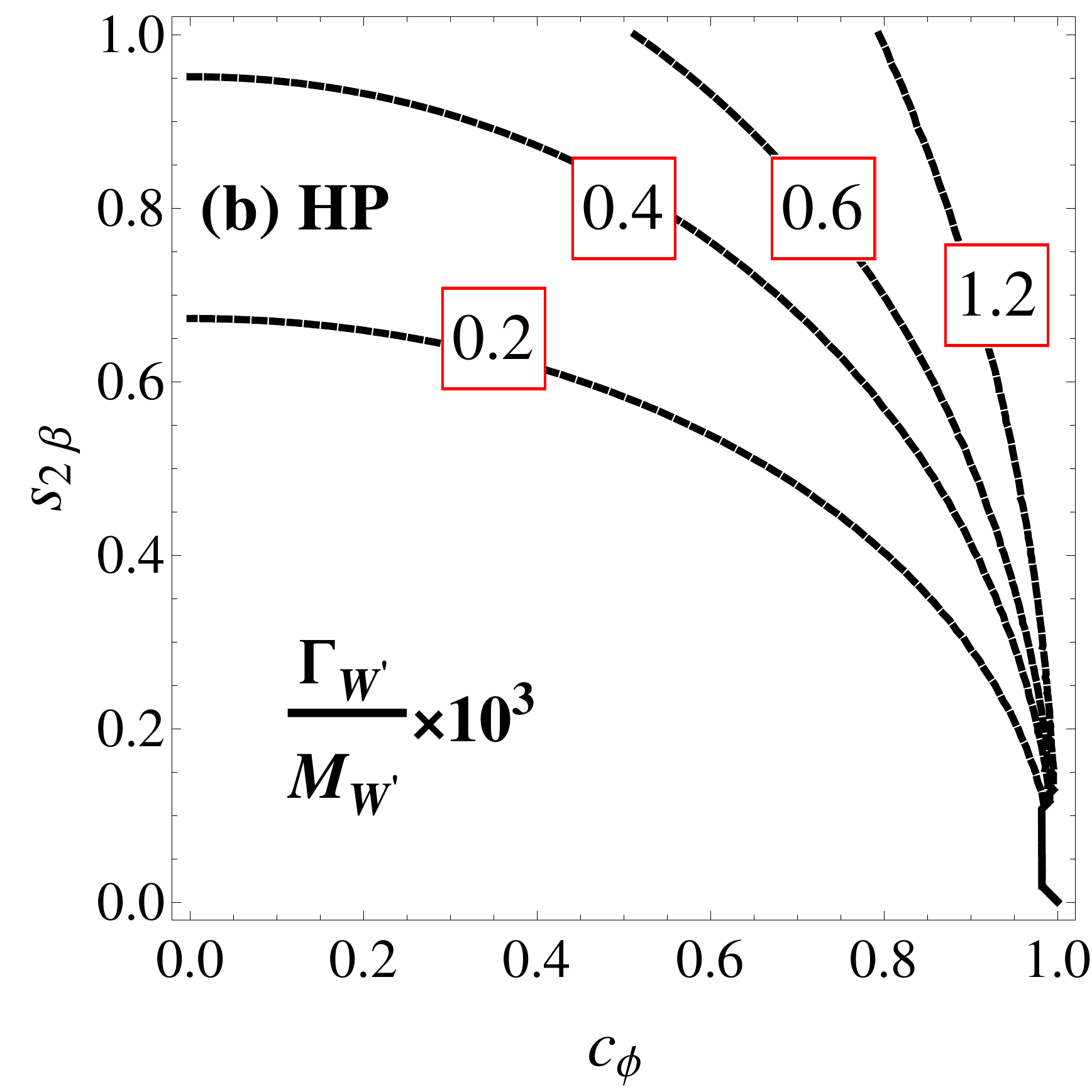}
\caption{\it The total width $\Gamma_{W^\prime}$ (a)  and $\Gamma_{W^\prime}/M_{W^\prime}$ (b) in the plane of $c_\phi$ and $s_{2\beta}$ in the Hadro-Phobic doublet model. 
}\label{HPw1}
\end{figure}

Figure~\ref{HPw1} displays the contour of the total width $\Gamma_{W^\prime}$ (a) and the ratio $\Gamma_{W^\prime}/M_{W^\prime}$ (b) in the plane of $c_\phi$ and $s_{2\beta}$. In the most of the parameter space, the $W^\prime$ width is around 1~GeV for a 2~TeV $W^\prime$. Therefore, the NWA is a good approximation to describe the production and decay of $W^\prime$ in the Hadro-Phobic model.

\begin{figure}
\includegraphics[width=0.32\textwidth]{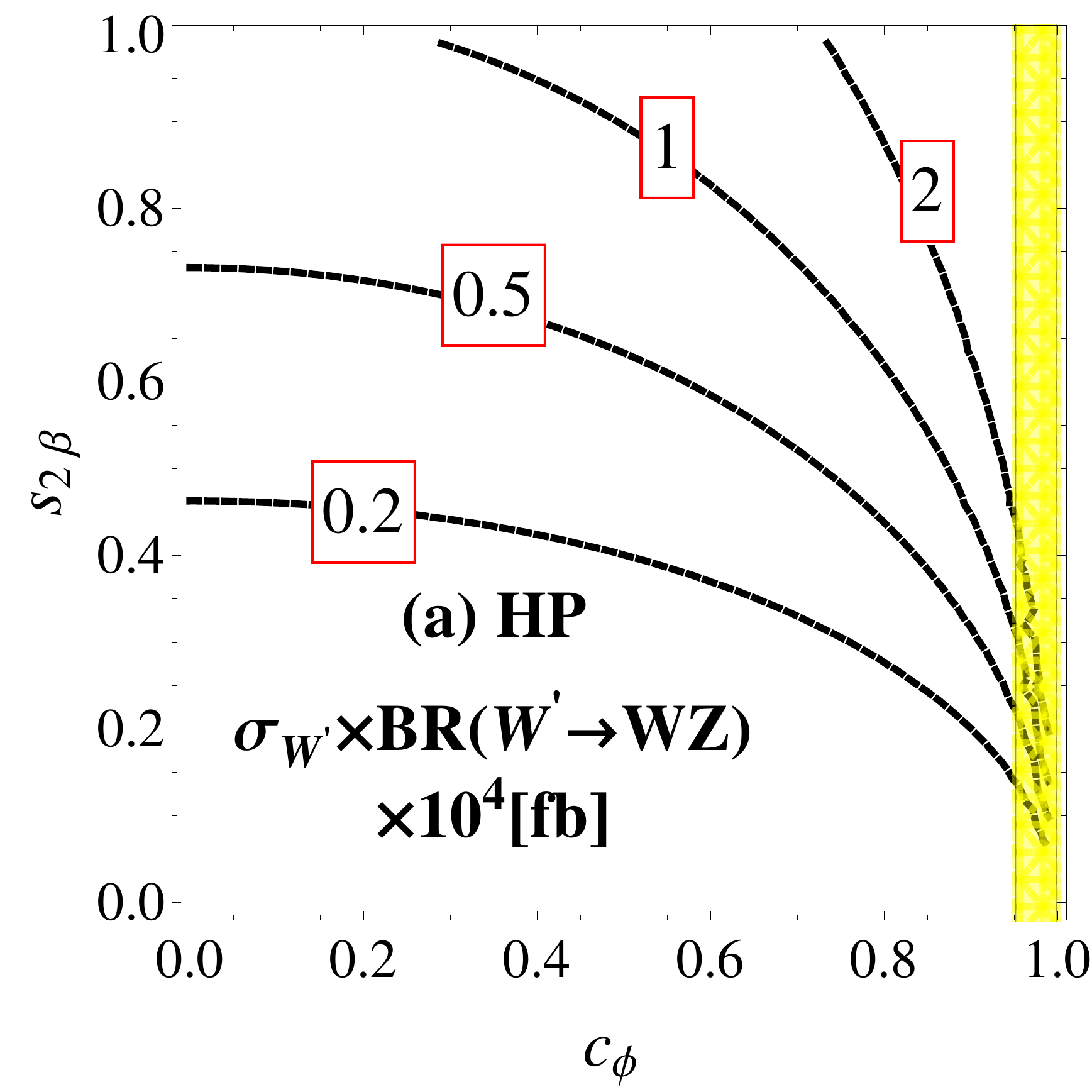}
\includegraphics[width=0.32\textwidth]{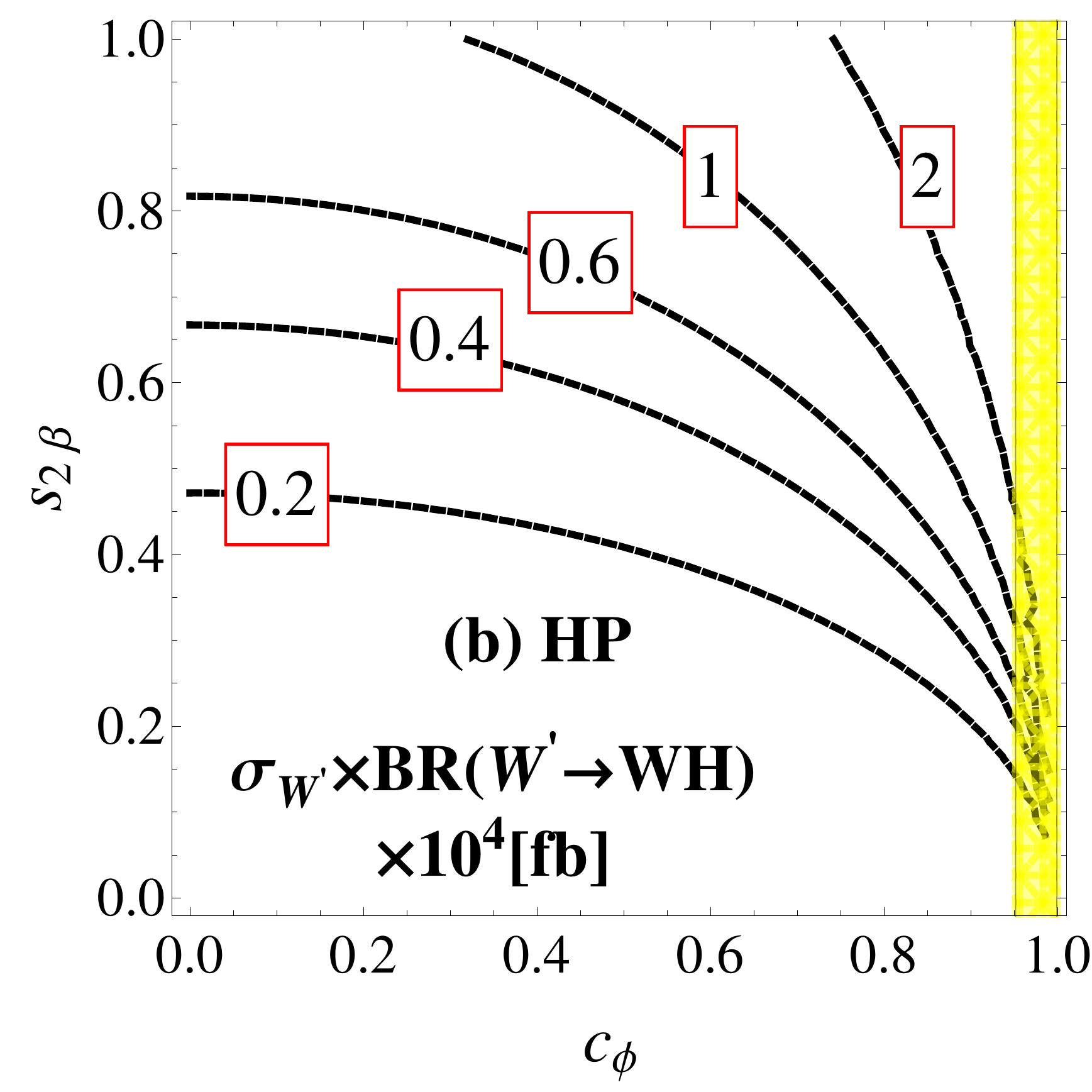}\\
\includegraphics[width=0.32\textwidth]{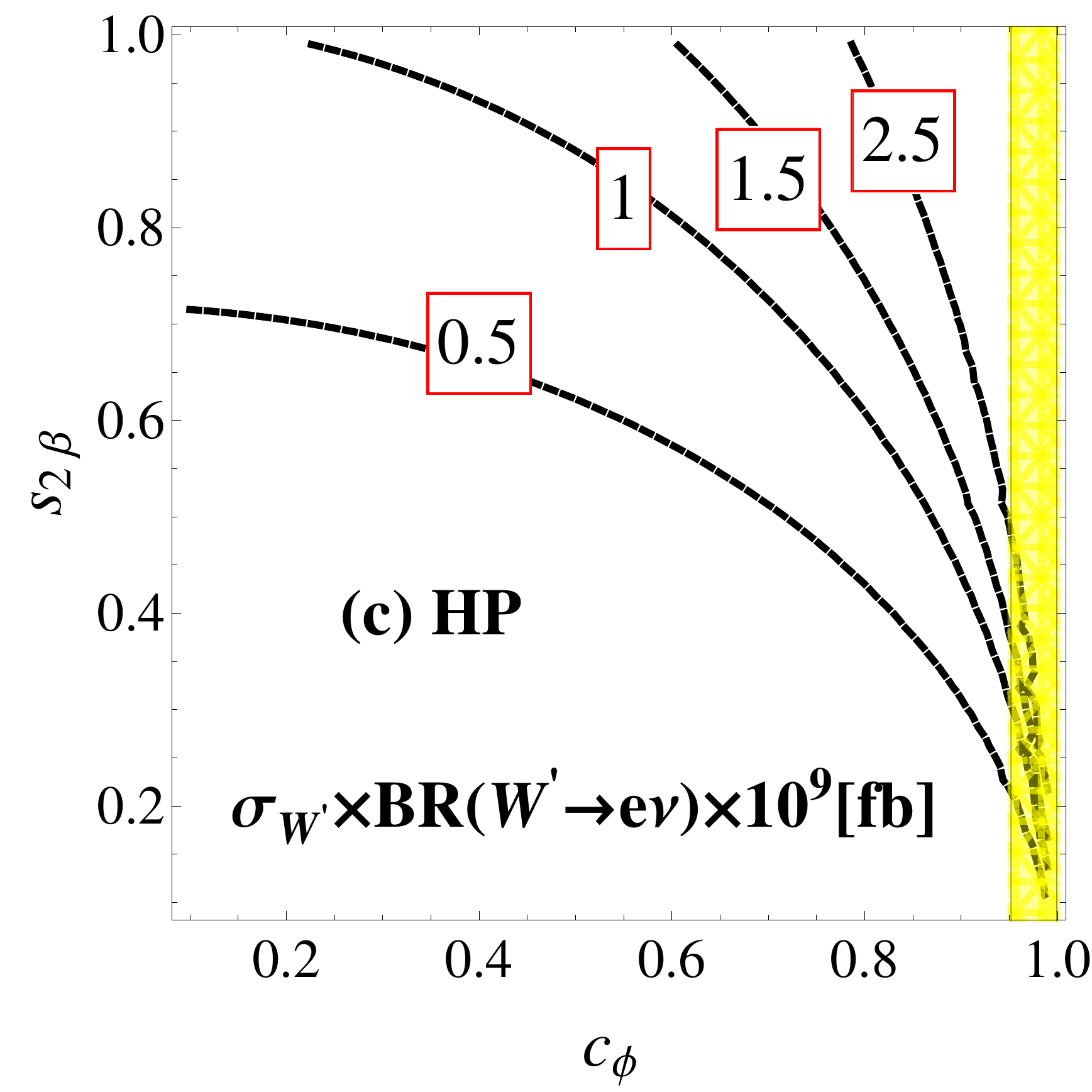}
\includegraphics[width=0.32\textwidth]{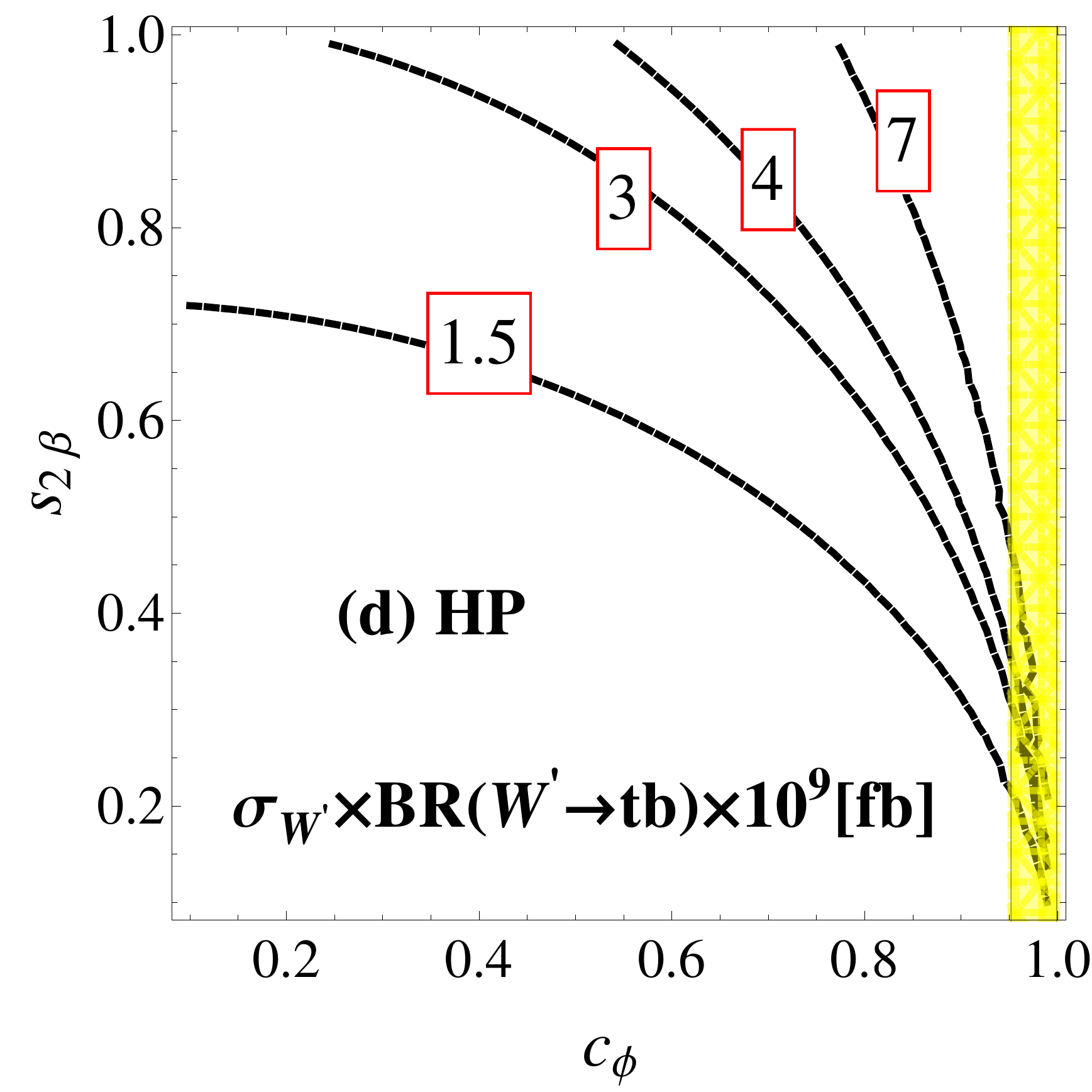}
\caption{\it The contours of the cross section (a) $\sigma(W^\prime) \times {\rm BR}(W^\prime \to WZ)$, (b) $\sigma(W^\prime)\times {\rm BR}(W^\prime \to WH)$, (c) $\sigma(W^\prime)\times {\rm BR}(W^\prime \to e\nu)$ and (d) $\sigma(W^\prime)\times {\rm BR}(W^\prime \to tb)$ in the plane of $c_\phi$ and $s_{2\beta}$ in the Hadro-Phobic doublet model. All the cross sections are in the unit of fb. The yellow shaded region is required for $M_{W^\prime}\simeq M_{Z^\prime}$.} 
\label{HPw2}
\end{figure}

As the gauge couplings of $W^\prime$ to the SM quarks are highly suppressed, the production cross section of $W^\prime$ in the Hadro-Phobic model is much smaller than those in the Left-Right and Lepton-Phobic models. Figure~\ref{HPw2}  displays the contour of the cross section of $\sigma({W^\prime})\times {\rm BR}(W^\prime \to WZ/WH/e\nu/tb)$ in the plane of $c_\phi$ and $s_{2\beta}$. The yellow shaded region is required for $M_{W^\prime}\simeq M_{Z^\prime}$. The cross sections of $pp \to W^\prime\to WZ$ and $pp  \to W^\prime \to WH$ are around $10^{-4}$ fb. Since the $W^\prime$ boson couples to the SM leptons/quarks through the mixing of $W$-$W^\prime$, the branching ratio of $W^\prime$ decaying into lepton/quark final states are highly suppressed, yielding $\sigma({W^\prime})\times {\rm BR}(W^\prime \to e\nu/tb/jj)\sim 10^{-9}~{\rm fb}$. 
It is clear that, in all the parameter space, the cross section of the $WZ$ mode is much smaller than $1~{\rm fb}$ such that it cannot explain the $WZ$ excess. 

\subsubsection{The $Z^\prime$ constraints}

\begin{figure}
\includegraphics[width=0.32\textwidth]{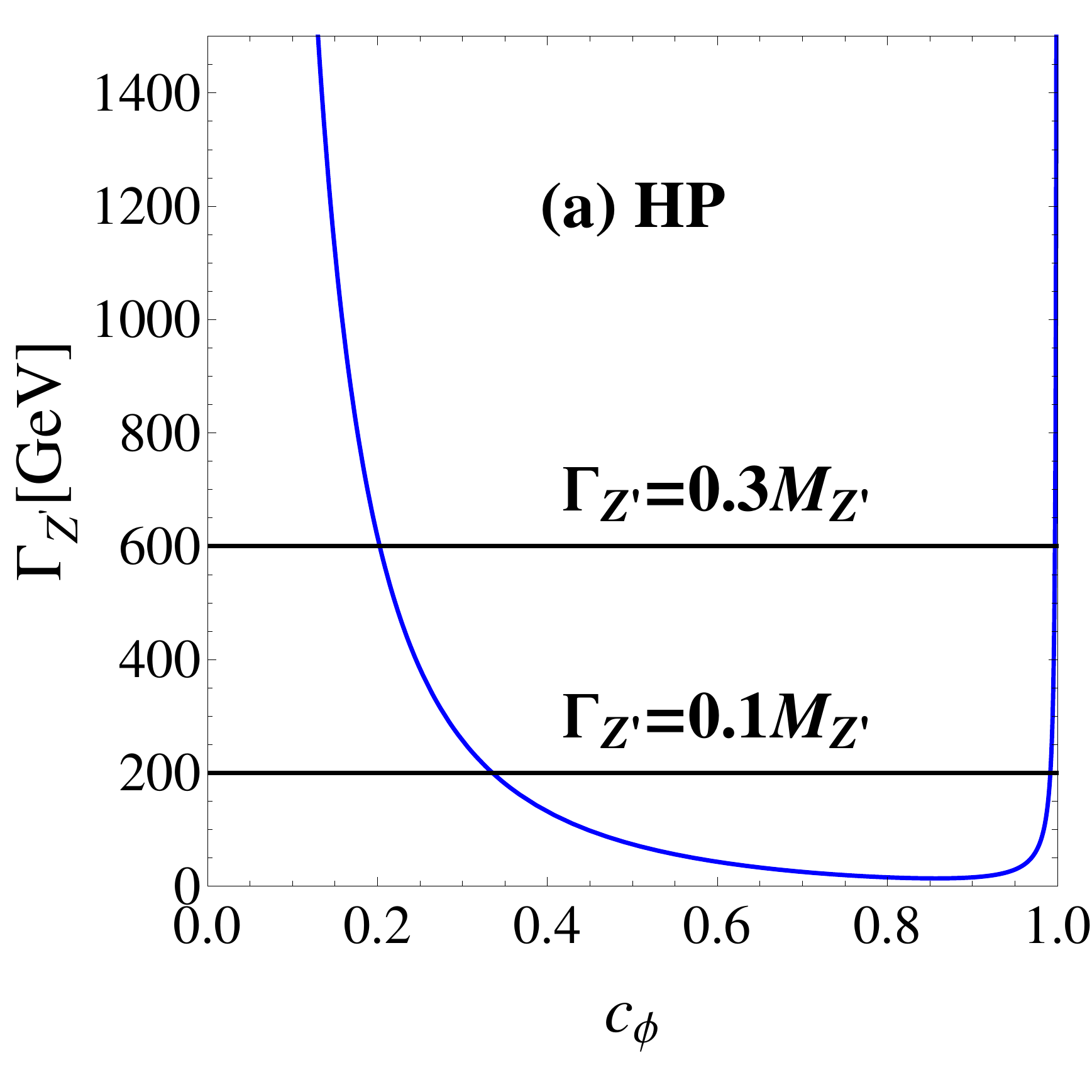}
\includegraphics[width=0.32\textwidth]{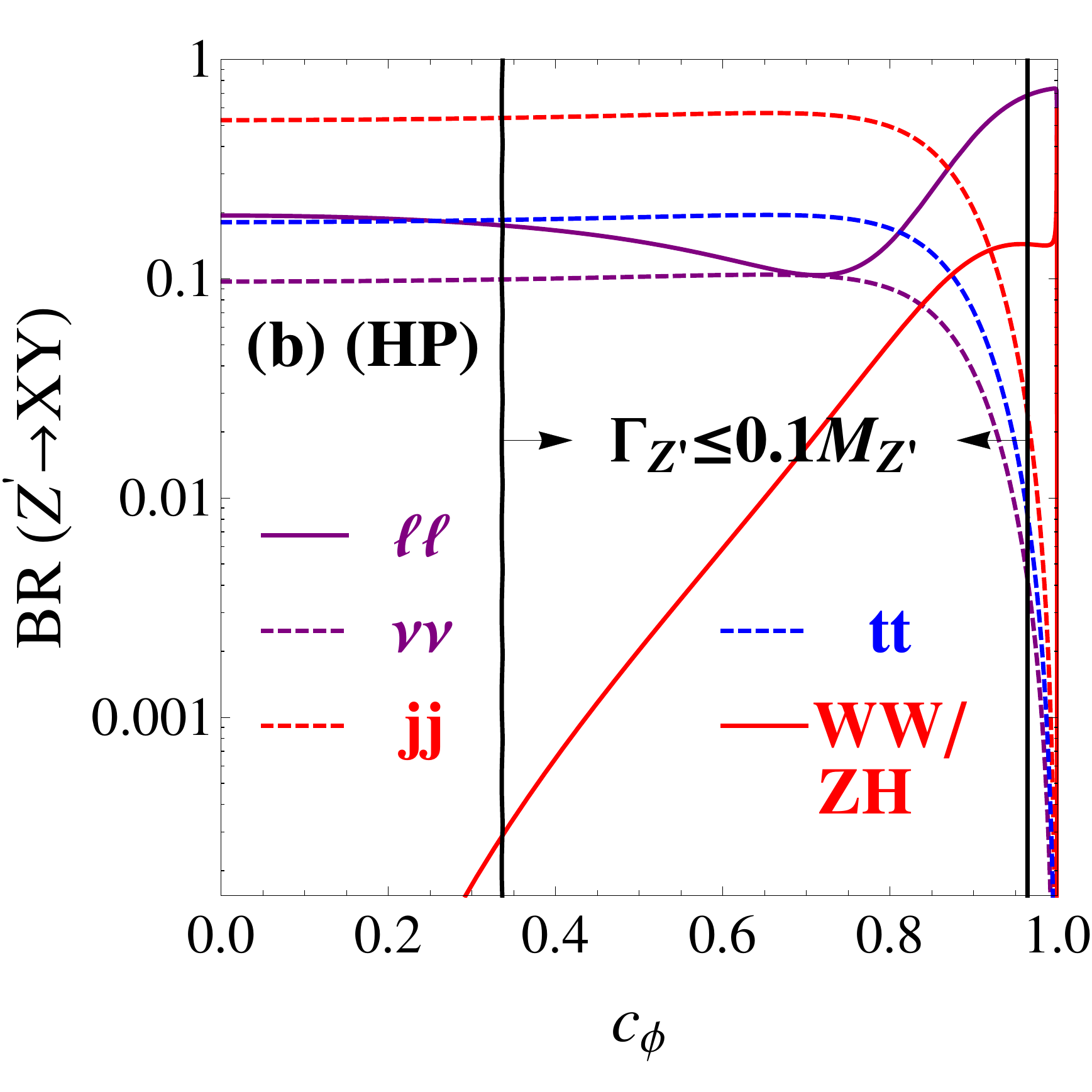}\\
\caption{\it The total width $\Gamma_{Z^\prime}$ (a)  and the branching ratios of the $Z^\prime$ decay (b) as a function of $c_{\phi}$ in Hadro-Phobic model.}\label{HPz1}
\end{figure}

\begin{figure}
\includegraphics[width=0.32\textwidth]{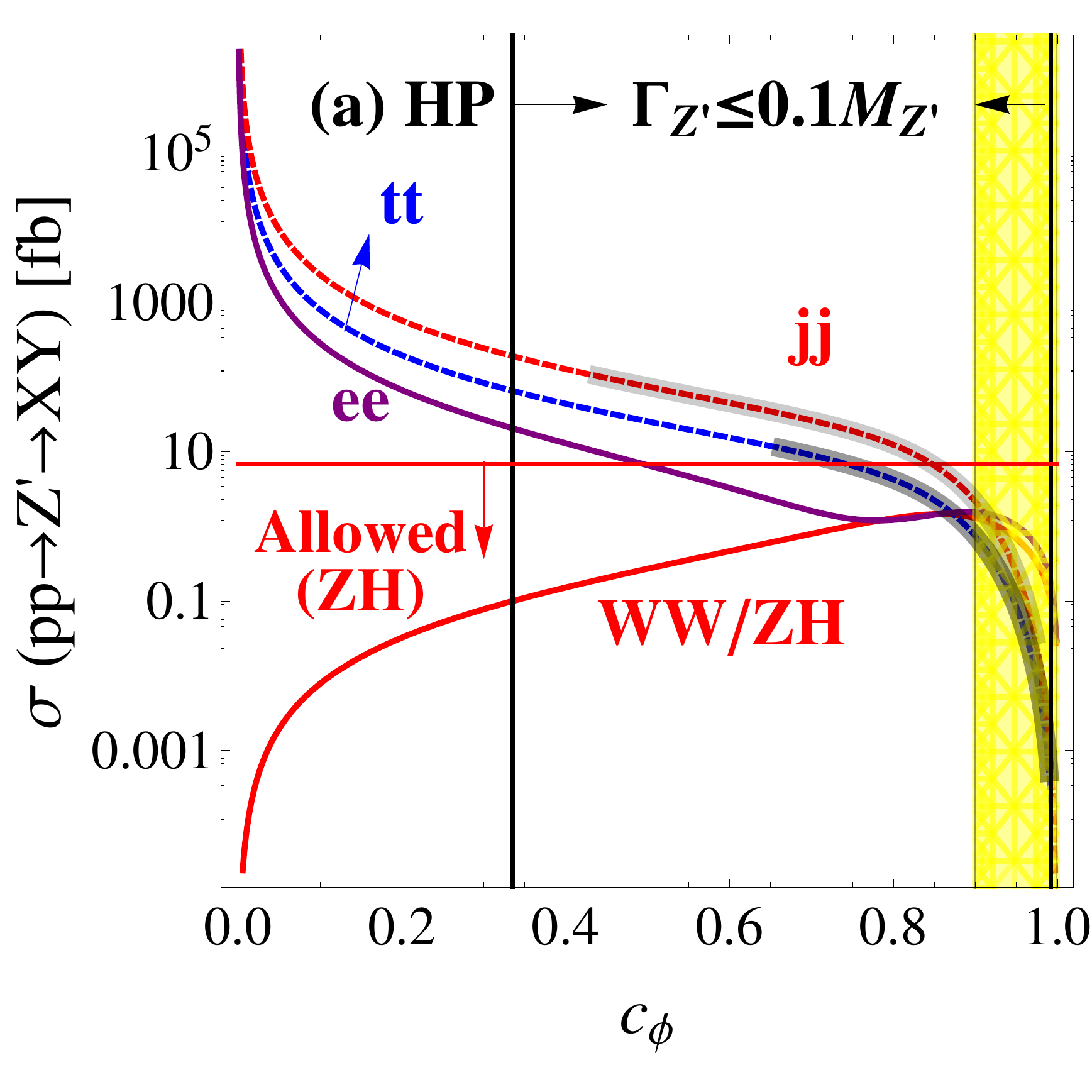}
\includegraphics[width=0.32\textwidth]{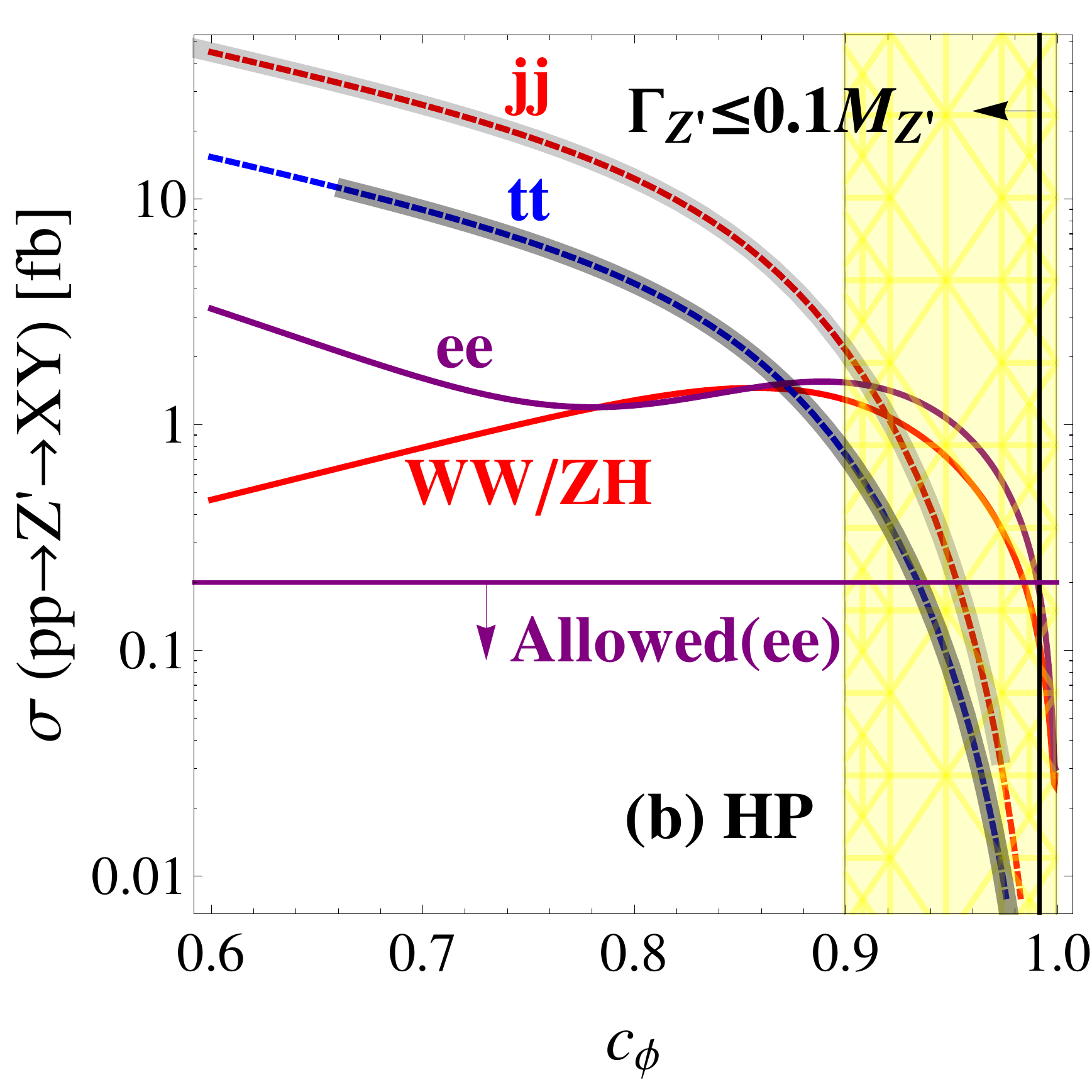}\\
\caption{\it The contours of the cross section of $\sigma(Z^\prime) \times {\rm BR}(Z^\prime \to XY)$, where $X$ and $Y$ denote the SM particles in the $Z^\prime$ decay as a function of $c_\phi$ in the Hadro-Phobic model. The shaded bands are corresponding to the allowed regions by the current experimental data. The yellow shaded region is required for $M_{W^\prime}\simeq M_{Z^\prime}$. 
}\label{HPz2}
\end{figure}

Now we consider the phenomenology of the $Z^\prime$ boson in the Hadro-Phobic doublet model. We require $\Gamma(Z^\prime)\leq 0.1 M_{Z^\prime}$, which leads to $0.34 \leq c_\phi \leq 0.99$; see Fig.~\ref{HPz1}(a). Figure~\ref{HPz1}(b) displays the decay branching ratios of $Z^\prime$. We note that the branching ratio of $Z^\prime \to jj$ and $Z^\prime \to t\bar{t} $ is suppressed for a large $c_\phi$ as one can see from Eq.~\eqref{zpf}.

In Fig.~\ref{HPz2} we present the cross section $\sigma(Z^\prime) \times {\rm BR}(Z^\prime \to XY)$ as a function of $c_\phi$. The curves show the theoretical predictions while the shaded band along each curve is allowed by current experimental data. The yellow shaded region is required for $M_{W^\prime}\simeq M_{Z^\prime}$.
The current bound on $\sigma(Z^\prime)\times {\rm BR}(Z^\prime \to t\bar{t})$ mode demands $0.66\leq c_\phi \leq 1$; see the blue-dashed curve with the $tt$ label. The di-jet constraint is slightly weaker than the $tt$ constraint. 
There's no parameter space to explain the $WW$ excess.
Furthermore, the cross section $\sigma(Z^\prime) \times {\rm BR}(Z^\prime \to ee)$ is above the current experimental constraint; see Fig.~\ref{HPz2}(b) for details. Thus, we conclude that it cannot explain the $WW$ excess in the Hadro-Phobic model.

\subsection{Fermio-Phobic doublet model}

\subsubsection{The $W^\prime$ constraints}

\begin{figure}
\includegraphics[width=0.32\textwidth]{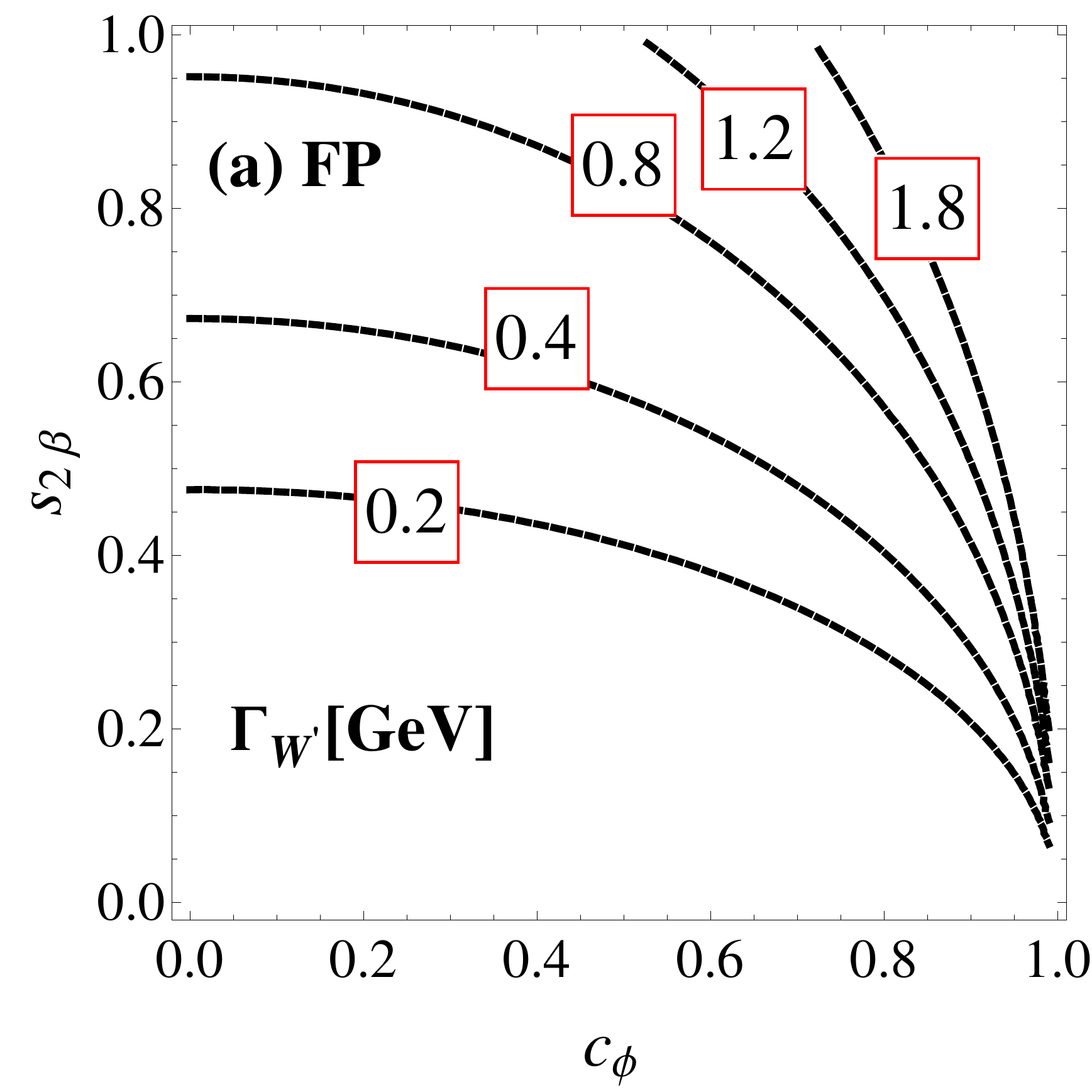}
\includegraphics[width=0.32\textwidth]{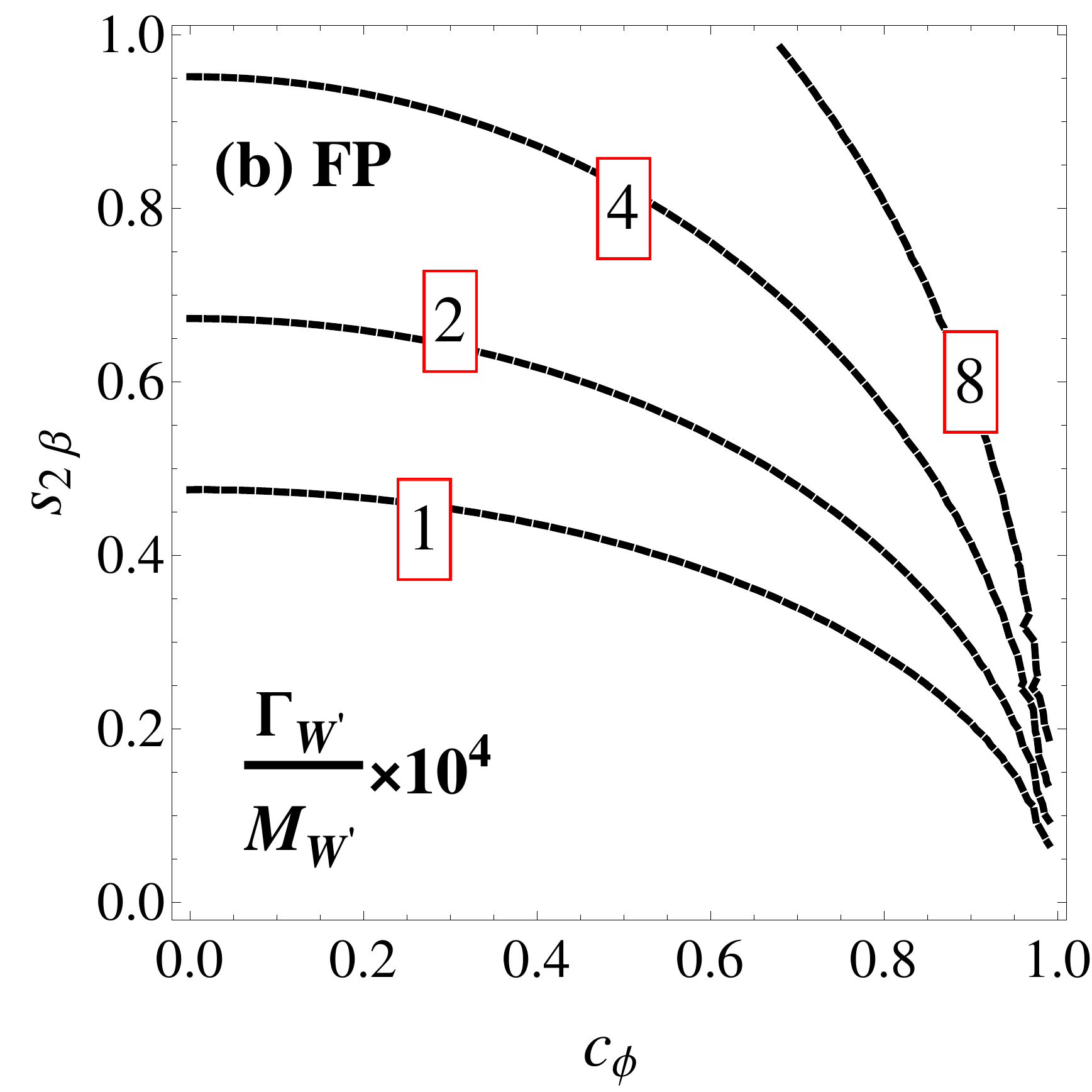}
\caption{\it The total width $\Gamma_{W^\prime}$ (a)  and $\Gamma_{W^\prime}/M_{W^\prime}$ (b) in the plane of $c_\phi$ and $s_{2\beta}$ in the Fermio-Phobic doublet model. 
}\label{FPw1}
\end{figure}

Finally, we examine the Fermio-Phobic doublet model in which both the SM quark and lepton doublets are gauged only under $SU(1)_1$; see Table~\ref{tb:models}. The gauge couplings of $W^\prime$ to SM fermions are suppressed due to the fact that the SM fermions are not gauged under gauge group $SU(2)_2$. The $W^\prime$ width in the Fermio-Phobic model is less than the $W^\prime$ width in the Lepto-Phobic and Hadro-Phobic models. Figure~\ref{FPw1} displays the contour of the total width $\Gamma_{W^\prime}$ and $\Gamma_{W^\prime}/M_{W^\prime}$ in the plane of $c_\phi$ and $s_{2\beta}$. Again, the NWA is a good approximation in the Fermio-Phobic doublet model.

\begin{figure}
\includegraphics[width=0.32\textwidth]{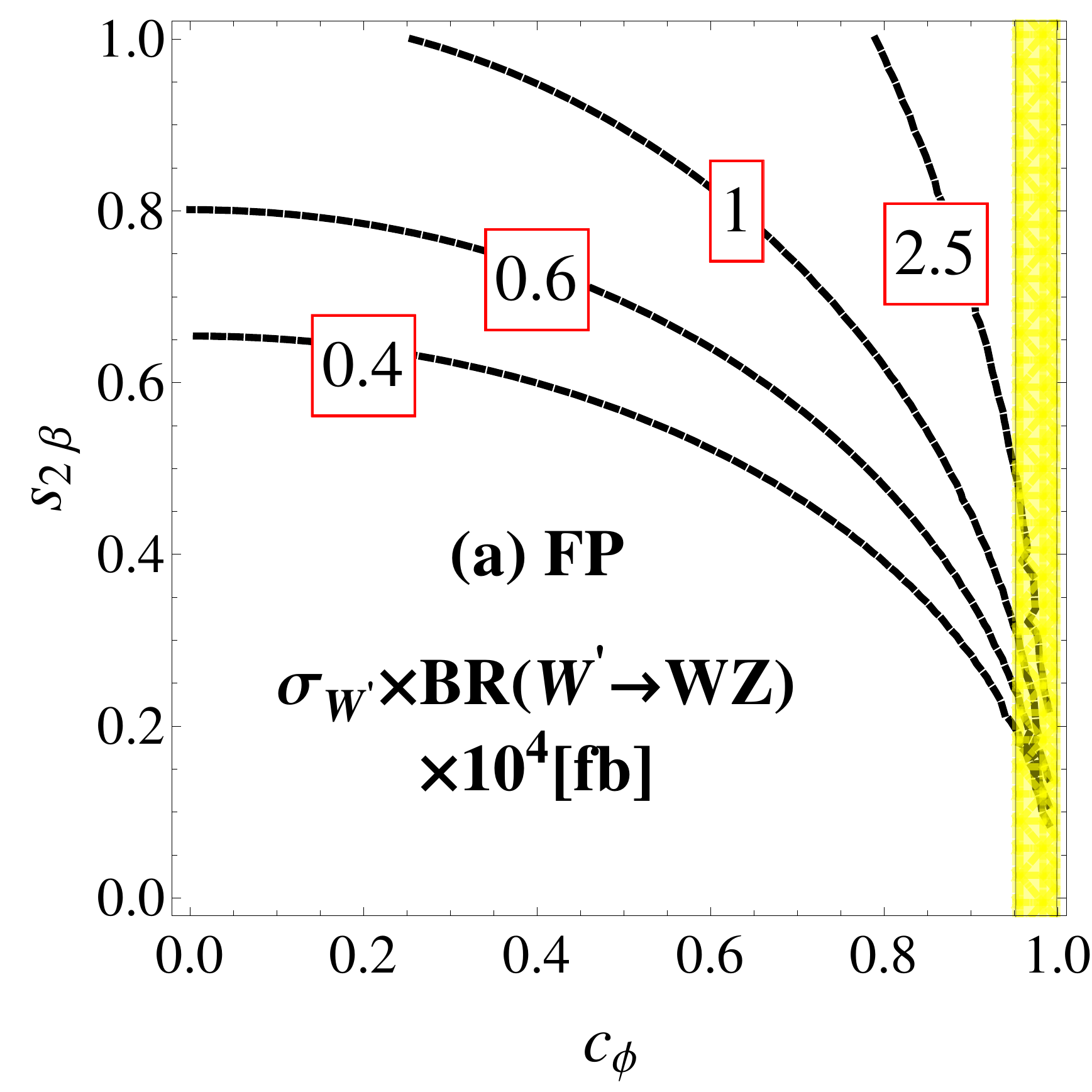}
\includegraphics[width=0.32\textwidth]{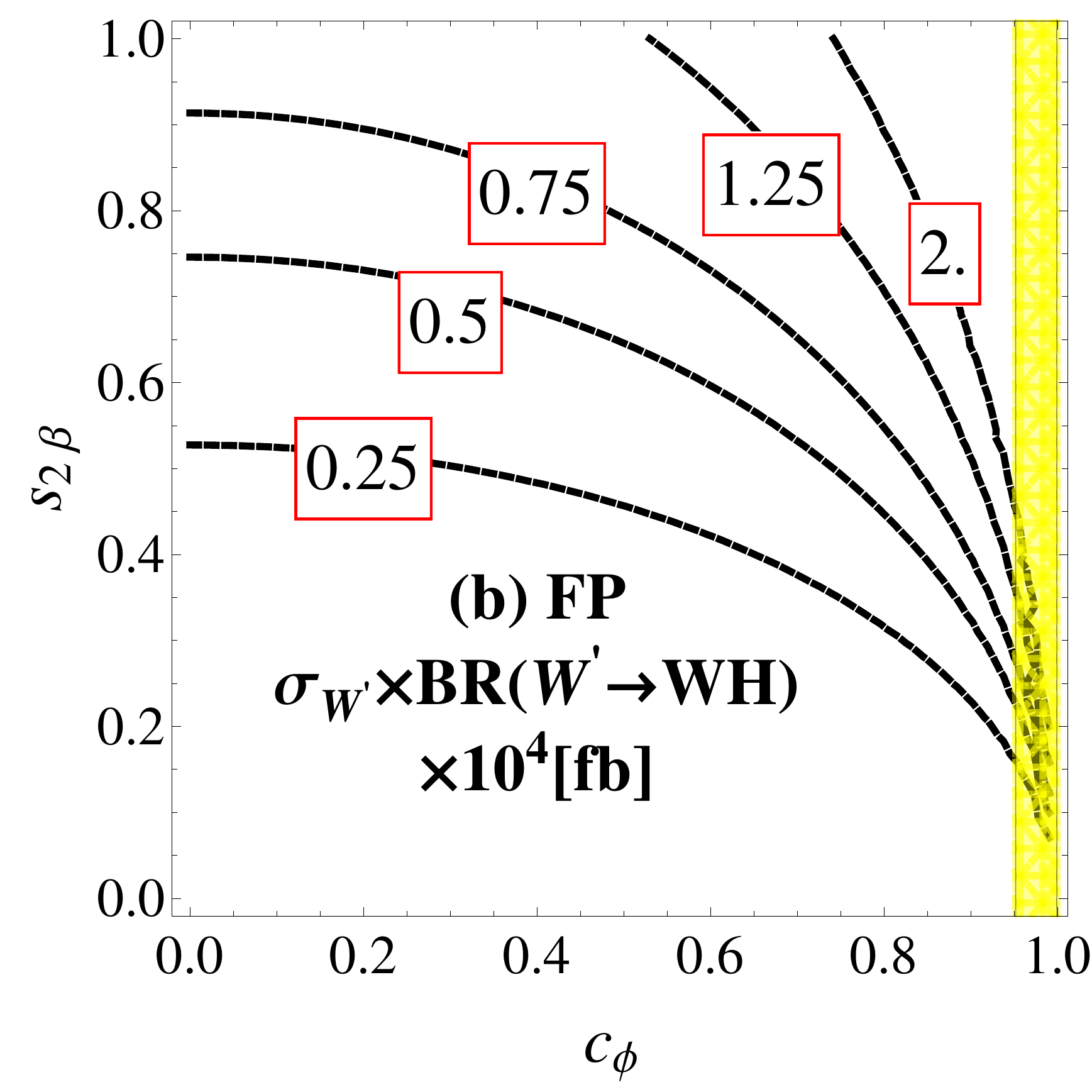}\\
\includegraphics[width=0.32\textwidth]{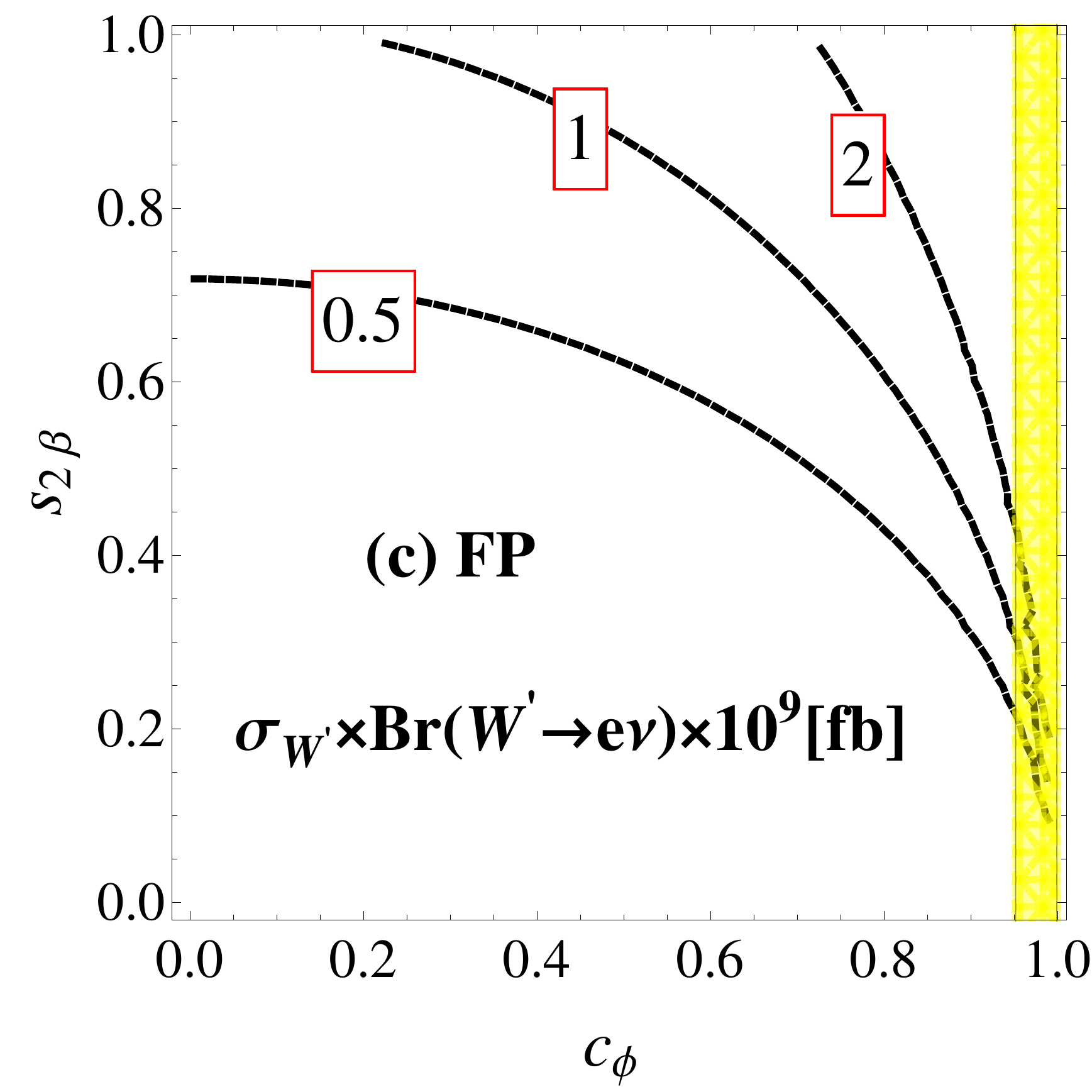}
\includegraphics[width=0.32\textwidth]{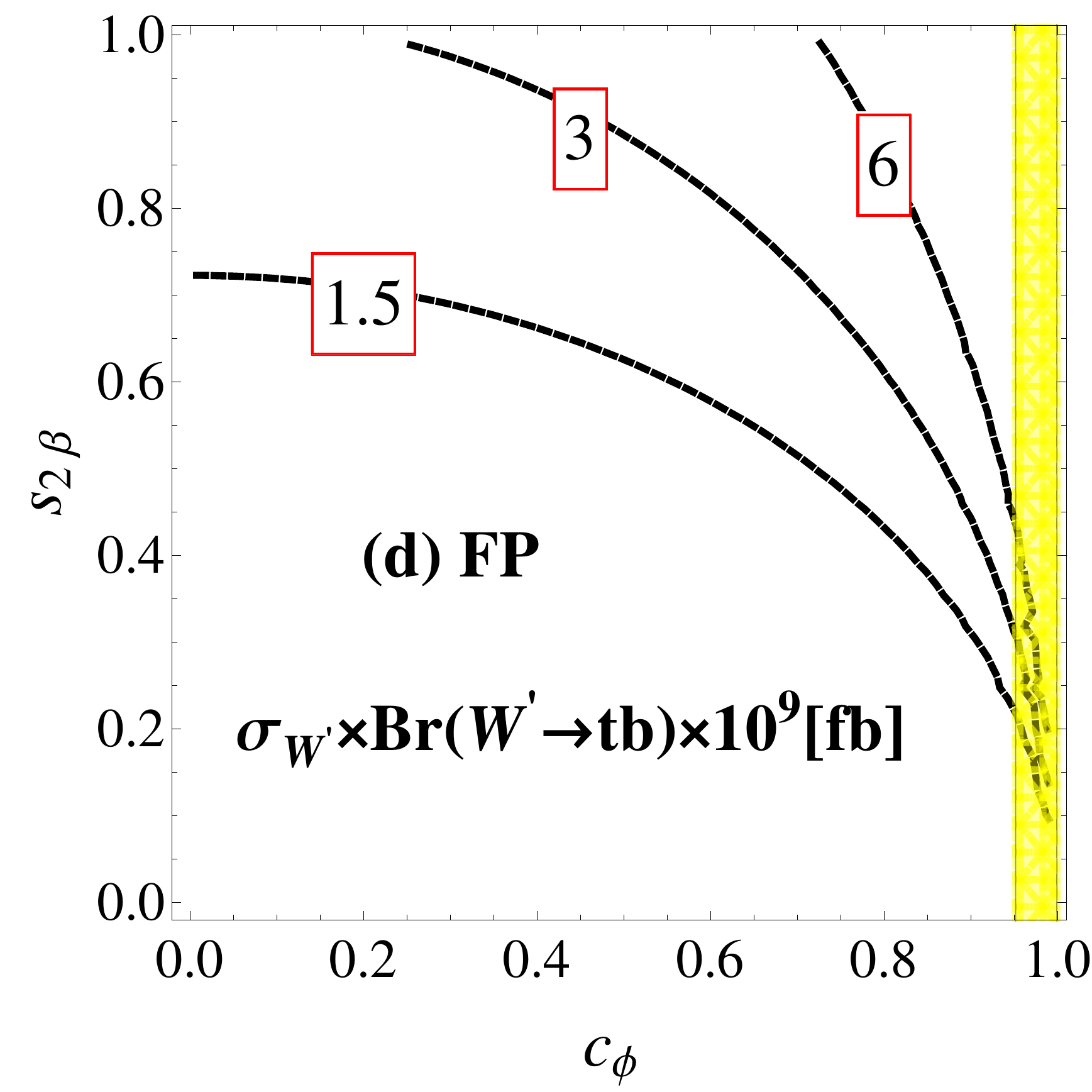}
\caption{\it The contours of the cross section (a) $\sigma(W^\prime) \times {\rm BR}(W^\prime \to WZ)$, (b) $\sigma(W^\prime)\times {\rm BR}(W^\prime \to WH)$, (c) $\sigma(W^\prime)\times {\rm BR}(W^\prime \to e\nu)$ and (d) $\sigma(W^\prime)\times {\rm BR}(W^\prime \to tb)$ in the plane of $c_\phi$ and $s_{2\beta}$ in the Fermio-Phobic doublet model. The yellow shaded region is required for $M_{W^\prime}\simeq M_{Z^\prime}$.
}\label{FPw2}
\end{figure}

The production cross section of $W^\prime$ in the model is much smaller than the cross section in the Left-Right and Lepton-Phobic models. It is, however, comparable to the Hadro-Phobic model. Figure~\ref{FPw2}  displays the contour of the cross section of $\sigma({W^\prime})\times {\rm BR}(W^\prime \to WZ/WH/e\nu/tb)$ in the plane of $c_\phi$ and $s_{2\beta}$. The yellow shaded region is required for $M_{W^\prime}\simeq M_{Z^\prime}$. Owing to the suppress of the production rate, the typical value of cross section in $WZ$ and $WH$ modes are around $10^{-4}$ fb. The branching ratios of $W^\prime$ decay to lepton/quark final states are suppressed dramatically due to the $W$-$W^\prime$ mixing and leads to $\sigma({W^\prime})\times {\rm BR}(W^\prime \to e\nu/tb/jj)\sim10^{-9}~{\rm fb}$. It is clear that the cross section at the all parameter space  is much smaller than $1~{\rm fb}$ such that it cannot explain the $WZ$ excess.

\subsubsection{The $Z^\prime$ constraints}

\begin{figure}\centering
\includegraphics[width=0.32\textwidth]{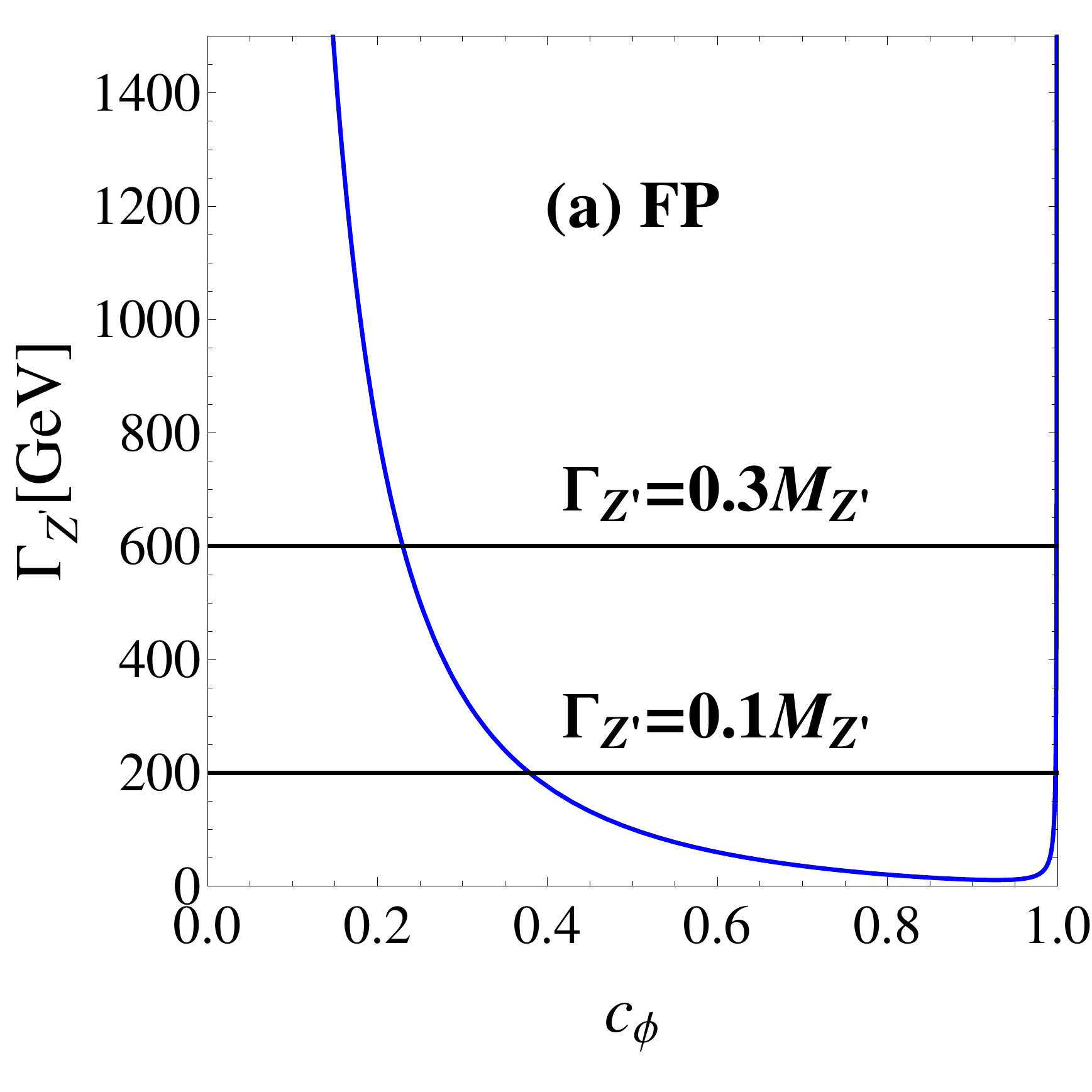}
\includegraphics[width=0.32\textwidth]{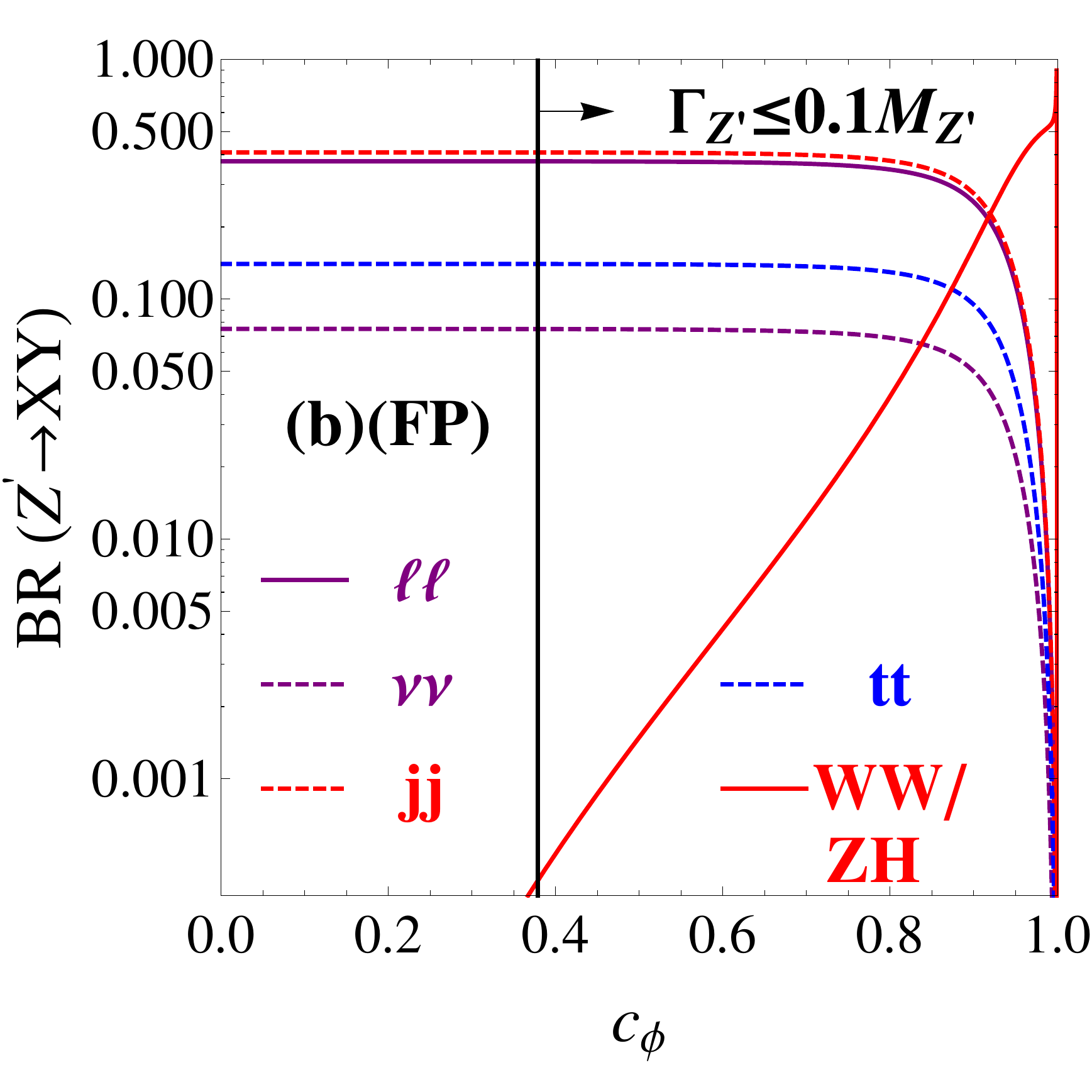}\\
\caption{\it The total width (a)  and the branching ratios of $Z^\prime$ decays (b) as a function of $c_{\phi}$ in Fermio-Phobic model. 
}\label{FPz1}
\end{figure}

In the Fermio-Phobic doublet model, the $Z^\prime$ couples to the SM fermions via the $U(1)_X$ component and the coupling strength is large in the region of $c_\phi \sim 0$ where $g_X\gg g_2$. We require $\Gamma(Z^\prime)\leq 0.1 M_{Z^\prime}$, which leads to $ c_\phi \geq 0.38$; see Fig.~\ref{FPz1}(a). Figure~\ref{FPz1}(b) displays the branching ratios of all the decay modes of $Z^\prime$. We note that the branching ratio of $Z^\prime \to WW$ and $Z^\prime \to ZH $ is highly enhanced for a large $c_\phi$, e.g. ${\rm BR}(Z^\prime \to WW/ZH) > 0.1$ when $c_\phi > 0.85$, which is different from other BP-I models. 
It is owing to the fact that the the decay rate of $W^\prime$ to SM fermions is highly suppressed when $c_\phi \to 1$ in this model. 

\begin{figure}\centering
\includegraphics[width=0.32\textwidth]{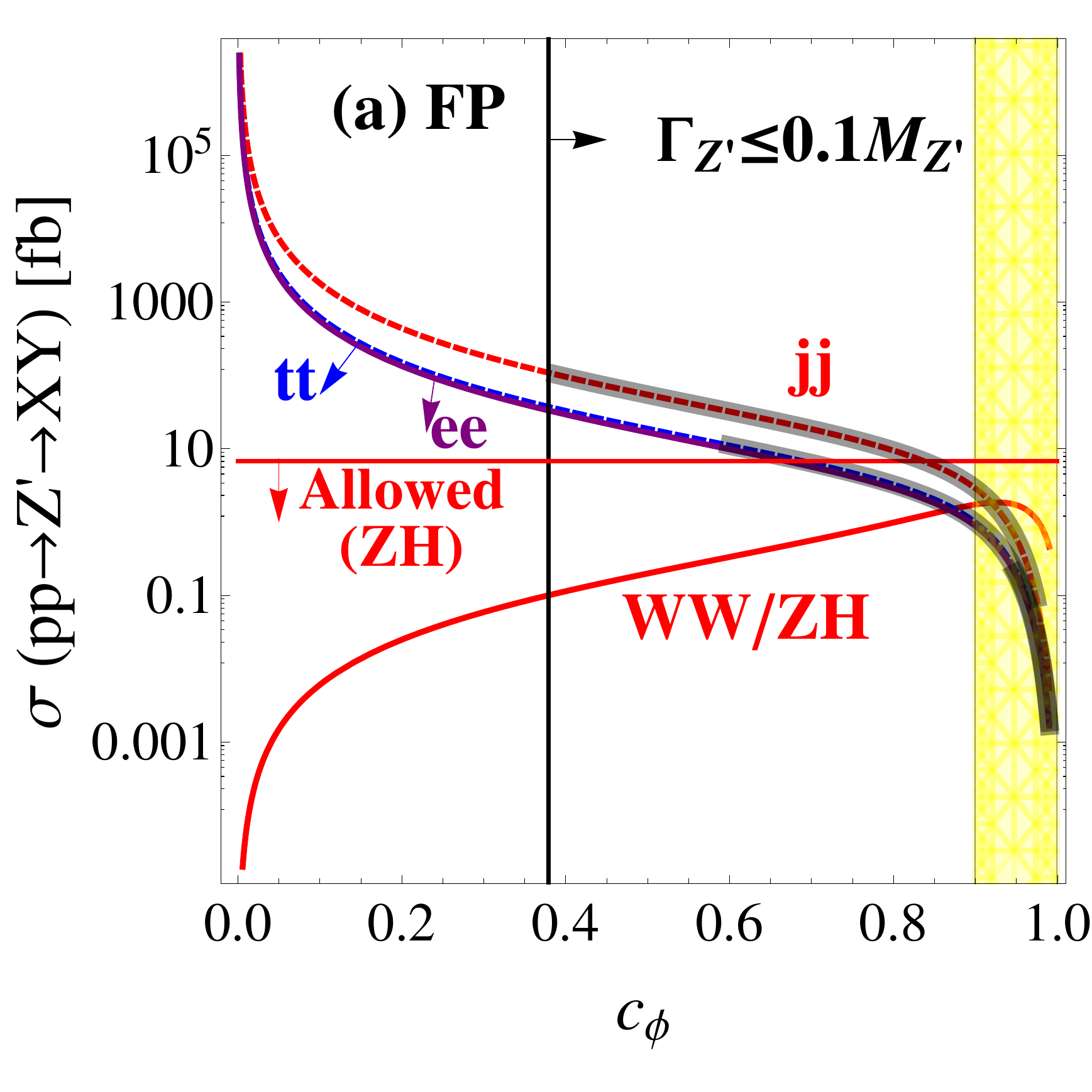}
\includegraphics[width=0.32\textwidth]{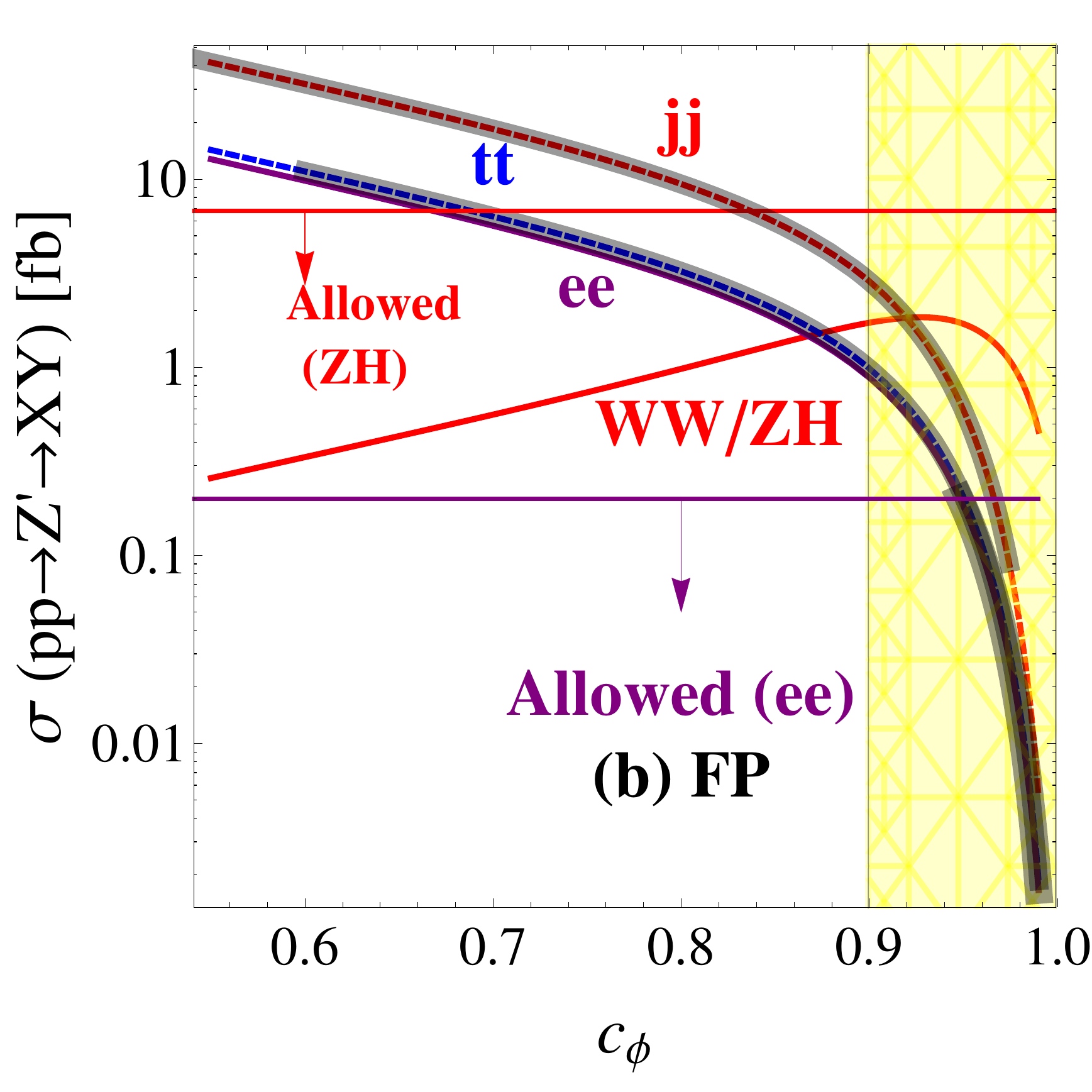}\\
\caption{\it The cross section contours of $\sigma(Z^\prime) \times {\rm BR}(Z^\prime \to XY)$, where $X$ and $Y$ denote the SM particles in the $Z^\prime$ decay as a function of $c_\phi$ in the Fermio-Phobic model. The yellow shaded region is required for $M_{W^\prime}\simeq M_{Z^\prime}$.
}\label{FPz2}
\end{figure}

In Fig.~\ref{FPz2} we present the cross section $\sigma(Z^\prime) \times {\rm BR}(Z^\prime \to XY)$, where $X$ and $Y$ denote the SM particles in the $Z^\prime$ decay, as a function of $c_\phi$. The  curves show the theoretical predictions while the shaded bands are allowed by current experimental data. The yellow shaded region is required for $M_{W^\prime}\simeq M_{Z^\prime}$.
The current bound on $\sigma(Z^\prime)\times {\rm BR}(Z^\prime \to t\bar{t})$ mode, denoted as $tt$ in the figure,
demands $0.6\leq c_\phi \leq 1$. The di-jet constraint is slightly weaker than the $tt$ constraint. 
The whole parameter space satisfies the current bound on $\sigma(Z^\prime)\times {\rm BR}(Z^\prime \to ZH)$, but cannot explain the excess of $WW$.
Again, the leptonic decay mode imposes much tighter constraint as $\sigma(Z^\prime)\times {\rm BR}(Z^\prime \to e^+e^-)\leq 0.2~{\rm fb}$ by the current measurements~\cite{Aad:2014cka,Khachatryan:2014fba}, which requires $c_\phi>0.95$. Thus we conclude that the Fermio-Phobic doublet model cannot explain the $WW$ excess. 

\section{G(211) Models: Breaking pattern II}\label{221b}

In the BP-II, $U(1)_X$ is identified as the $U(1)_Y$ of the SM. The first stage of symmetry breaking $SU(2)_{1}\times SU(2)_{2}\to SU(2)_{L}$ occurs at the TeV scale, which is owing to a scalar bi-doublet $\Phi\sim(2,\bar{2},0)$ with only one VEV $u$. The subsequent breaking of $SU(2)_{L}\otimes U(1)_{Y}\to U(1)_{\rm em}$ at the electroweak scale is generated by a Higgs doublet $H\sim(2,1,1/2)$ with a VEV $v$. The explicit forms of the bi-doublet and doublet as well as their vacuum expectation values are given as follows:
\begin{align}
&\Phi=
\begin{pmatrix}\phi^{0} &\sqrt{2}\phi^{+} \\ \sqrt{2}\phi^{-} & \phi^{0} \end{pmatrix},
&\VEV{\Phi} &= \mfrac{1}{2}
\begin{pmatrix}u & 0 \\ 0 & u \end{pmatrix}, \nn \\
&H = \begin{pmatrix} h^+ \\ h^0\end{pmatrix},
&\VEV{H} &=
\mfrac{1}{\sqrt{2}}
\begin{pmatrix} 0 \\ v \end{pmatrix}.
\end{align}
In the BP-II, the couplings of the three gauge groups are 
\begin{eqnarray}
g_{1}=\frac{e}{s_{W}c_{\phi}},\quad g_{2}=\frac{e}{s_{W}s_{\phi}},\quad g_{X}=\frac{e}{c_{W}},
\end{eqnarray}
where $\phi=\arctan(g_2/g_1)$ is the mixing angle. 
After the symmetry breaking both $W^\prime$ and $Z^\prime$ bosons obtain their masses and are degenerated at the tree level,
\bea
M_{{W^{\prime}}^{\pm}}^{2}=M_{Z^{\prime}}^{2}=\frac{e^{2}v^{2}}
{4s_{W}^{2}s_{\phi}^{2}c_{\phi}^{2}}\left(x+s_{\phi}^{4}\right)\,. 
\label{mvp_bp2}
\eea
The gauge couplings of $W^\prime$ and $Z^\prime$ to the SM Higgs boson and gauge bosons are generated after the second stage of the symmetry breaking, which are given as follows,
\bea
H~W_{\nu}~W_{\rho}^{\prime}  &\quad:\quad&  \frac{1}{2}\frac{e^{2}s_{\phi}}{s_{W}^{2}c_{\phi}}vg_{\nu\rho}\biggl[1+\frac{s_{\phi}^{2}\left(c_{\phi}^{2}-s_{\phi}^{2}\right)}{x}\biggr], \nn\\
H~Z_{\nu}~Z_{\rho}^{\prime} &:& \frac{1}{2}\frac{e^{2}s_{\phi}}{c_{W}s_{W}^{2}c_{\phi}}vg_{\nu\rho}\biggl[1-\frac{s_{\phi}^{2}\left(s_{\phi}^{2}c_{W}^{2}-c_{\phi}^{2}\right)}{xc_{W}^{2}}\biggr], \nn\\
W_{\mu}^{+}~W_{\nu}^{\prime-}~Z_{\rho}&:& \frac{ec_{\phi}s_{\phi}^{3}}{xc_{W}s_{W}}, \nn\\
W_{\mu}^{+}~W_{\nu}^{-}~Z_{\rho}^{\prime} &:& \frac{ec_{\phi}s_{\phi}^{3}}{xs_{W}}.
\eea
In BP-II the bosonic decays of $W^\prime/Z^\prime$ in the limit of $x\gg 1$ and $M_{W^\prime}\gg m_{W/Z/H}$ are correlated as follows
\beq
\frac{{\rm BR}(W^\prime \to WZ)}{{\rm BR}(W^\prime \to WH)} \sim  1\quad, \quad\frac{{\rm BR}(Z^\prime \to WW)}{{\rm BR}(Z^\prime \to ZH)} \sim 1. 
\eeq

The couplings of the $W^\prime$ bosons to the SM fermions in the BP-II are 
\begin{align}
g_L^{W^\prime\bar{f}f^\prime} &=\frac{es_{\phi}}{\sqrt{2}s_{W}c_{\phi}}\gamma^{\mu}\left(1+\frac{s_{\phi}^{2}c_{\phi}^{2}}{x}\right), &
g_R^{W^\prime\bar{f}f^\prime} &= 0,\nn \\
g_L^{W^\prime\bar{F}F^\prime} &=-\frac{ec_{\phi}}{\sqrt{2}s_{W}s_{\phi}}\gamma^{\mu}\left(1-\frac{s_{\phi}^{4}}{x}\right), &
g_R^{W^\prime\bar{F}F^\prime} &= 0.
\label{BPIIWF}
\end{align}
while those of the $Z^\prime$ boson are
\begin{align}
g_L^{Z^{\prime}\bar{f}f} &= \frac{e}{s_{W}}\gamma^{\mu}\left[\frac{s_{\phi}}{c_{\phi}}T_{3}^1\left(1+\frac{s_{\phi}^{2}c_{\phi}^{2}}{xc_{W}^{2}}\right)-\frac{s_{\phi}}{c_{\phi}}\frac{s_{\phi}^{2}c_{\phi}^{2}}{xc_{W}^{2}}s_{W}^{2}Q\right], \nn\\
g_R^{Z^{\prime}\bar{f}f} &=-\frac{e}{s_{W}}\gamma^{\mu}\left(\frac{s_{\phi}}{c_{\phi}}\frac{s_{\phi}^{2}c_{\phi}^{2}}{xc_{W}^{2}}s_{W}^{2}Q\right),\nn\\
g_L^{Z^{\prime}\bar{F}F} &=-\frac{e}{s_{W}}\gamma^{\mu}\left[\frac{c_{\phi}}{s_{\phi}}T_{3}^2\left(1-\frac{s_{\phi}^{4}}{xc_{W}^{2}}\right)+\frac{c_{\phi}}{s_{\phi}}\frac{s_{\phi}^{4}}{xc_{W}^{2}}s_{W}^{2}Q\right], \nn\\
g_R^{Z^{\prime}\bar{F}F} &=-\frac{e}{s_{W}}\gamma^{\mu}\left(\frac{c_{\phi}}{s_{\phi}}\frac{s_{\phi}^{4}}{xc_{W}^{2}}s_{W}^{2}Q\right),
\end{align}
where $f$ represents the fermions are gauged under $SU(2)_1$ while $F$ the fermions gauged under $SU(2)_2$. 

Next we consider Un-unified model and Non-universal/Top-Flavor model, and discuss their implications in the production of $W^\prime/Z^\prime$ and their decay modes of the $WZ/WW/WH/ZH$ pair at the LHC.

\subsection{Un-unified model}

\subsubsection{The $W^\prime$ constraints}

\begin{figure}\centering
\includegraphics[width=0.32\textwidth]{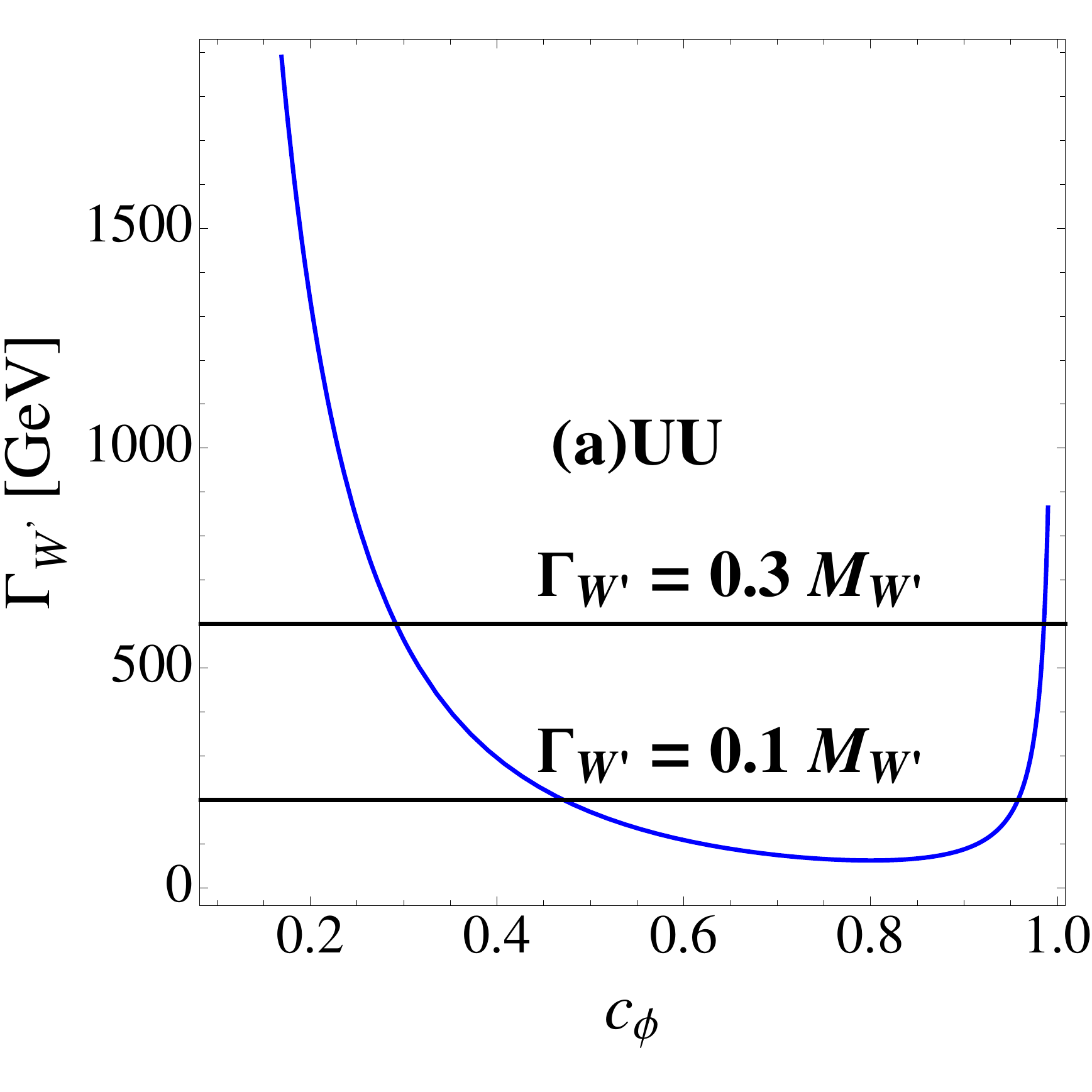}
\includegraphics[width=0.32\textwidth]{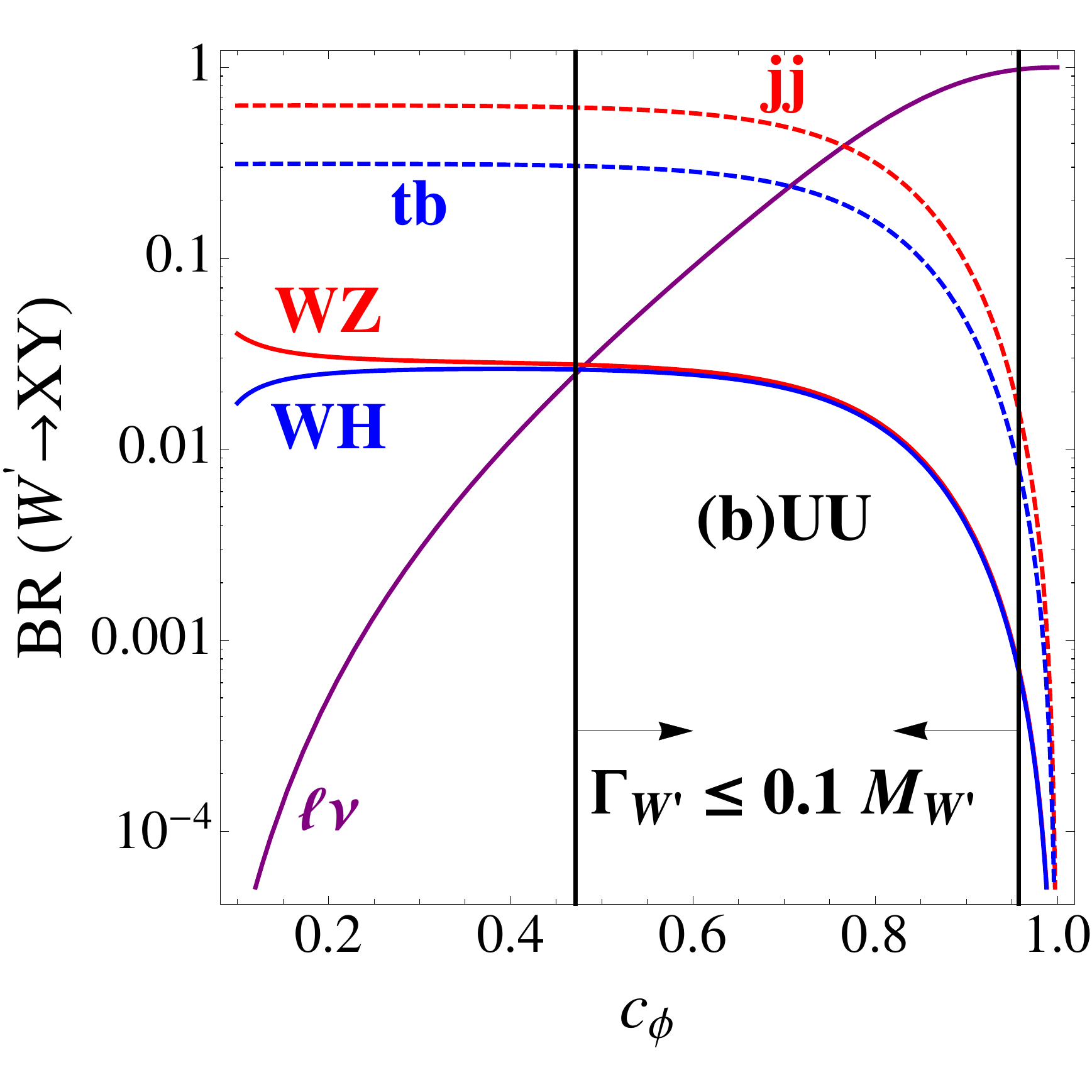}
\includegraphics[width=0.32\textwidth]{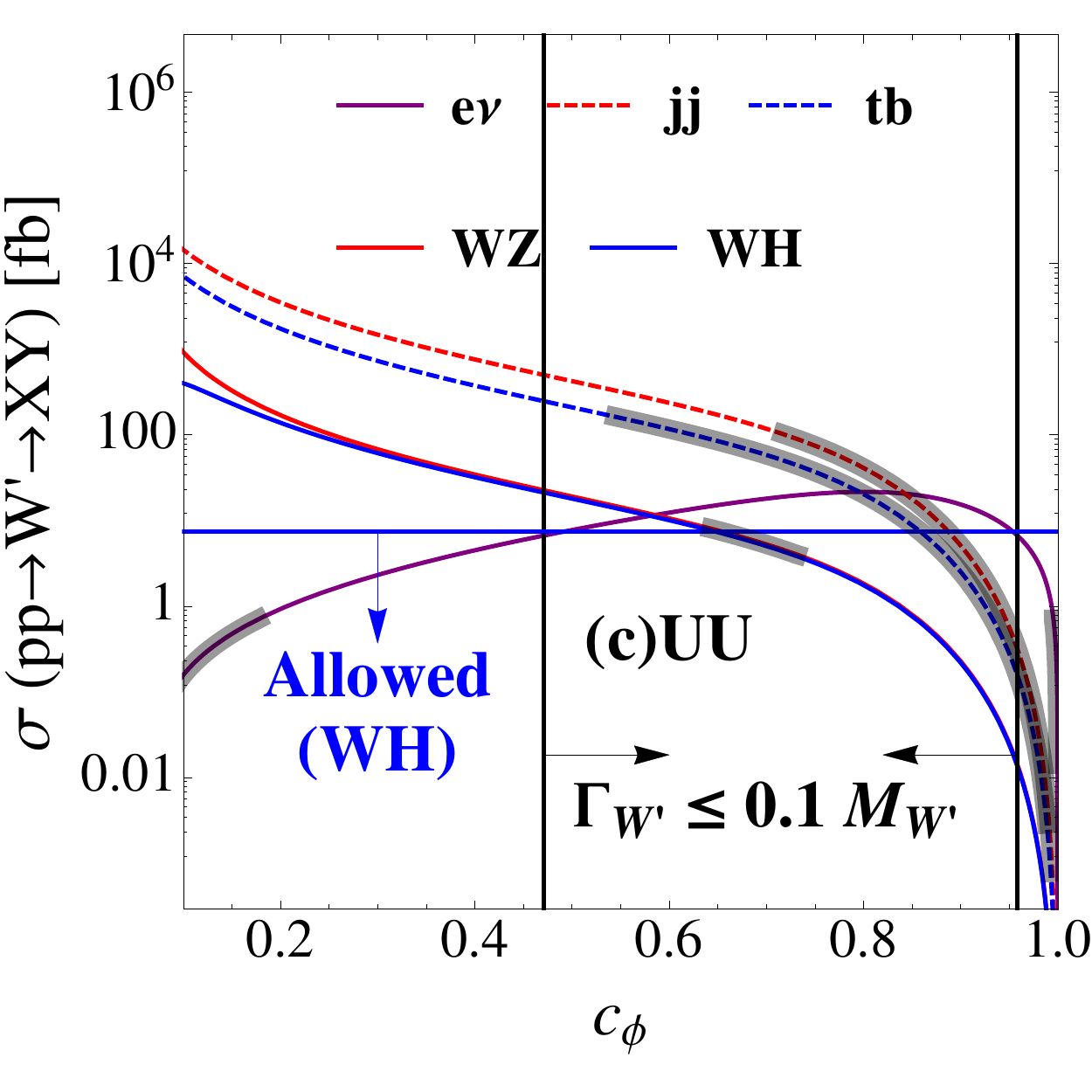}
\caption{\it (a) The total width $\Gamma_{W'}$ as a function of $c_\phi$ in the Un-unified (UU) model of BP-II. (b) The decay branching ratio $BR~(W'\to XY)$ as a function of $c_\phi$. (c) The cross section $\sigma~(pp\to W'\to XY)$ as a function of $c_\phi$ at the LHC Run-1. The shaded band of each curve satisfies the current experiment data.
}\label{UUW}
\end{figure}

We begin with the Un-unified model in which the left-handed quarks are gauged under $SU(2)_1$ while the lepton doublets gauged under $SU(2)_2$. Figure~\ref{UUW}(a) shows the total width $\Gamma_{W'}$ as a function of $c_\phi$. The $W^\prime$ couples to the SM quarks and leptons strongly in the region of  $c_\phi \sim 0$  and $c_\phi \sim 1$, respectively.  That yields a wide width of $W^\prime$. In order to validate the NWA, we demand $\Gamma_{W'}\le 0.1M_{W'}$ which is presented by the black horizontal line. It requires $0.47\leq c_\phi \leq 0.96$. 

The branching ratios of $W'$ are plotted in Fig.~\ref{UUW}(b). For a large $c_\phi$, the branching ratio of $W^\prime \to jj/tb$ are suppressed while the branching ratio of $W^\prime \to l\nu$ is enhanced. Such a behavior can be understood from the gauge coupling of $W^\prime$ to the SM fermions; see Eq.~\eqref{BPIIWF}. The coupling of $W^\prime$ to the SM quarks is proportional to $\tan\phi$, while for the leptons, the gauge coupling is proportional to $\cot\phi$. The branching ratios of $W^\prime \to WZ/WH$ can reach $\sim 0.01$ for most of the parameter space in the model. Figure~\ref{UUW}(c) shows the cross sections of $\sigma(W^\prime)\times {\rm BR}(W^\prime\to XY)$. The shaded bands are consistent with current experimental data.  In order to explain the $WZ$ excess, one needs $0.64<c_\phi <0.73$. However, the $jj$ mode requires $c_\phi > 0.72$. There is a tension between the $WZ$ mode and the $jj$ mode. The negative searching result of the $WH$ mode demands $c_\phi>0.65$~. It is possible to satisfy the $WZ$, $jj$ and $WH$ modes within $2\sigma$ confidential level.  

We also plot the cross section of the leptonic decay in Fig.~\ref{UUW}(c). Unfortunately, the cross section of $\sigma(W^\prime)\times {\rm BR}(W^\prime \to e \nu)$ in the region of $c_\phi\sim 0.4-0.7$ is far beyond the current experimental limit. In order to explain the $WZ$ excess in the Un-unified model, one has to extend the model to reduce the leptonic decay mode.

\subsubsection{The $Z^\prime$ constraints}

\begin{figure}
\includegraphics[width=0.32\textwidth]{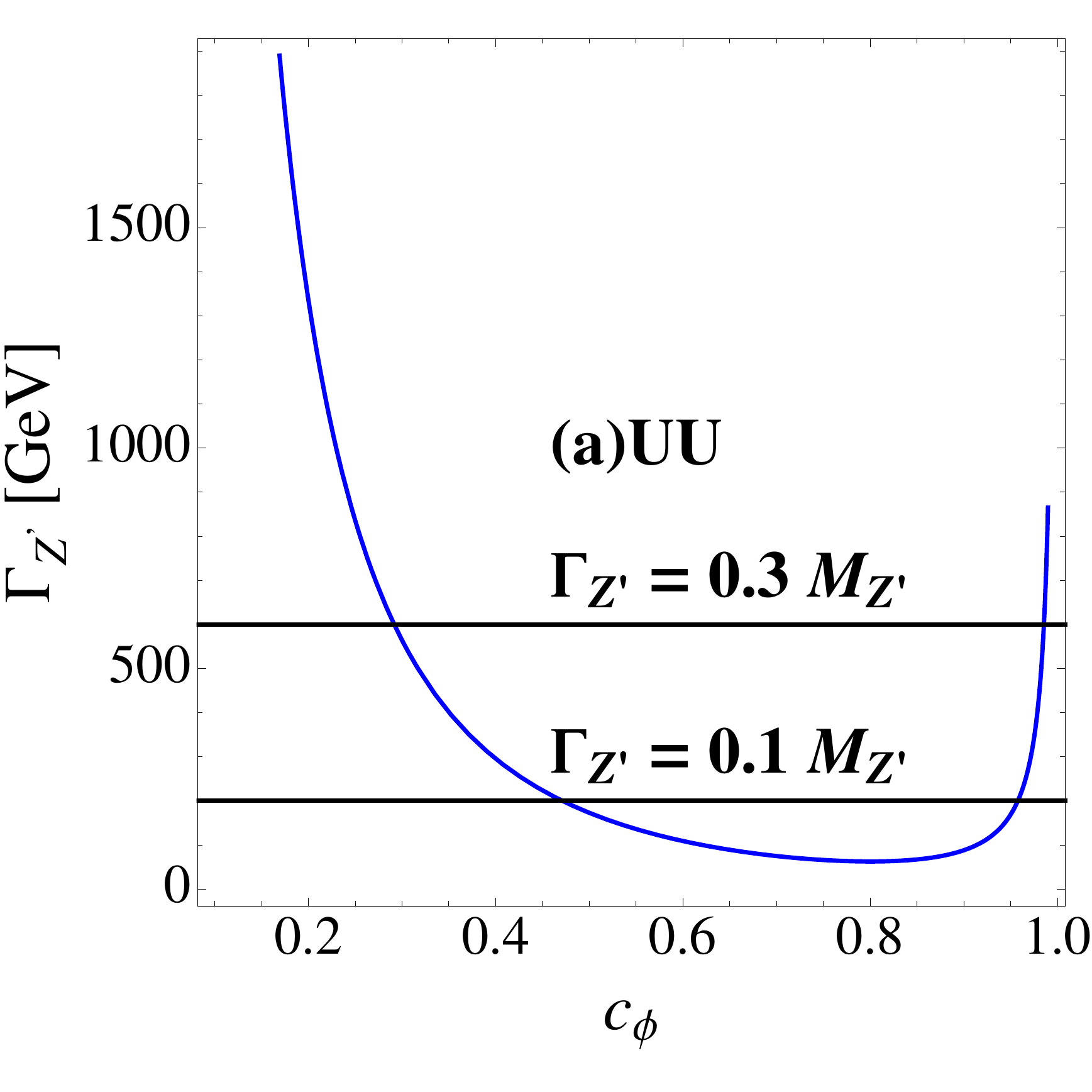}
\includegraphics[width=0.32\textwidth]{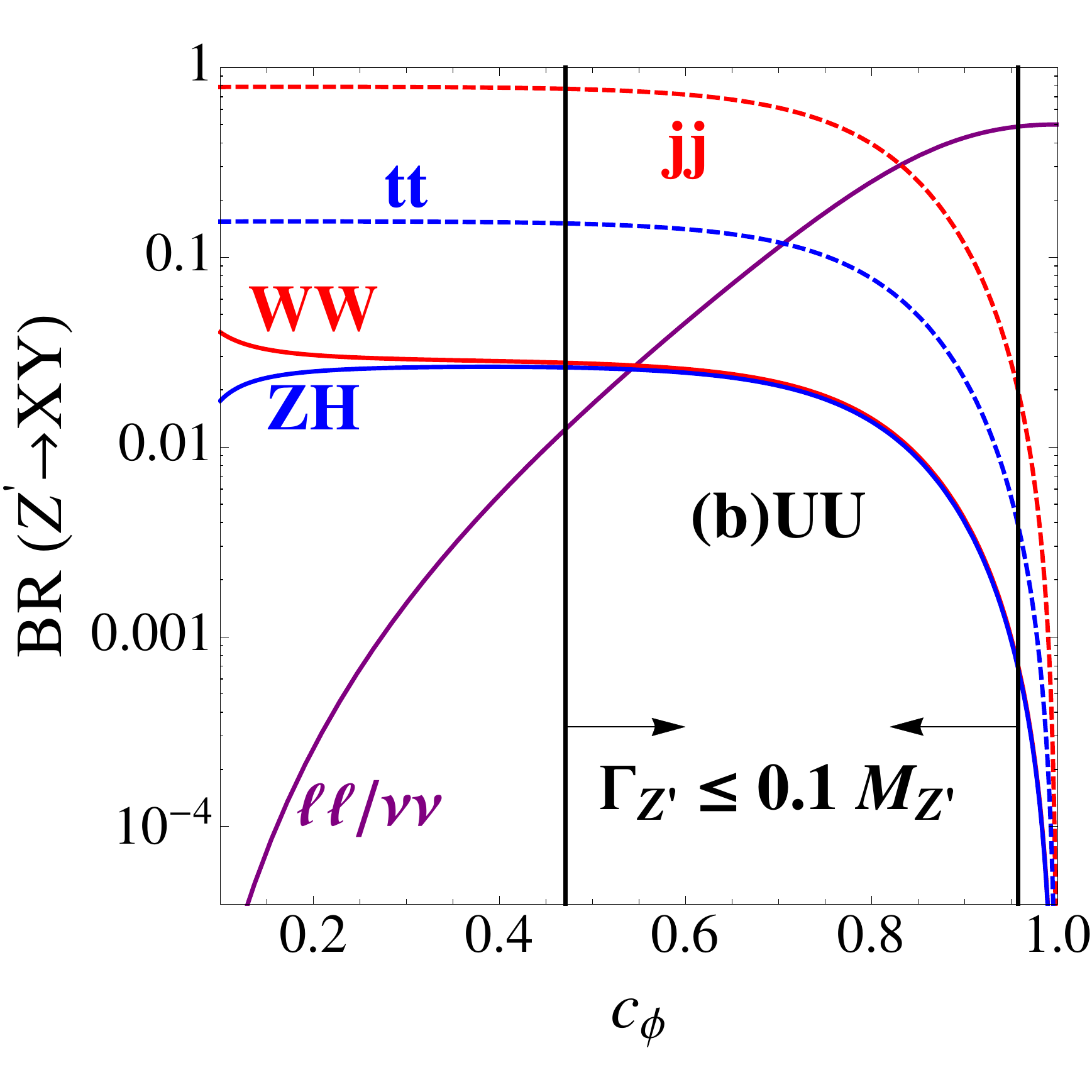}
\includegraphics[width=0.32\textwidth]{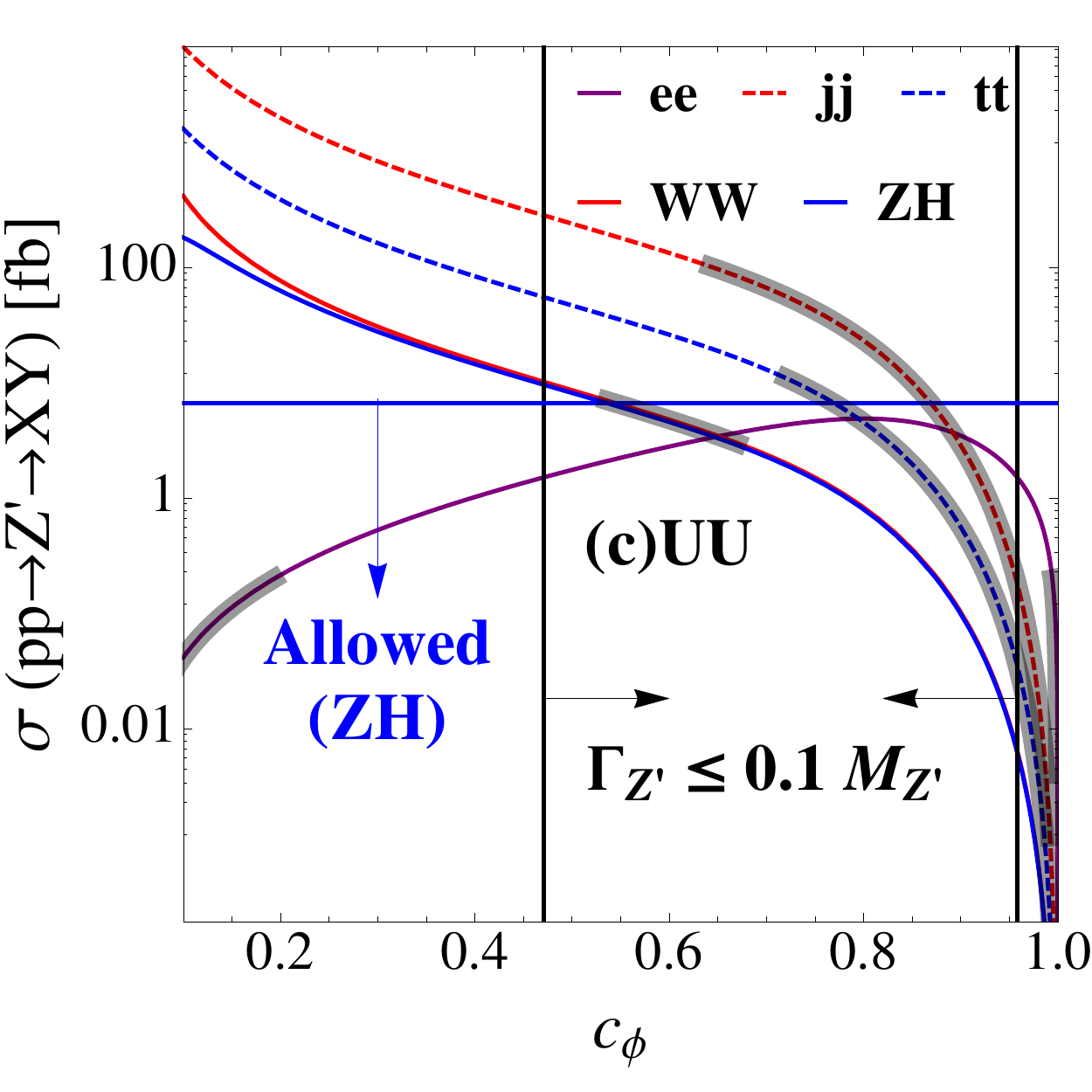}
\caption{\it (a)~The total width $\Gamma_{Z'}$ as a function of $c_\phi$ in the Un-unified (UU) model of BP-II. (b)~The decay branching ratio $BR~(Z'\to XY)$ as a function of $c_\phi$ in the Un-unified (UU) model of BP-II. (c)~The cross section $\sigma~(pp\to Z'\to XY)$ at LHC Run-1 as a function of $c_\phi$ in the Un-unified (UU) model of BP-II. The shaded band of each curve satisfies the current experiment.
}\label{UUZ}
\end{figure}

Figure~\ref{UUZ} shows the total width $\Gamma_{Z'}$ (a) and decay branching ratios of $Z^\prime$ (b) as a function of $c_\phi$. We also demand the narrow width constraint $\Gamma_{Z'}\le 0.1M_{Z'}$, which also requires $0.47\leq c_\phi \leq 0.96$. In analogue with $W^\prime$,  the branching ratios of $Z^\prime \to jj$ and $Z^\prime \to t\bar{t}$ are suppressed, while the branching ratio of $Z^\prime \to ll/\nu\nu$ are enhanced in the region of  large $c_\phi$. Note that the branching ratios of $W^\prime \to WZ/WH$ are independent on the variable $c_\phi$ in the range $0.3\leq c_\phi\leq 0.7$, which is about 0.03. Figure~\ref{UUW}(c) shows the cross section of various decay modes of $Z^\prime$.  We observe a tension between the $WW$ mode and the $jj$ mode. Again, the leptonic decay mode imposes much tighter constraint as $\sigma(Z^\prime)\times {\rm BR}(Z^\prime \to e^+e^-)\leq 0.2~{\rm fb}$ by the current measurements~\cite{Aad:2014cka,Khachatryan:2014fba}, which requires $c_\phi<0.19$. 
Similar to the case of $W^\prime$ boson, it is also possible to explain the $WW$ excess if there exists some mechanism to decrease the leptonic decay mode of $Z^\prime$ boson.

\subsection{Non-universal model}

\subsubsection{The $W^\prime$ constraints}

\begin{figure}
\includegraphics[width=0.32\textwidth]{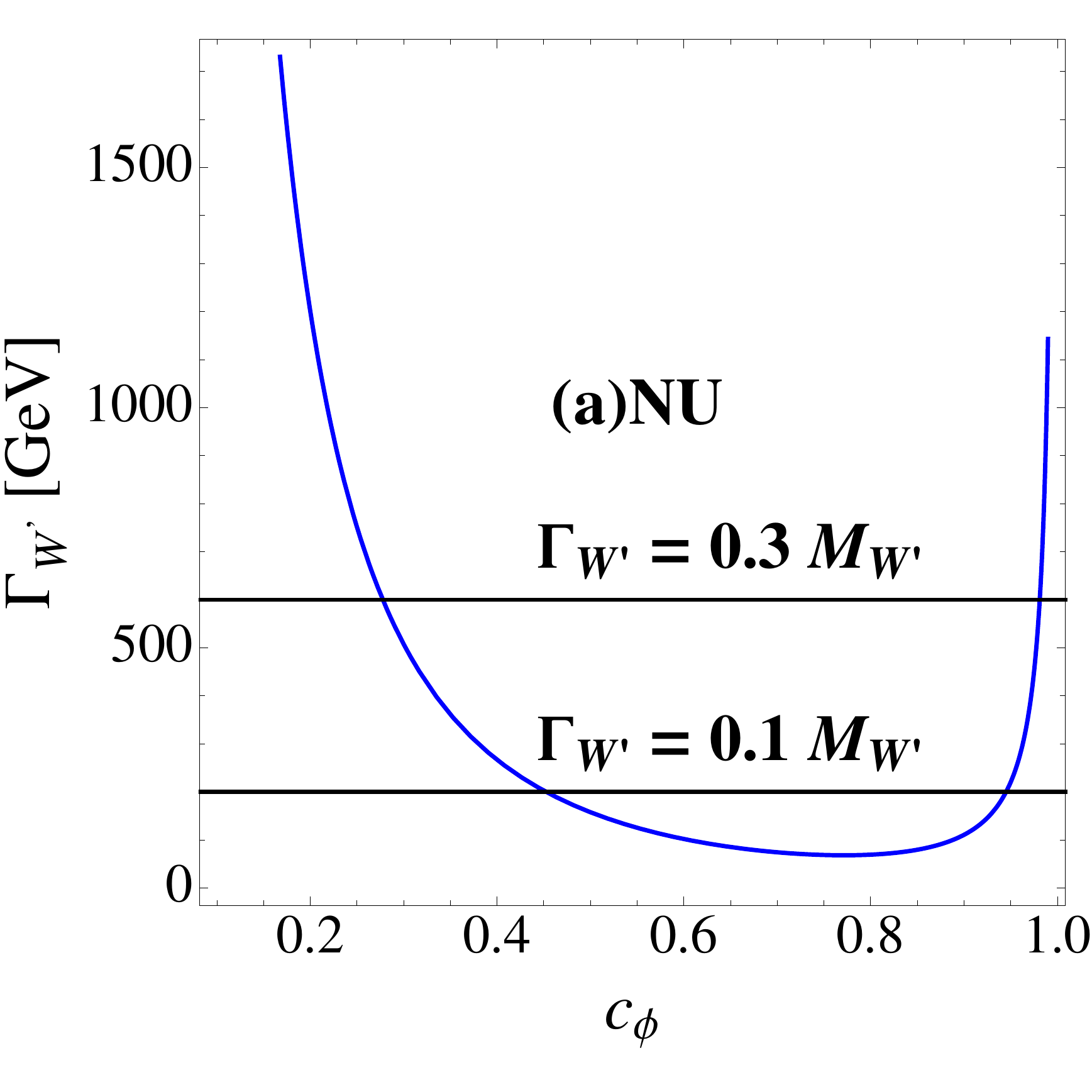}
\includegraphics[width=0.32\textwidth]{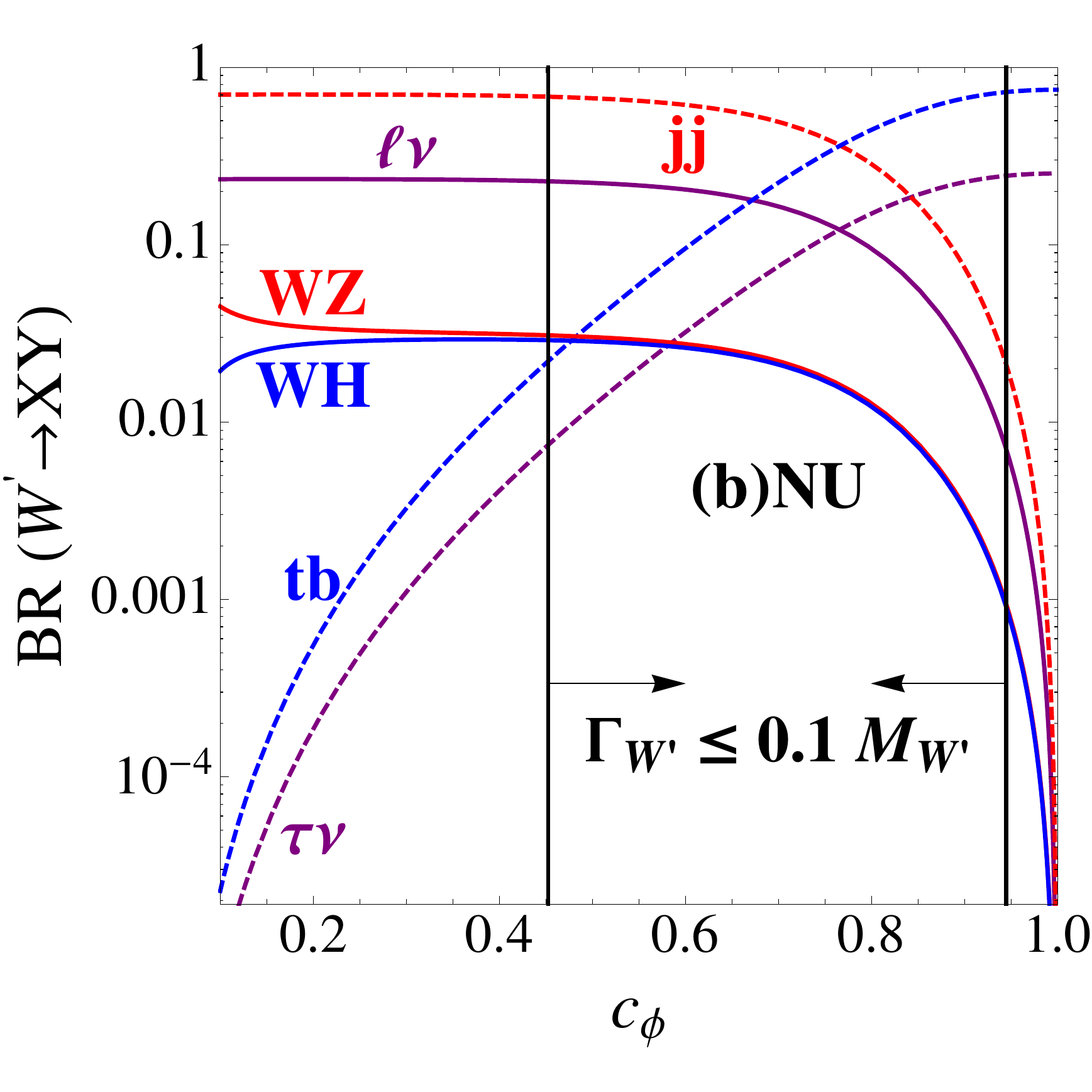}
\includegraphics[width=0.32\textwidth]{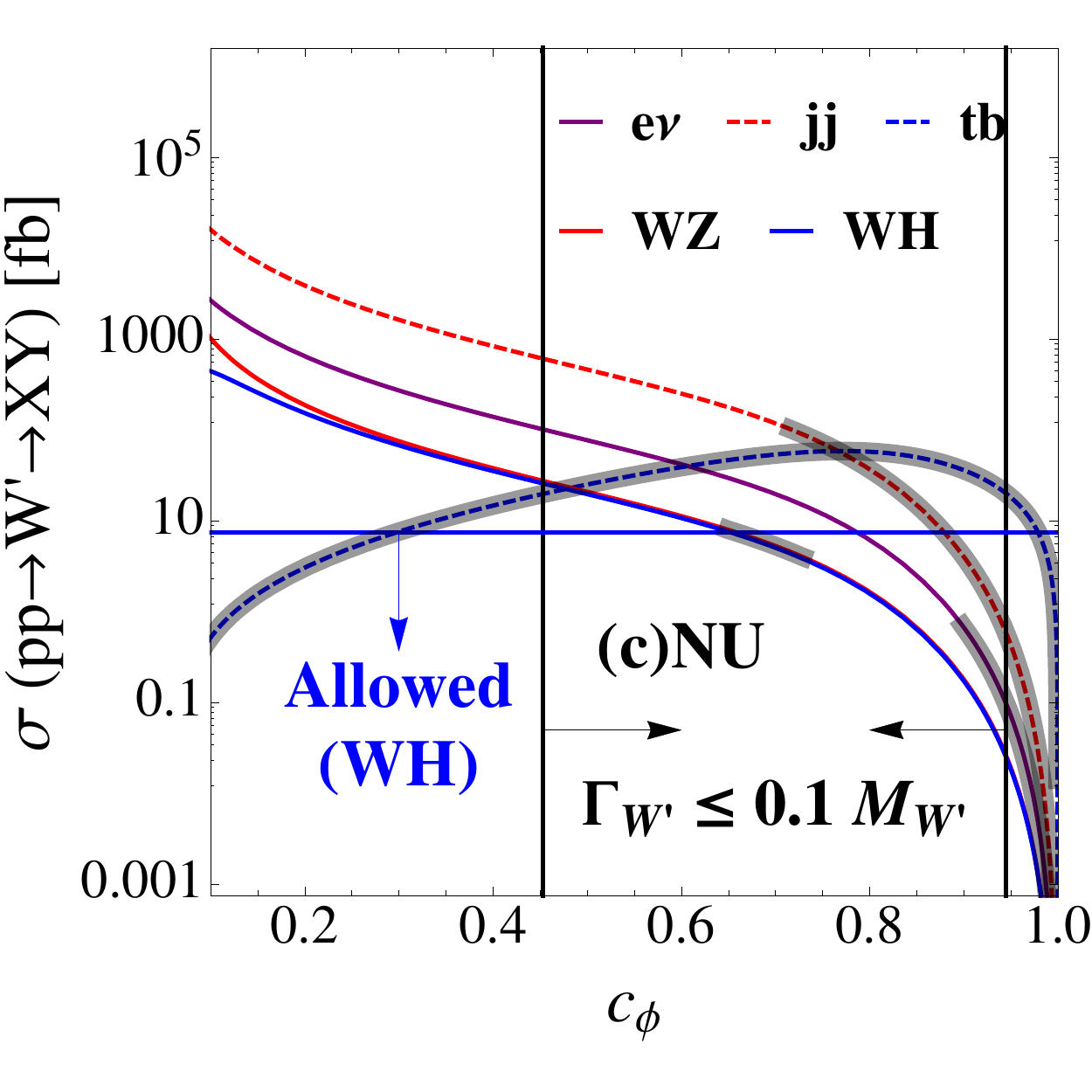}
\caption{\it (a) The total width $\Gamma_{W'}$ as a function of $c_\phi$ in the Nonuniversal (NU) model of BP-II. (b) The decay branching ratio ${\rm BR}(W^\prime\to XY)$ as a function of $c_\phi$. (c) The cross section $\sigma(pp\to W^\prime\to XY)$ versus $c_\phi$ at the LHC Run-1 in the NU model. The shaded band along each curve satisfies the current experimental data.}
\label{NUW}
\end{figure}

The Non-universal model is often named as the Top-Flavor model. In the model, the left-handed fermions of the first two generations are gauged under $SU(2)_1$, while the left-handed fermions of the third generation are gauged under $SU(2)_2$; see Table~\ref{tb:models} for the detail charge assignments.  The $W^\prime$ couples strongly to the first two generation fermions in the region of $c_\phi \sim 0$ and to the third generation fermions in the region of $c_\phi \sim 1$. Figure~\ref{NUW}(a) displays the decay width of $W^\prime$ versus $c_\phi$.  In order to validate the NWA, we demand $\Gamma_{W'}\leq 0.1M_{W'}$ which is presented by the black-dashed horizontal line. It requires $0.45\leq c_\phi \leq 0.95$. The branching ratios of the $W^\prime$ decays are also plotted in Fig.~\ref{NUW}(b). Here we separate the first two generation of the SM fermions from the third generation. The $\ell\nu$ mode includes the first two generation of leptons ($e\nu$ and $\mu\nu$). For a large $c_\phi$, the branching ratio of $W^\prime \to\ell\nu$ and $W^\prime\to jj$ are suppressed while the branching ratio of $W^\prime \to \tau\nu$ and $W^\prime\to tb$ are enhanced. It is owing to the fact that the gauge couplings of $W^\prime$ to the first two generation fermions are proportional to $\tan\phi$, while the gauge couplings to the third generation fermions are proportional to $\cot\phi$; see Eq.~\eqref{BPIIWF}.

The branching ratios of $W^\prime \to WZ/WH$ is about $0.01$ for most of the parameter space. Figure~\ref{NUW}(c) shows the cross sections of $\sigma(W^\prime)\times {\rm BR}(W^\prime\to XY)$. The shaded bands are consistent with current experimental data.  The $WZ$ excess prefers $0.65<c_\phi <0.73$. However, there is a tension between the $WZ$ mode and the $jj$ mode  as the $jj$ mode requires $c_\phi > 0.72$. The negative searching result of the $WH$ mode demands $c_\phi>0.66$~. It is possible to satisfy the $WZ$, $jj$ and $WH$ modes within $2\sigma$ confidential level. 

Unfortunately, the cross section of $\sigma(W^\prime)\times {\rm BR}(W^\prime \to e \nu)$ in the region of $c_\phi\sim 0.4-0.7$ is far beyond the current experimental limit; see the purple solid curve in Fig.~\ref{NUW}(c). 
In order to explain the $WZ$ excess, one needs to introduce new ingredients into the Non-universal model to reduce the leptonic decay modes of the $W^\prime$ boson.

\subsubsection{The $Z^\prime$ constraints}

\begin{figure}
\includegraphics[width=0.32\textwidth]{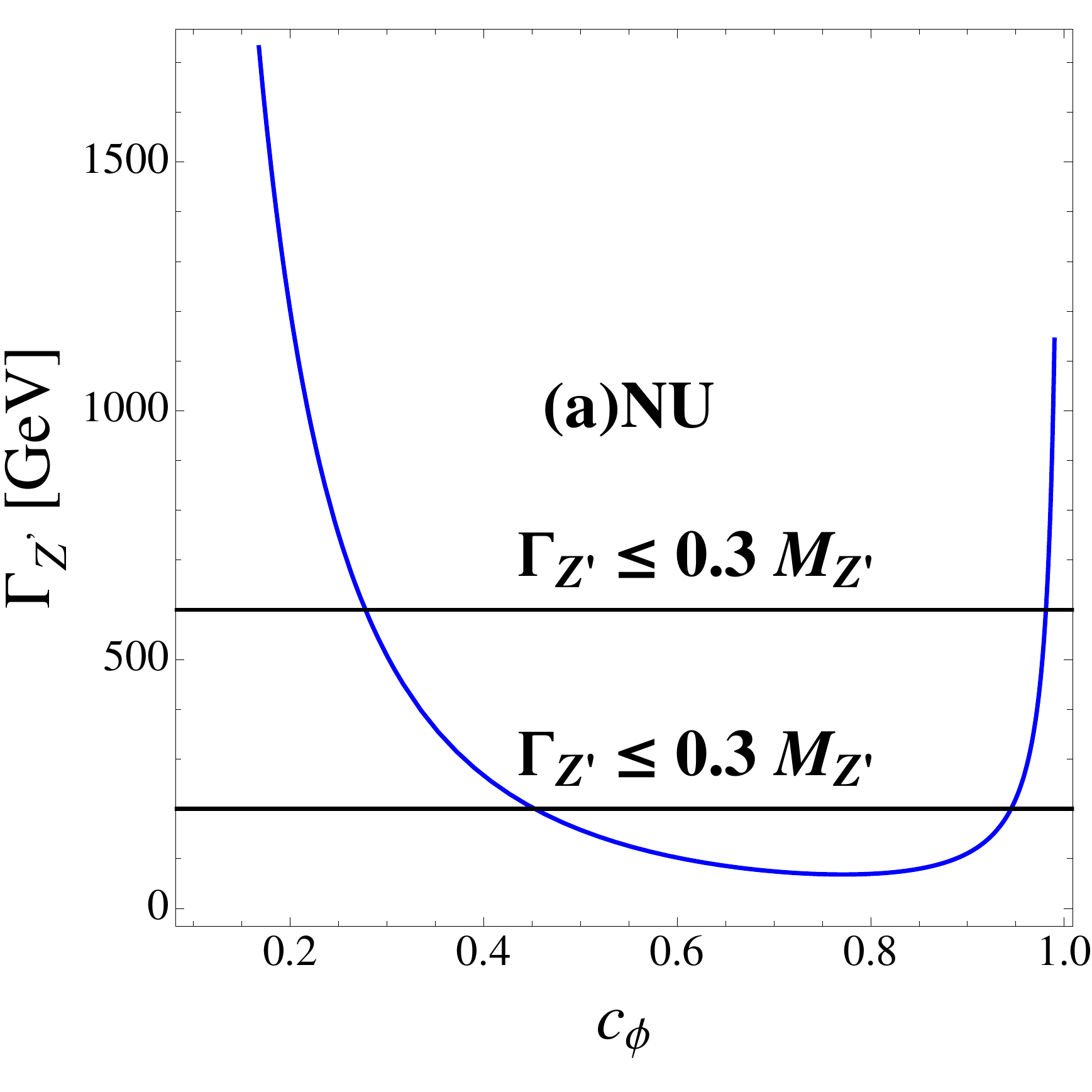}
\includegraphics[width=0.32\textwidth]{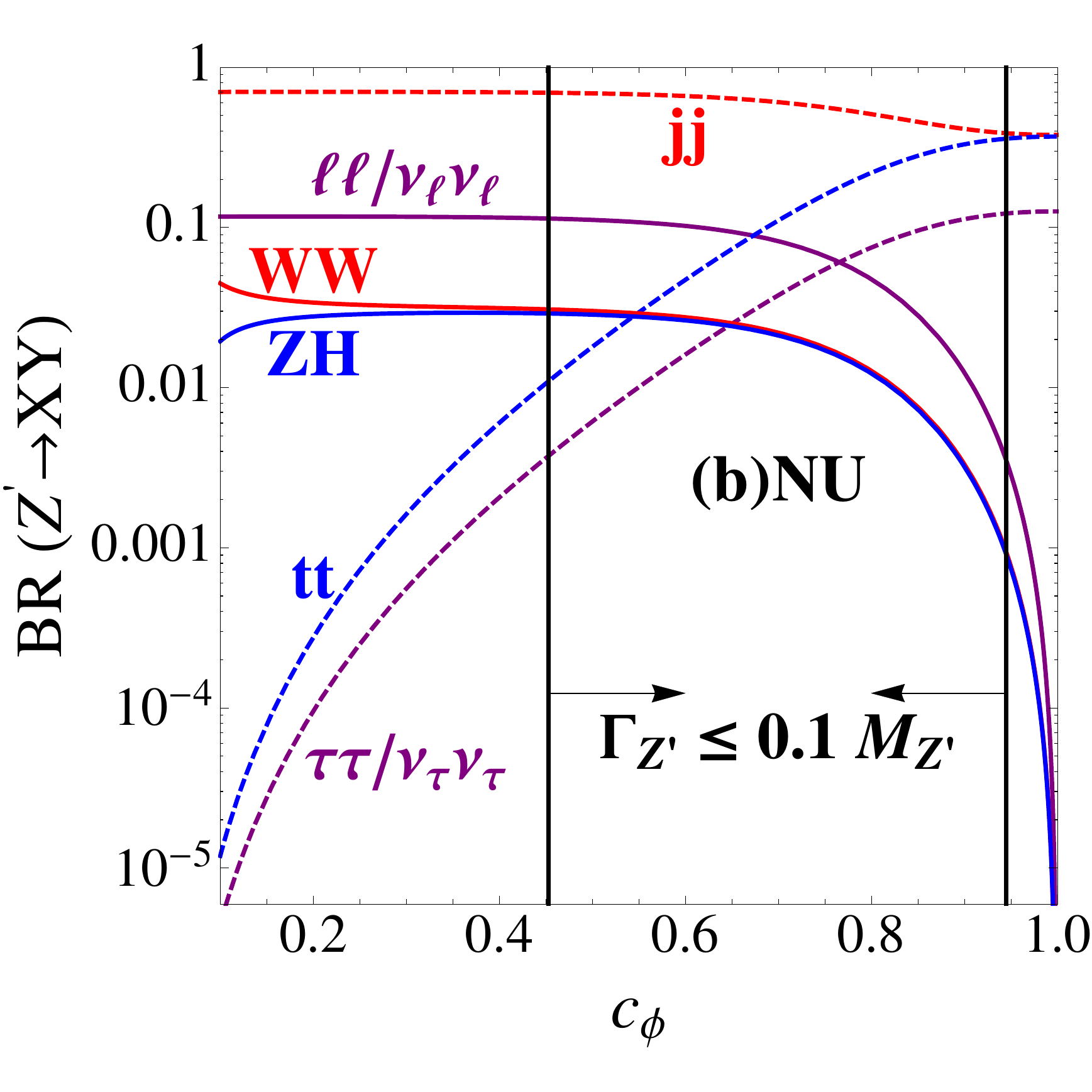}
\includegraphics[width=0.32\textwidth]{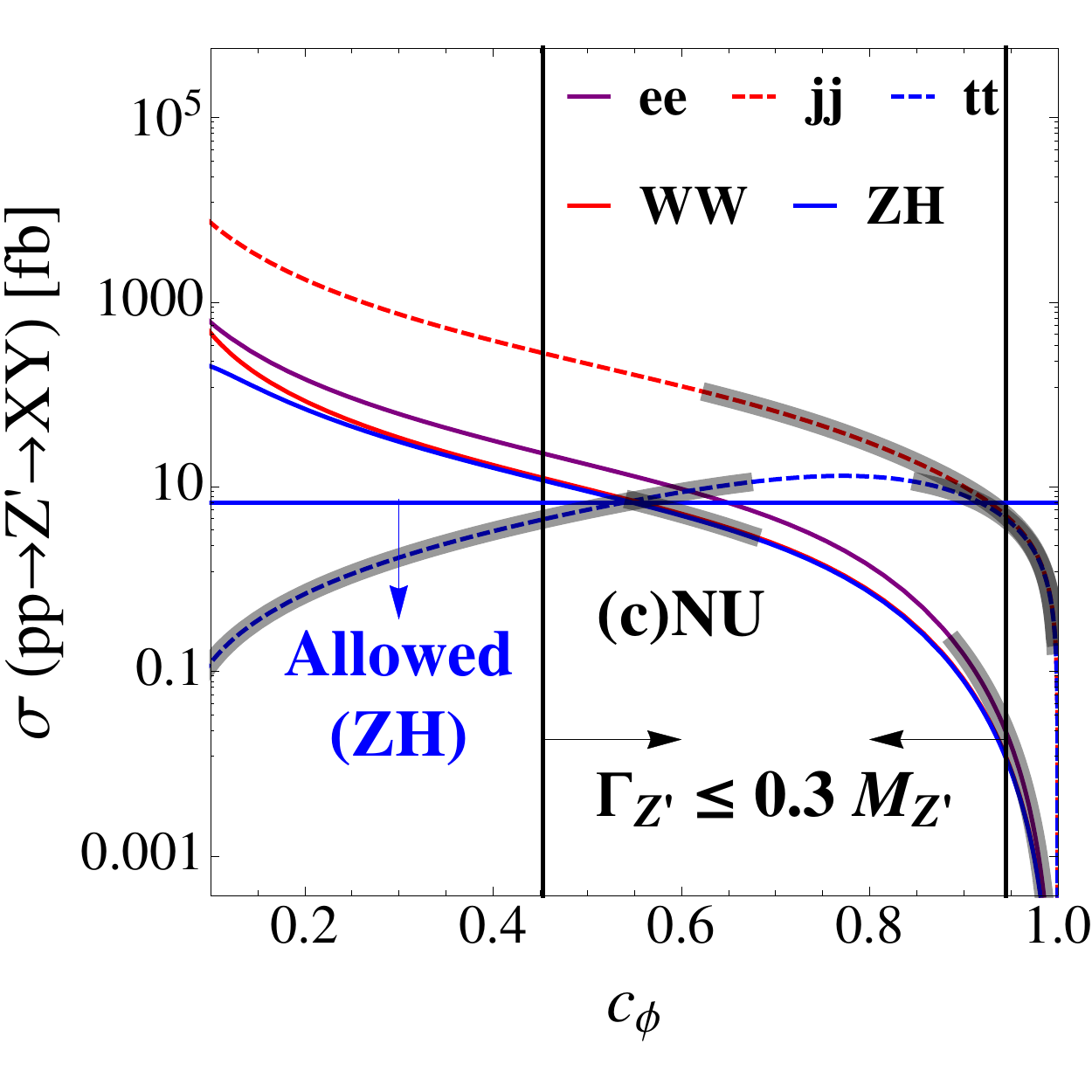}
\caption{\it (a) The total width $\Gamma_{Z'}$ versus $c_\phi$ in the Nonuniversal (NU) model of BP-II. (b) The decay branching ratio ${\rm BR}(Z^\prime\to XY)$ as a function of $c_\phi$. (c) The cross section $\sigma(pp\to Z^\prime\to XY)$ versus $c_\phi$ at the LHC Run-1. The shaded band along each curve satisfies the current experimental data.}
\label{NUZ}
\end{figure}

Figure~\ref{NUZ} shows the total width $\Gamma_{Z'}$ (a) and decay branching ratios of $Z^\prime$ (b) as a function of $c_\phi$. We also demand the narrow width constraint $\Gamma_{Z'}\le 0.1M_{Z'}$ which also requires $0.45\leq c_\phi \leq 0.95$. Here, the $\ell\ell$ mode sums over the electron ($e$) and muon ($\mu$) while the $\nu\nu$ mode sums over the first two generation neutrinos.

We first notice that the $jj$ mode dominates over the other modes in the entire parameter space of $c_\phi$. The branching ratio of $Z^\prime \to \ell\ell/\nu_{\ell}\nu_{\ell}$ is suppressed in the region of large $c_\phi$. On the other hand, the branching ratios of $Z^\prime \to tt$ and $Z^\prime \to \tau\tau/\nu_{\tau}\nu_{\tau}$ are enhanced for a large $c_\phi$. The branching ratios of $W^\prime \to WZ/WH$ are not sensitive to $c_\phi$ in the range $0.3\leq c_\phi\leq 0.7$, which is about 0.02. Figure~\ref{NUZ}(c) shows the cross section of various decay modes of $Z^\prime$.  We observe a tension between the $WW$ mode and the $jj$ mode. Again, the leptonic decay mode imposes much tighter constraint as $\sigma(Z^\prime)\times {\rm BR}(Z^\prime \to e^+e^-)\leq 0.2~{\rm fb}$ by the current measurements~\cite{Aad:2014cka,Khachatryan:2014fba}, which requires $c_\phi>0.89$~. 
Again, it requires to decrease the branching ratio of the leptonic decay mode in order to explain the $WW$ excess in the Non-universal model.

\section{G(331) model}\label{331}
 
Another simple non-Abelian extension of the SM gauge group is the so-called 331 model which exhibits a gauge structure of $SU(3)_C\otimes SU(3)_L\otimes U(1)_X$~\cite{Buras:2014yna,Buras:2012dp,Ninh:2005su, Martinez:2012ni, Montalvo:2012qg,Alves:2011mz,Montalvo:2008cx,CiezaMontalvo:2008sa,CiezaMontalvo:2008ew,Soa:2007zz,VanSoa:2006ea,CiezaMontalvo:2006zt,VanSoa:2008bm,Coutinho:2013lta,Martinez:2009ik,RamirezBarreto:2007mt,RamirezBarreto:2006tn,Coutinho:1999hf,Alves:2012yp,Alvares:2012qv,Kelso:2013nwa,Barreto:2013paa}. The electroweak symmetry is broken spontaneously as follows,
\beq
SU(3)_L\times U(1)_X\rightarrow SU(2)_L\times U(1)_Y\rightarrow U(1)_{\rm em},
\eeq
by three scalar triplets $\rho$, $\eta$ and $\chi$ with vacuum expectation values as follows,
\beq
\left<\rho\right>=\frac{1}{\sqrt{2}} \left(\begin{array}{c}
0  \\
v_\rho   \\
0  \\
\end{array}\right),\quad
\left<\eta\right>=\frac{1}{\sqrt{2}} \left(\begin{array}{c}
v_\eta  \\
0   \\
0  \\
\end{array}\right),\quad
\left<\chi\right>=\frac{1}{\sqrt{2}} \left(\begin{array}{c}
0  \\
0   \\
v_\chi  \\
\end{array}\right).
\eeq
The $\chi$ triplet is responsible for the first step of symmetry breaking, while the $\rho$ and $\eta$ triplets are responsible for the second step of symmetry breaking.

The electric charge is defined as $Q=T_3+Y=T_3+\beta T_8+X$ where $T_i$ ($i=1\sim 8$) are eight Gell-Mann Matrices and $X$ is the quantum number associated with 
$U(1)_X$. The parameter $\beta$ stands for the different definitions of the hypercharge $Y$ or $Q$.

At the first step of spontaneously symmetry breaking at the TeV scale, three new gauge bosons $Y$, $V$ and $Z'$ obtain their masses. The $W$ and $Z$ bosons are massive after the second step of symmetry breaking at the electroweak scale. Neglecting the small mixing of $Z'$ and $Z$, the mass eigenstates of those gauge bosons can be written in terms of the $SU(3)_L$ and $U(1)_X$ gauge eigenstates $W^i_\mu$ ($i=1\sim 8$) and $X_\mu$ as follows:
\bea
&&Y^{\pm Q_Y}_\mu=\frac{1}{\sqrt{2}}(W^4_\mu \mp iW^5_\mu), \qquad V^{\pm Q_V}_\mu=\frac{1}{\sqrt{2}}(W^6_\mu \mp iW^7_\mu), \nn\\
&&Z^\prime_\mu =-s_{331}W^8_\mu+c_{331}X_\mu, \qquad~W^\pm_\mu=\frac{1}{\sqrt{2}}(W^1_\mu \mp iW^2_\mu), \nn\\
&&Z_\mu=\frac{1}{\sqrt{g^2+g^2_Y}} \left[gW^3_\mu-g_Y \left(c_{331}W^8_\mu+s_{331}X_\mu\right)\right],
\eea
where $s_{331}$ and $c_{331}$ are the sine and cosine of  the 331 mixing angle, respectively, $g_Y$ is the coupling strength of $U(1)_Y$. They can be written in terms of the $SU(3)_L$ and $U(1)_X$ 
coupling constants $g$ and $g_X$ as follows:
\beq
s_{331}=\frac{g}{\sqrt{g^2+\beta^2g_X^2}},
\quad c_{331}=\frac{\beta g_X}{\sqrt{g^2+\beta^2g_X^2}},
\quad g_Y=\frac{gg_X}{\sqrt{g^2+\beta^2g_X^2}}.
\eeq

\begin{figure}
\includegraphics[width=0.32\textwidth]{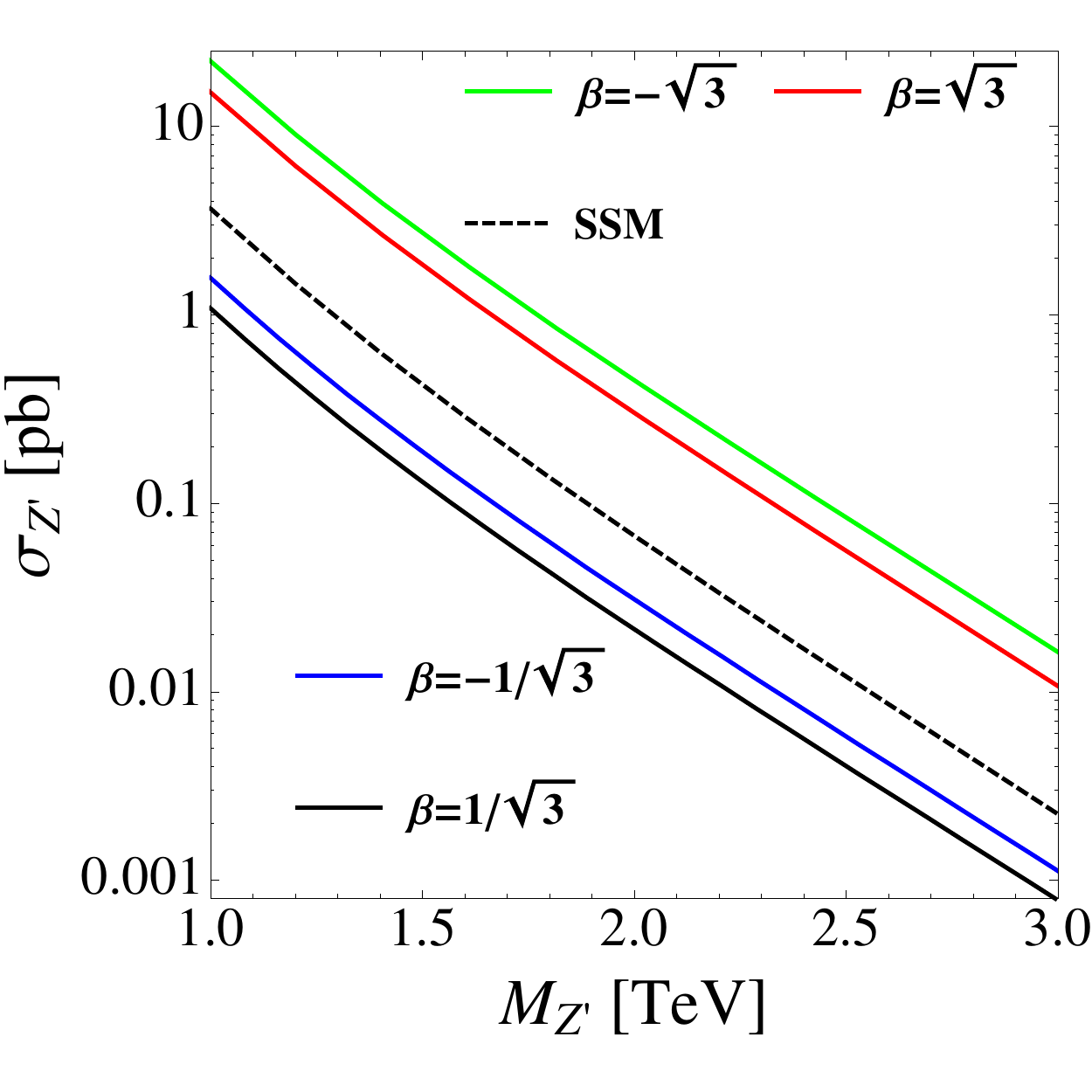}
\caption{\it The cross section of $Z^\prime$ production versus $M_{Z^\prime}$ for different choices of $\beta$ in the $G(331)$ model at the LHC Run-1.  For comparison the production cross section of a sequential $Z^\prime$ boson is also plotted (black-dotted curve).}
\label{fig:331a}
\end{figure}

Owing to the gauge symmetry,  the trilinear gauge couplings of $Y(V)WZ$ and $Z^\prime ZZ$ are absent in the $G(331)$ model. It is difficult to explain the excesses observed by the ATLAS collaboration. The $Z^\prime$ can couple to the $WW/ZH$ pair through the mixing with the $Z$ boson. The mixing angle is~\cite{Buras:2014yna},
\beq
\sin\theta_{ZZ'}=\frac{c_W^2}{3}\sqrt{f(\beta)}\left(3\beta\frac{s_W^2}{c_W^2}+\sqrt{3}\alpha\right)\frac{m_Z^2}{M_{Z'}^2},
\eeq
where
\beq
f(\beta)=\frac{1}{1-(1+\beta^2)s_W^2},\qquad -1<\alpha=\frac{v_-^2}{v_+^2}<1,
\eeq
with $v_+^2=v_\eta^2+v_\rho^2$ and $v_-^2=v_\eta^2-v_\rho^2$. Thus the branching ratios of $Z^\prime \to WW$ and $Z^\prime \to ZH$ are sensitive to $\alpha$.

Figure~\ref{fig:331a} displays the cross section of the $Z^\prime$ production in the $G(331)$ model at the LO for various choices of $\beta$ parameter.  See Ref.~\cite{Buras:2012dp} for the couplings of $Z^\prime$ to the SM fermions. For a 2~TeV resonance, the production cross sections $ \sigma(Z^\prime)$ are $300~{\rm fb}$ for $\beta = \sqrt{3}$, $454~{\rm fb}$ for $\beta = -\sqrt{3}$, $21~{\rm fb}$ for $\beta = +1/\sqrt{3}$ and $31~{\rm fb}$ for $\beta = -1/\sqrt{3}$.

\begin{figure}
\includegraphics[width=0.32\textwidth]{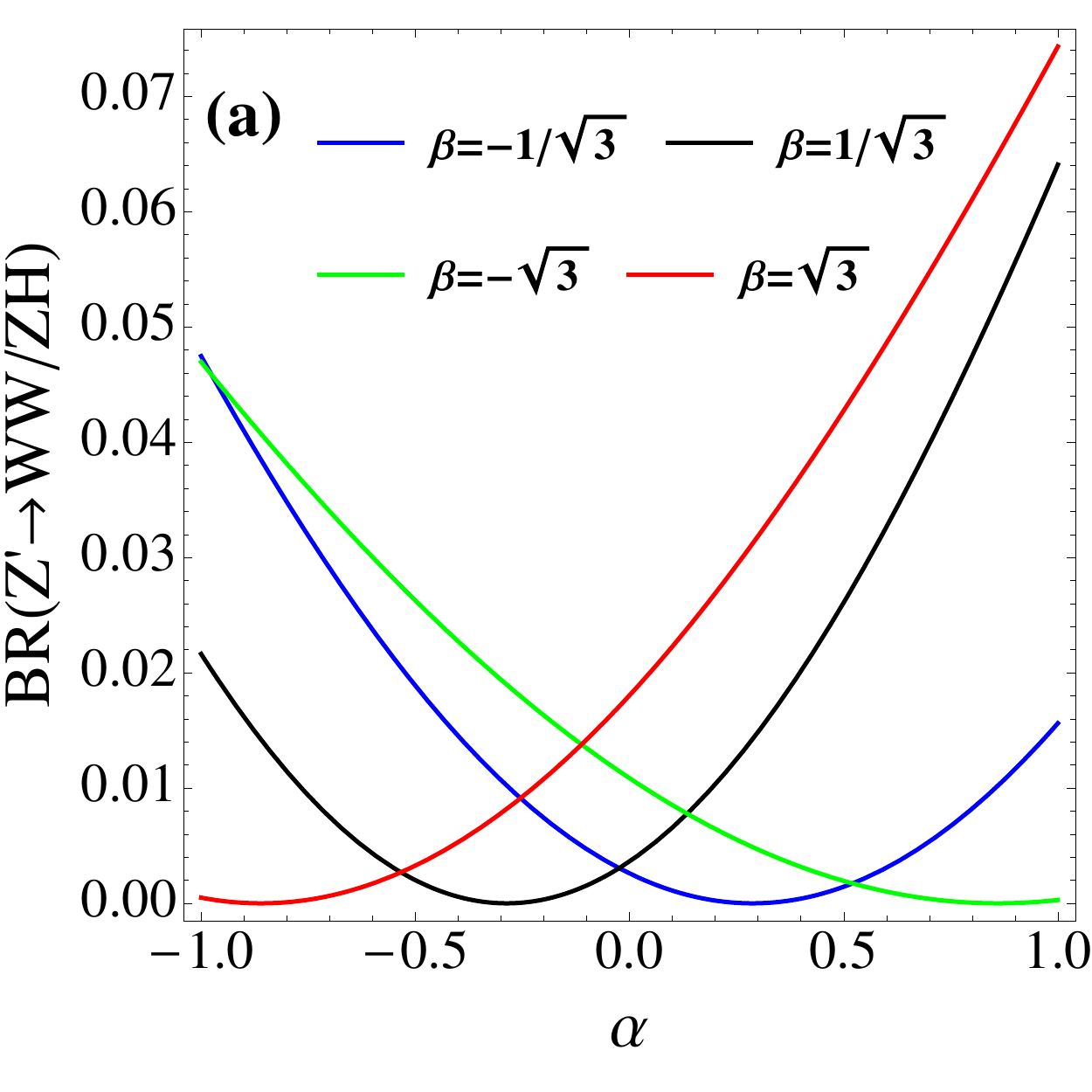}
\includegraphics[width=0.32\textwidth]{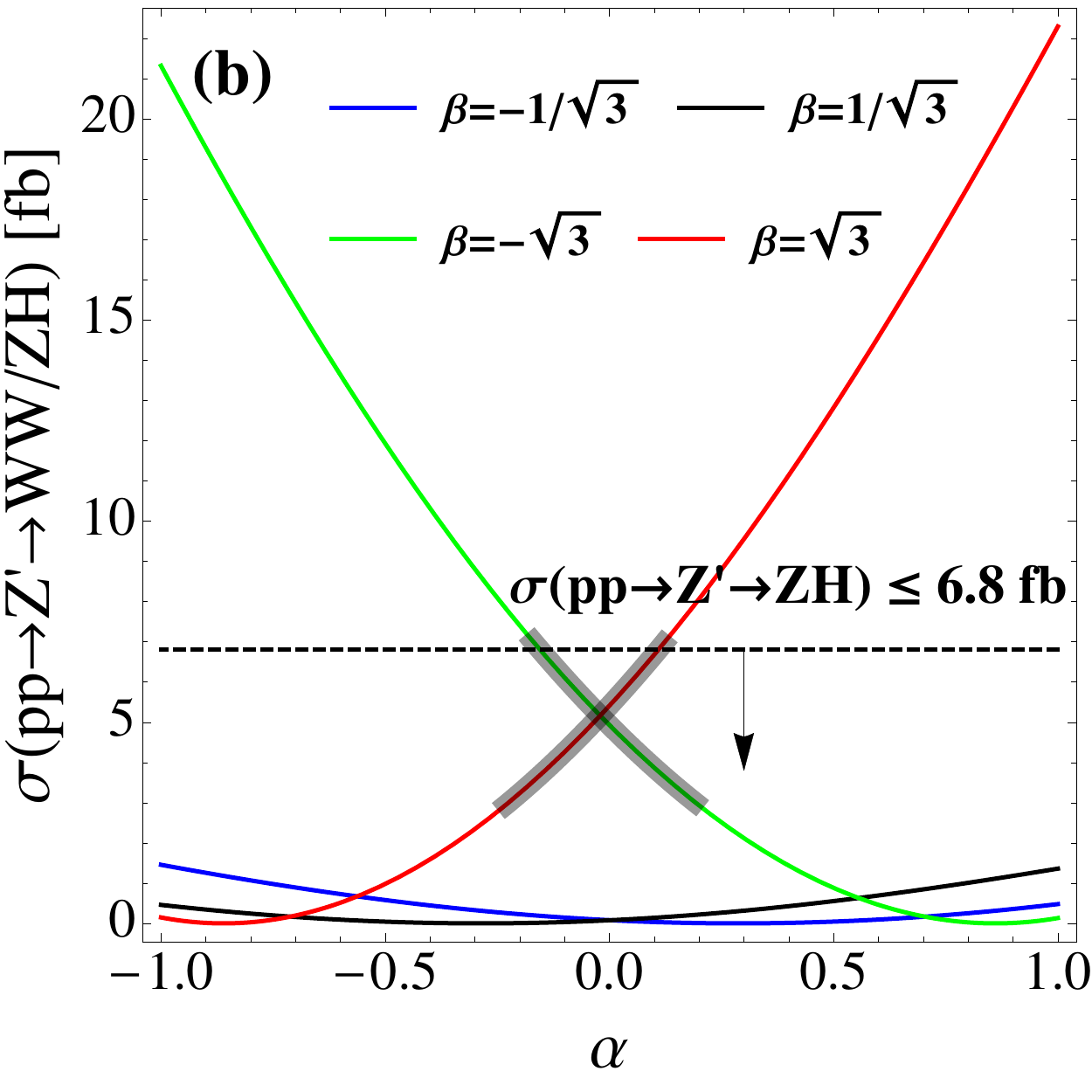}
\caption{\it (a) The branching ratio ${\rm BR}(Z^\prime \to WW/ZH)$ as a function of $\alpha$ for different choices of $\beta$. (b) The cross section $\sigma(Z^\prime)\times {\rm BR}(Z^\prime \to WW/ZH)$ as a function of $\alpha$ for different choices of $\beta$ at the LHC Run-1. The shaded bands along the curves represent the parameter space that could explain the $WW$ excess. The black-dashed horizontal line shows the upper limit of $ZH$.
}
\label{fig:331b}
\end{figure}

We first consider the decay mode of $Z^\prime \to WW$ in the $G(331)$ model. Figure~\ref{fig:331b}(a) displays the  branching ratios of ${\rm BR}(Z^\prime \to WW/ZH)$ for the four choices of $\beta$. The branching ratios are sensitive to the $\alpha$ parameter. Figure~\ref{fig:331b}(b) displays the cross section of $\sigma(pp \to Z^\prime \to WW/ZH)$ versus $\alpha$. The shaded bands along the curves of $\beta=-\sqrt{3}$ and $\beta=\sqrt{3}$ denote the region that is compatible with the $WW$ excess, where $-0.17\leq \alpha\le 0.19$ and $-0.23\leq \alpha\le 0.12$ for $\beta=-\sqrt{3}$ and $\beta=\sqrt{3}$ respectively. The current exclusion limit, $\sigma(pp \to Z^\prime \to ZH)\leq 6.8{\rm fb}$, is shown as the black-dashed horizontal curve.

\begin{figure}[]
\includegraphics[width=0.32\textwidth]{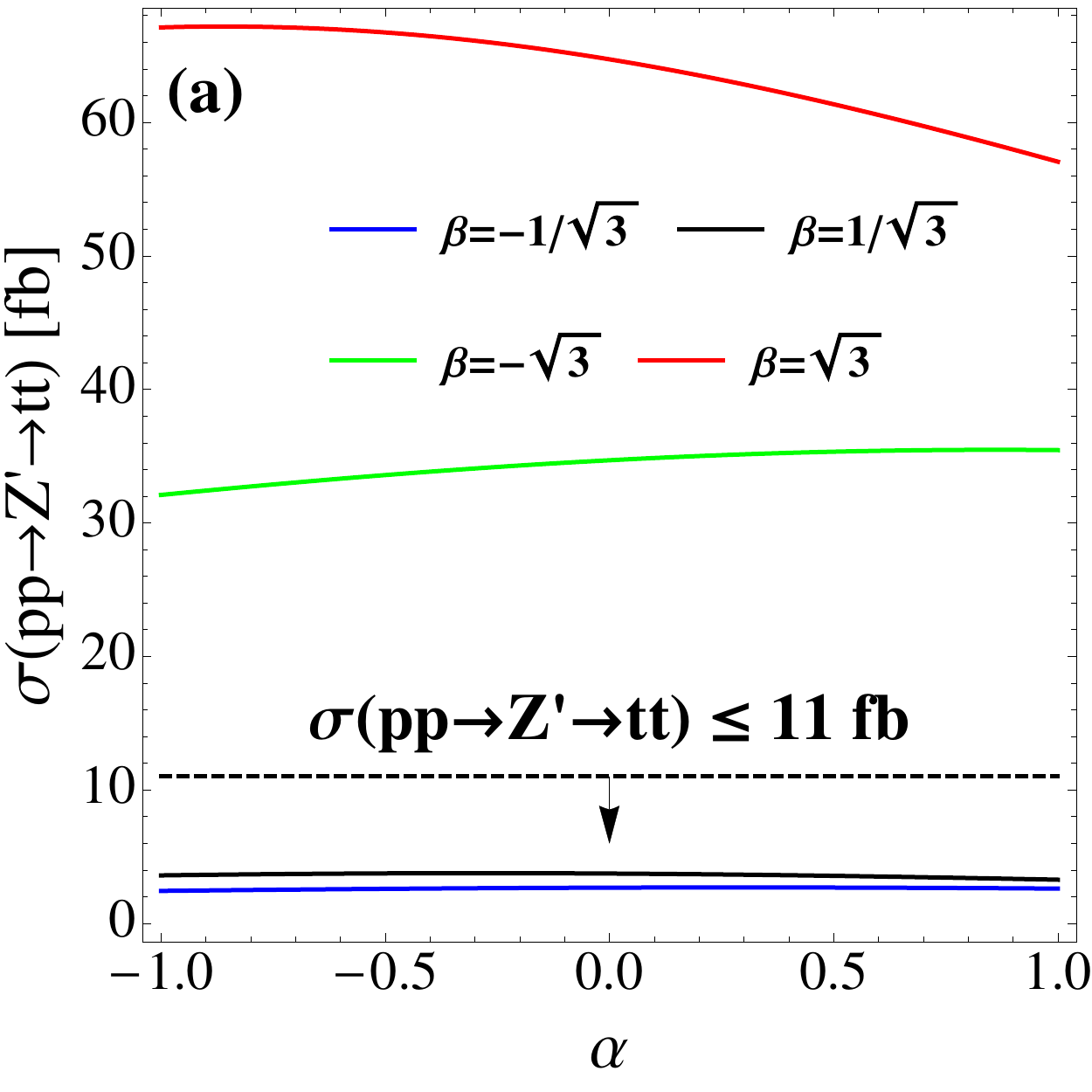}
\includegraphics[width=0.32\textwidth]{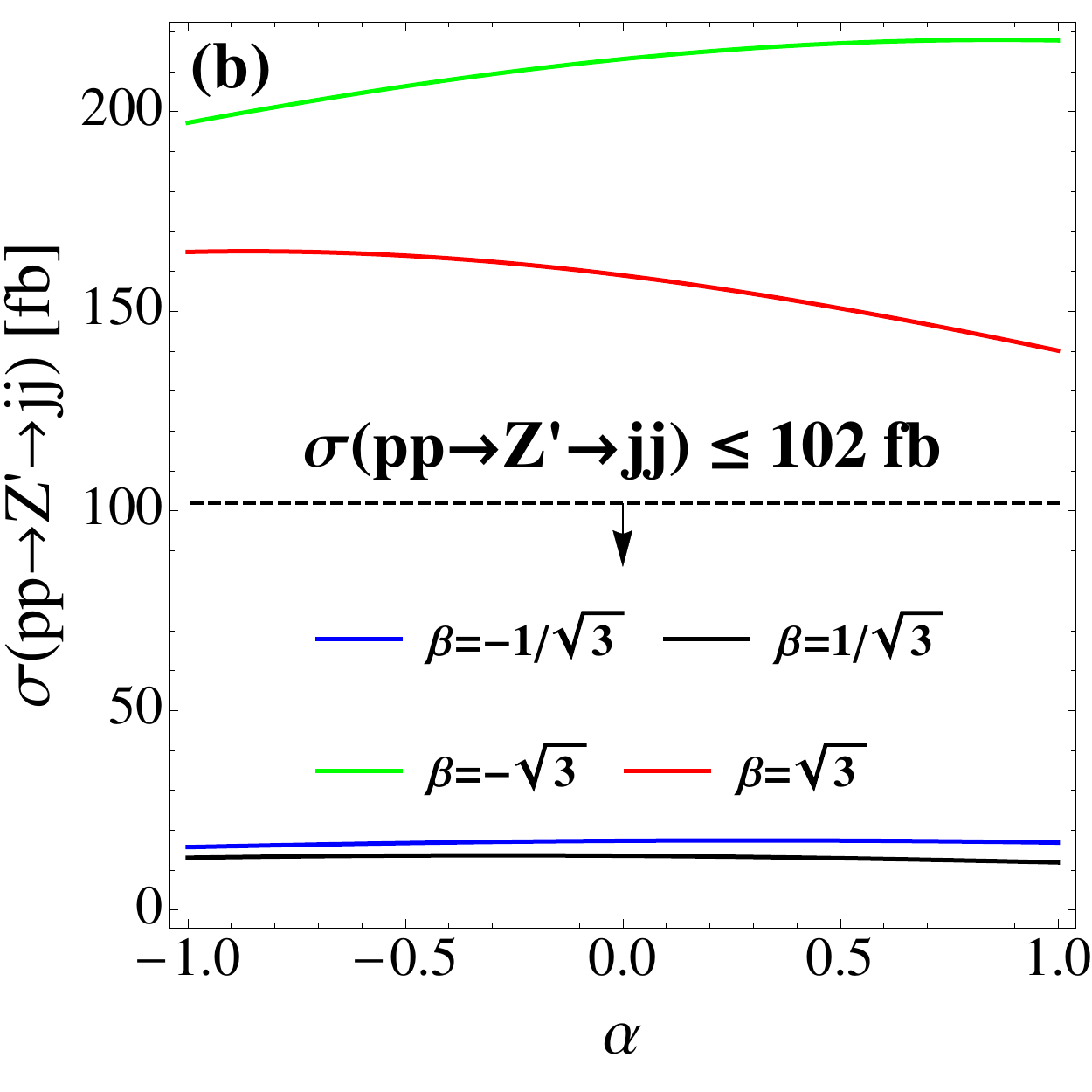}
\includegraphics[width=0.32\textwidth]{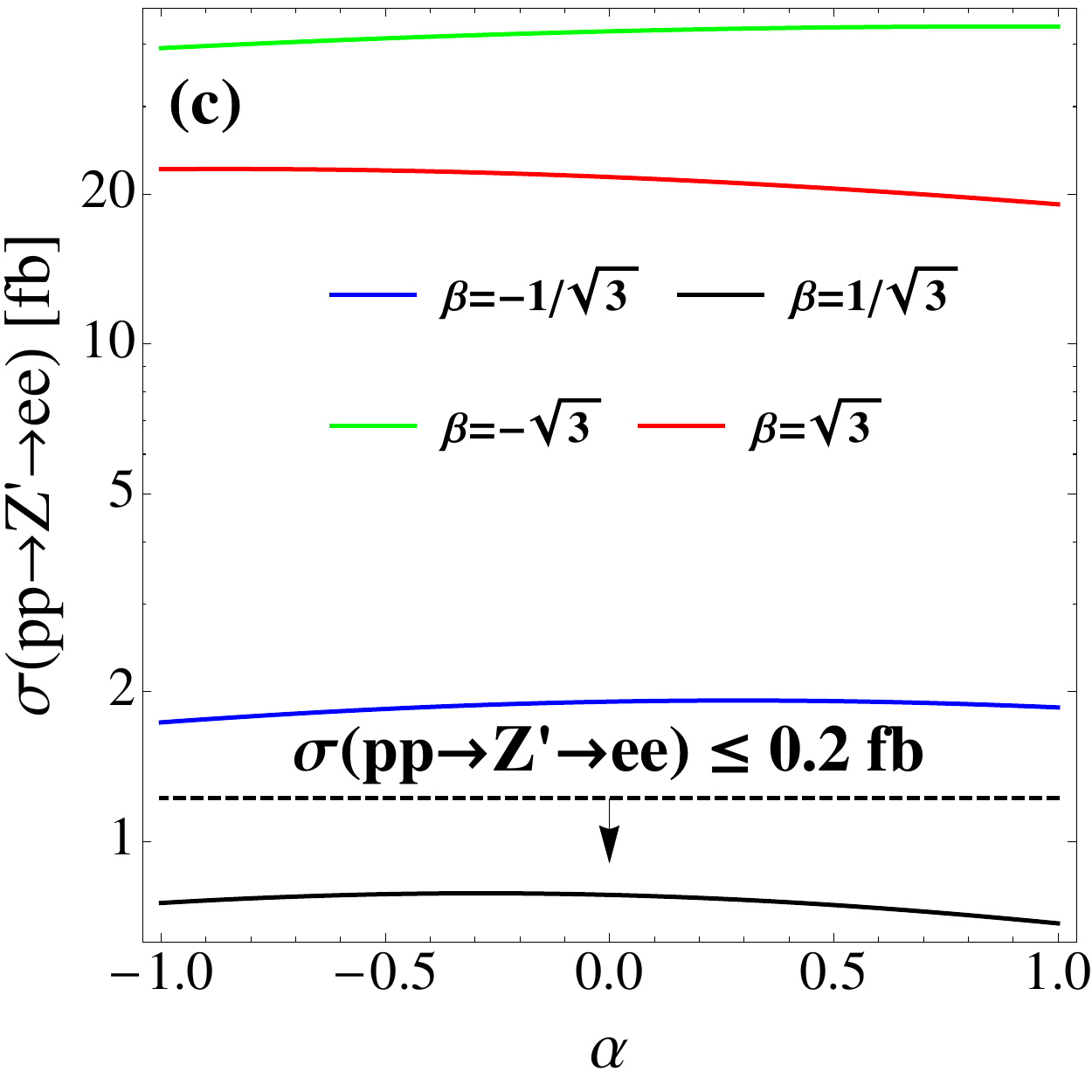}
\caption{\it The cross section of $\sigma(pp \to Z^\prime\to t\bar{t})$ (a),  $\sigma(pp \to Z^\prime\to jj)$ (b) and $\sigma(pp \to Z^\prime\to e\bar{e})$ (c) as a function of $\alpha$ in the $G(331)$ model. The current experimental limits are also displayed. 
}
\label{fig:331c}
\end{figure}

Other decay modes of the $Z^\prime$ boson are also checked in this work. Figure~\ref{fig:331c} shows the cross section of $Z^\prime$ production with its subsequent decays into the SM quarks and leptons, i.e. (a) $\sigma(pp\to Z^\prime \to t\bar{t})$, (b) $\sigma(pp \to Z^\prime \to jj)$ and (c) $\sigma(pp \to Z^\prime \to e^+ e^-)$. The current experiment bounds are also plotted in the figure. The choices of $\beta=\pm \sqrt{3}$ yield a large cross section which exceeds the current limits. Even though the choices of $\beta=\pm 1/\sqrt{3}$ are allowed, they cannot explain the 2.6$\sigma$ excess in the $WW$ channel.

\section{Summary and Discussion}\label{summary}

The excesses around 2 TeV in the diboson invariant mass distribution invoke excitement among theorists recently. We examine the possibility of explaining the resonances as extra gauge bosons. Two simple extensions of the SM gauge symmetry are explored. One is named as the $G(221)$ model with a gauge structure of $SU(2)_1\times SU(2)_2 \times U(1)_X$, the other is called $G(331)$ model with $SU(3)_C\times SU(3)_L\times U(1)_X$ symmetry. Extra gauge bosons emerge after the symmetry is broken down to the SM gauge symmetry at the TeV scale in the breaking pattern (BP) listed as follows: (i) $SU(2)_L\times SU(2)_2\times U(1)_X \to SU(2)_L \times U(1)_Y$ (BP-I); (ii) $SU(2)_1\times SU(2)_2\times U(1)_Y \to SU(2)_L \times U(1)_Y$ (BP-II); (iii) $SU(3)_L\times U(1)_X \to SU(2)_L \times U(1)_Y$. The SM symmetry is further broken at the electroweak scale. We consider several new physics models which can be classified by the symmetry breaking pattern: (i) the Left-Right (LR), Lepto-Phobic (LP), Hadro-Phobic (HP), Fermio-Phobic (FP) models; (ii) the Un-unified (UU) model and the Non-universal (NU)model, (iii) $G(331)$ model with $\beta=\pm \sqrt{3}$ and $\beta=\pm 1/\sqrt{3}$. 
The phenomenology of $W^\prime$ and $Z^\prime$ bosons in the above NP models is explored at the LHC Run-1. All the decay modes of $W^\prime/Z^\prime$ are included, e.g. $W^\prime \to jj/t\bar{b}/\ell\nu/WZ/WH$ and $Z^\prime \to \ell\ell/\nu\nu/jj/t\bar{t}/WW/ZH$.

Firstly, we examine the possibility of interpreting the $WZ$ excess as a 2~TeV $W^\prime$ boson in those NP models.  The parameter spaces compatible with the experimental data  are summarized in Table~\ref{tab:wprime}. For those $G(221)$ models, a large $s_{2\beta}$ is favored to induce a large branching ratio of $W^\prime \to WZ/WH$.  For illustration we choose $s_{2\beta}\sim 1$ in Table~\ref{tab:wprime}.  In the Left-Right model the parameter of $0.68 \leq c_\phi \leq 0.81$ is compatible with both the $WZ$ excess and $WH/jj/tb/e\nu$ upper limits, but it predicts $2.47~{\rm TeV}<M_{Z^\prime}<2.94~{\rm TeV}$ which is in contradiction with the $WW$ excess around 2~TeV. 
In the Lepto-Phobic model the parameter of $0.68<c_\phi<0.81$ satisfies the $WZ$ excess and all other experimental  bounds, but it predicts $2.47~{\rm TeV}<M_{Z^\prime}<2.94~{\rm TeV}$ which is also in contradiction with the $WW$ excess around 2~TeV. 
It is still difficult to judge whether or not the $WW$ excess exists at the moment. If the $2.6\sigma$ deviation in the $WW$ pair turns out to be from the fluctuation of the SM backgrounds,  then the $3.4\sigma$ excess in the $WZ$ pair can be interpreted as the $W^\prime$ boson in both Left-Right and Lepto-Phobic models.
In the Hadro-Phobic and Fermio-Phobic models the production cross section of $W^\prime$ is too small to explain the $WZ$ excess. In the Un-unified model, we require $\Gamma_{W^\prime}\leq 0.1 M_{W^\prime} $ to validate the NWA which yields $0.47<c_\phi <0.96$.  The parameter of $0.72<c_\phi<0.73$ could address on the $WZ$ excess and the $WH/tb/jj$ limits, but it comes into conflict with the tight constraint from the $e  \nu$ mode ($c_\phi<0.18$). A similar result also holds for the Non-universal model. 
It is hard to explain the $WW$ excess in the Un-unified and Non-universal models unless one can extend the models to introduce a mechanism to reduce the leptonic decays of the $Z^\prime$ boson. In the $G(331)$ model, the $W^\prime$-$W$-$Z$ and $Z^\prime$-$Z$-$Z$ couplings are forbidden by symmetry, therefore, it does not affect the $W^\prime$ phenomenology at all. 

Secondly, we examine the possibility of interpreting the $WW$ excess as a 2~TeV $Z^\prime$ boson in those NP models.  The parameter spaces compatible with the experimental data are summarized in Table~\ref{tab:zprime}. 
In the Left-Right model we require $\Gamma_{Z^\prime}\leq 0.1 M_{Z^\prime} $ to validate the NWA which yields $0.23<c_\phi <0.96$.  The parameter of $0.9 \leq c_\phi \leq 0.95$ could satisfy the $WW$ excess and $ZH$ limit at the 95\% confidence level. It has a tension with the $jj$ mode which demands $0.13<c_\phi<0.91$ but predicts too large cross section of $pp\to Z^\prime \to e^+ e^-$ to respect the current experimental bound.  A similar result is found in the Lepto-Phobic model. In the Hadro-Phobic and Fermio-Phobic models, the $WW$ excess cannot be explained due to the small production cross section of $Z^\prime$. In the Un-unified model the parameter of $0.64<c_\phi<0.67$ satisfies the $WW$ excess and the $ZH/jj$ mode but is in conflict with the $ee/tt$ mode. 
In the Non-universal model the parameter of $0.63<c_\phi<0.67$ is compatible with the $WW$ excess and the $ZH/tt/jj$ limits but it violates the $ee$ limit. 
However, if one extend the current models to decrease the branching ratio of the $Z^\prime$ leptonic decays, it is still possible to explain the $WW$ excess in the $G(221)$ models except the Hadro-Phobic and Fermio-Phobic models.

\begin{table}
\caption{\it The parameter space of $c_\phi$ obtained from the processes of $pp \to W^\prime \to XY$ at the LHC Run-1 in various $G(221)$ models. The $W^\prime$ mass is fixed to be 2~TeV. In the $G(221)$ model with BP-I,  $M_{Z^\prime}\simeq M_{W^\prime}/c_\phi$ and $s_{2\beta} \sim 1$. The $G(331)$ models are not shown as they do not exhibit the $W^\prime$-$W$-$Z$ and $W^\prime$-$W$-$H$ couplings. The symbol $\times$ means no parameter space compatible with the current experimental limits. The symbol $\surd$ means all the parameter spaces are allowed. }
\label{tab:wprime}
\begin{tabular}{|c|c|c|c|c|c|c|c|c|}
\hline 

 &  & $WZ$ & $WH$ & $e\nu$ & $tb$ & $jj$ & NWA & $M_{W^\prime}\simeq M_{Z^\prime}$ \tabularnewline
\hline 
\hline 
\multirow{4}{*}{\tabincell{c}{$G(221)$\\ (BP-I)} } & LR & (0.68, 0.9) & (0, 0.88) & $\surd$ & (0, 0.91) & (0, 0.81) & $\surd$ & \multirow{4}{*}{(0.95, 1)} \tabularnewline
\cline{2-8} 
 & LP & (0.68, 0.9) & (0, 0.88) & $\surd$  & (0, 0.91) & (0, 0.81) & $\surd$ &   \tabularnewline
\cline{2-8} 
 & HP & $\times$ & \multicolumn{5}{c|}{$\surd$} &    \tabularnewline
\cline{2-8} 
 & FP & $\times$ & \multicolumn{5}{c|}{$\surd$} &  \tabularnewline
\hline 
\multirow{2}{*}{\tabincell{c}{$G(221)$\\ (BP-II)}} & UU & (0.64, 0.73) & (0.65, 1) & (0, 0.18) & (0.54, 1) & (0.72, 1) & (0.47, 0.96) & \multirow{2}{*}{$\surd$} \tabularnewline
\cline{2-8} 
 & NU & (0.65, 0.73) & (0.66, 1) & (0.9, 1) & $\surd$ & (0.72, 1) & (0.45, 0.95) &   \tabularnewline
 \hline
\end{tabular}
\end{table}

\begin{table}
\caption{\it The parameter space of $c_\phi$ obtained from the processes of $pp \to Z^\prime \to XY$ at the LHC Run-1 in various $G(221)$ models. The $Z^\prime$ mass is fixed to be 2~TeV. In the $G(221)$ model with BP-I,  $M_{W^\prime}\simeq c_\phi M_{Z^\prime}$. Shown in the $G(331)$ models is the parameter space of $\alpha$. The symbol $\times$ means no parameter space compatible with the current experimental limits. The symbol $\surd$ means all the parameter space is allowed by the current data. }
\label{tab:zprime}
\begin{tabular}{|c|c|c|c|c|c|c|c|c|}
\hline 
 &  & $WW$ & $ZH$ & $ee$ & $tt$ & $jj$ & NWA & $M_{W^\prime}\simeq M_{Z^\prime}$\tabularnewline
\hline 
\hline
\multirow{4}{*}{\tabincell{c}{$G(221)$\\ (BP-I)}} & LR & (0.9, 0.95) & (0, 0.95) & $\times$ & (0.16, 0.88) & (0.13, 0.91) & (0.23, 0.96) & \multirow{4}{*}{(0.9, 1)}\tabularnewline
\cline{2-8} 
 & LP & (0.89, 0.95) & (0, 0.95) & (0.99,1) & (0.13, 0.88) & (0.1, 0.91) & (0.29, 0.96) & \tabularnewline
\cline{2-8} 
 & HP & $\times$ & $\surd$ & $\times$ & (0.66, 1) & (0.44, 1) & (0.34, 0.99) & \tabularnewline
\cline{2-8}
 & FP & $\times$ & $\surd$ & (0.95, 1) & (0.6, 1) & (0.39, 1) & (0.38, 1) & \tabularnewline
\hline 
\multirow{2}{*}{\tabincell{c}{$G(221)$\\ (BP-II)}} & UU & (0.54, 0.67) & (0.53, 1) & (0, 0.19) & (0.72, 1) & (0.64, 1) & (0.47, 0.96) & \multirow{2}{*}{$\surd$}\tabularnewline
\cline{2-8} 
 & NU & (0.55, 0.67) & (0.55, 1) & (0.89, 1) & \tabincell{c}{(0, 0.67) \\ (0.86, 1)} & (0.63, 1) & (0.45, 0.95) & \tabularnewline
\hline 
\multirow{4}{*}{$G(331)$} 
 & $\beta=-\frac{1}{\sqrt{3}}$ & $\times$ & $\surd$ & $\times$ & \multicolumn{3}{c|}{$\surd$} & \multirow{4}{*}{\tabincell{c}{Not \\Applicable}}\tabularnewline
\cline{2-8}
 & $\beta=+\frac{1}{\sqrt{3}}$ & $\times$ & \multicolumn{5}{c|}{$\surd$} & \tabularnewline
\cline{2-8} 
 & $\beta=-\sqrt{3}$ & (-0.16, 0.16) & (-0.15, 1) & \multicolumn{3}{c|}{$\times$} & $\surd$ & \tabularnewline
\cline{2-8}
 & $\beta=+\sqrt{3}$ & (-0.2, 0.11) & (-1, 0.11) & \multicolumn{3}{c|}{$\times$} & $\surd$ & \tabularnewline
 \hline
\end{tabular}
\end{table}

In the $G(331)$ models the $Z^\prime$-$W$-$W$ and $Z^\prime$-$Z$-$H$ couplings arise from the $Z$-$Z^\prime$ mixing which leads to a rich $Z^\prime$ phenomenology. We note that the choice of $\beta=\pm 1/\sqrt{3}$ cannot produce an enough cross section of $Z^\prime$ production to explain the $WW$ excess. The parameter of $-0.17 < \alpha < 0.19$ for $\beta=-\sqrt{3}$ and of $-0.23<\alpha < 0.12$ for $\beta=+\sqrt{3}$ could explain the $WW$ excess and satisfy the $ZH$ limit. However, the parameter space cannot satisfy the $ee/tt/jj$ limits.

In summary, we study in this work several new physics models with the simple non-abelian extension of the gauge structure, either $SU(2)_1\times SU(2)_2\times U(1)_X$ or $SU(3)_C\times SU(3)_L\times U(1)_X$. 
We note that one can explain the excesses in these new physics models if either the branching ratios of the leptonic and dijet modes in the Un-Unified and Non-Universal model could be reduced to satisfy the 
experimental bounds, or the $WW$ excess is found to be only a fluctuation of the 
backgrounds rather than the signal of a 2TeV $Z^\prime$ in the Left-Right and Lepto-Phobic 
model. 

\begin{acknowledgments}
We thank Jiang-Hao Yu and Hao Zhang for useful discussions. The work is supported in part by the National Science Foundation of China under Grand No. 11275009. 

\end{acknowledgments}

\appendix

\section{Decays of $V^\prime$ ($W^\prime$ and $Z^\prime$)}

For completeness, we present the analytical expression of the partial decay width of $W^\prime$ and $Z^\prime$ bosons. 
The partial decay width of $V' \to \bar{f}_{1}f_{2}$ is
\beq
\Gamma_{V'\to\bar{f_{1}}f_{2}}=\frac{M_{V'}}{24\pi}\beta_{0}\left[(g_{L}^{2}+g_{R}^{2})\beta_{1}+6g_{L}g_{R}\frac{m_{f_{1}}m_{f_{2}}}{M_{V'}^2}\right]\Theta(M_{V'}-m_{f_{1}}-m_{f_{2}})\,,
\label{eq:v_width}
\eeq
where 
\bea
\beta_{0}&=&\sqrt{1-2\frac{m_{f_{1}}^{2}+m_{f_{2}}^{2}}{M_{V'}^{2}}+\frac{(m_{f_{1}}^{2}-m_{f_{2}}^{2})^{2}}{M_{V'}^{4}}}, \nonumber \\
\beta_{1}&=&1-\frac{m_{f_{1}}^{2}+m_{f_{2}}^{2}}{2M_{V'}^{2}}-\frac{(m_{f_{1}}^{2}-m_{f_{2}}^{2})^{2}}{2 M_{V'}^{4}}.
\eea
The color factor is not included and
the top quark decay channel only opens when the $Z^\prime$ and $W^\prime$ masses are heavy. 

The partial decay width of $V'\to V_{1}V_{2}$ is
\beq
\Gamma_{V'\to V_{1}V_{2}}=\frac{M_{V'}^{5}}{192\pi M_{V_{1}}^{2}M_{V_{2}}^{2}}g_{V'V_{1}V_{2}}^{2}\beta_{0}^{3}\beta_{1}\Theta(M_{V'}-M_{V_{1}}-M_{V_{2}})\,,
\eeq
where 
\bea
\beta_{0}&=&\sqrt{1-2\frac{M_{V_{1}}^{2}+M_{V_{2}}^{2}}{M_{V'}^{2}}+\frac{(M_{V_{1}}^{2}-M_{V_{2}}^{2})^{2}}{M_{V'}^{4}}},\nonumber \\
\beta_{1}&=&1+10\frac{M_{V_1}^{2}+M_{V_2}^{2}}{2M_{V'}^{2}}+\frac{M_{V_{1}}^{4}+10M_{V_1}^{2}M_{V_{2}}^{2}+M_{V_{2}}^{4}}{M_{V'}^{4}}.
\eea
The partial decay width of $V'\to V_{1}H$ (where $V_1=W$ or $Z$ boson and $H$ is the lightest Higgs boson) is
\beq
\Gamma_{V'\to V_{1}H}=\frac{M_{V'}}{192\pi}\frac{g_{V'V_{1}H}^{2}}{M_{V_{1}}^{2}}\beta_{0}\beta_{1}\Theta(M_{V'}-M_{V_{1}}-m_H)\,,
\eeq
where 
\bea
\beta_{0}&=&\sqrt{1-2\frac{M_{V_{1}}^{2}+m_{H}^{2}}{M_{V'}^{2}}+\frac{(M_{V_{1}}^{2}-m_{H}^{2})^{2}}{M_{V'}^{4}}}, \nonumber \\
\beta_{1}&=&1+\frac{10M_{V_{1}}^{2}-2m_{H}^{2}}{2M_{V'}^{2}}+\frac{(M_{V_{1}}^{2}-m_{H}^{2})^{2}}{M_{V'}^{4}}.
\eea
Assuming the $W^\prime$ and $Z^\prime$ only decay to the SM particles, then
the total decay width of the $W^\prime$ boson is
\beq
\Gamma_{W^\prime,\rm tot}=3\Gamma_{W^\prime\to\bar{e}\nu}+2N_{C}\Gamma_{W^\prime\to\bar{u}d}+N_{C}\Gamma_{W^\prime\to\bar{t}b}+\Gamma_{W^\prime\to WZ}+\Gamma_{W^\prime\to WH}\,,
\eeq
while the width of the $Z^\prime$ boson is
\beq
\Gamma_{Z^\prime,\rm tot}=3\Gamma_{Z^\prime\to\bar{e}e}+3\Gamma_{Z^\prime\to\bar{\nu}\nu}+2N_{C}
\Gamma_{Z^\prime\to\bar{u}u}+3N_{C}\Gamma_{Z^\prime\to\bar{d}d}+N_{C}\Gamma_{Z^\prime\to\bar{t}t}+\Gamma_{Z^\prime\to WW}+\Gamma_{Z^\prime\to ZH}\,,
\eeq
where $N_C=3$ originates from summation of all possible color quantum number.

\bibliographystyle{apsrev}
\bibliography{reference}

\begin{thebibliography}{69}
\expandafter\ifx\csname natexlab\endcsname\relax\def\natexlab#1{#1}\fi
\expandafter\ifx\csname bibnamefont\endcsname\relax
  \def\bibnamefont#1{#1}\fi
\expandafter\ifx\csname bibfnamefont\endcsname\relax
  \def\bibfnamefont#1{#1}\fi
\expandafter\ifx\csname citenamefont\endcsname\relax
  \def\citenamefont#1{#1}\fi
\expandafter\ifx\csname url\endcsname\relax
  \def\url#1{\texttt{#1}}\fi
\expandafter\ifx\csname urlprefix\endcsname\relax\def\urlprefix{URL }\fi
\providecommand{\bibinfo}[2]{#2}
\providecommand{\eprint}[2][]{\url{#2}}

\bibitem[{\citenamefont{Aad et~al.}(2015{\natexlab{a}})}]{Aad:2015owa}
\bibinfo{author}{\bibfnamefont{G.}~\bibnamefont{Aad}} \bibnamefont{et~al.}
  (\bibinfo{collaboration}{ATLAS}) (\bibinfo{year}{2015}{\natexlab{a}}),
  \eprint{1506.00962}.

\bibitem[{\citenamefont{Khachatryan
  et~al.}(2014{\natexlab{a}})}]{Khachatryan:2014hpa}
\bibinfo{author}{\bibfnamefont{V.}~\bibnamefont{Khachatryan}}
  \bibnamefont{et~al.} (\bibinfo{collaboration}{CMS}), \bibinfo{journal}{JHEP}
  \textbf{\bibinfo{volume}{1408}}, \bibinfo{pages}{173}
  (\bibinfo{year}{2014}{\natexlab{a}}), \eprint{1405.1994}.

\bibitem[{\citenamefont{Khachatryan
  et~al.}(2014{\natexlab{b}})}]{Khachatryan:2014gha}
\bibinfo{author}{\bibfnamefont{V.}~\bibnamefont{Khachatryan}}
  \bibnamefont{et~al.} (\bibinfo{collaboration}{CMS}), \bibinfo{journal}{JHEP}
  \textbf{\bibinfo{volume}{1408}}, \bibinfo{pages}{174}
  (\bibinfo{year}{2014}{\natexlab{b}}), \eprint{1405.3447}.

\bibitem[{\citenamefont{Khachatryan
  et~al.}(2015{\natexlab{a}})}]{Khachatryan:2015bma}
\bibinfo{author}{\bibfnamefont{V.}~\bibnamefont{Khachatryan}}
  \bibnamefont{et~al.} (\bibinfo{collaboration}{CMS})
  (\bibinfo{year}{2015}{\natexlab{a}}), \eprint{1506.01443}.

\bibitem[{\citenamefont{Fukano et~al.}(2015)\citenamefont{Fukano, Kurachi,
  Matsuzaki, Terashi, and Yamawaki}}]{Fukano:2015hga}
\bibinfo{author}{\bibfnamefont{H.~S.} \bibnamefont{Fukano}},
  \bibinfo{author}{\bibfnamefont{M.}~\bibnamefont{Kurachi}},
  \bibinfo{author}{\bibfnamefont{S.}~\bibnamefont{Matsuzaki}},
  \bibinfo{author}{\bibfnamefont{K.}~\bibnamefont{Terashi}}, \bibnamefont{and}
  \bibinfo{author}{\bibfnamefont{K.}~\bibnamefont{Yamawaki}}
  (\bibinfo{year}{2015}), \eprint{1506.03751}.

\bibitem[{\citenamefont{Hisano et~al.}(2015)\citenamefont{Hisano, Nagata, and
  Omura}}]{Hisano:2015gna}
\bibinfo{author}{\bibfnamefont{J.}~\bibnamefont{Hisano}},
  \bibinfo{author}{\bibfnamefont{N.}~\bibnamefont{Nagata}}, \bibnamefont{and}
  \bibinfo{author}{\bibfnamefont{Y.}~\bibnamefont{Omura}}
  (\bibinfo{year}{2015}), \eprint{1506.03931}.

\bibitem[{\citenamefont{Franzosi et~al.}(2015)\citenamefont{Franzosi, Frandsen,
  and Sannino}}]{Franzosi:2015zra}
\bibinfo{author}{\bibfnamefont{D.~B.} \bibnamefont{Franzosi}},
  \bibinfo{author}{\bibfnamefont{M.~T.} \bibnamefont{Frandsen}},
  \bibnamefont{and} \bibinfo{author}{\bibfnamefont{F.}~\bibnamefont{Sannino}}
  (\bibinfo{year}{2015}), \eprint{1506.04392}.

\bibitem[{\citenamefont{Cheung et~al.}(2015)\citenamefont{Cheung, Keung, Tseng,
  and Yuan}}]{Cheung:2015nha}
\bibinfo{author}{\bibfnamefont{K.}~\bibnamefont{Cheung}},
  \bibinfo{author}{\bibfnamefont{W.-Y.} \bibnamefont{Keung}},
  \bibinfo{author}{\bibfnamefont{P.-Y.} \bibnamefont{Tseng}}, \bibnamefont{and}
  \bibinfo{author}{\bibfnamefont{T.-C.} \bibnamefont{Yuan}}
  (\bibinfo{year}{2015}), \eprint{1506.06064}.

\bibitem[{\citenamefont{Dobrescu and Liu}(2015)}]{Dobrescu:2015qna}
\bibinfo{author}{\bibfnamefont{B.~A.} \bibnamefont{Dobrescu}} \bibnamefont{and}
  \bibinfo{author}{\bibfnamefont{Z.}~\bibnamefont{Liu}} (\bibinfo{year}{2015}),
  \eprint{1506.06736}.

\bibitem[{\citenamefont{Aguilar-Saavedra}(2015)}]{Aguilar-Saavedra:2015rna}
\bibinfo{author}{\bibfnamefont{J.}~\bibnamefont{Aguilar-Saavedra}}
  (\bibinfo{year}{2015}), \eprint{1506.06739}.

\bibitem[{\citenamefont{Gao et~al.}(2015)\citenamefont{Gao, Ghosh, Sinha, and
  Yu}}]{Gao:2015irw}
\bibinfo{author}{\bibfnamefont{Y.}~\bibnamefont{Gao}},
  \bibinfo{author}{\bibfnamefont{T.}~\bibnamefont{Ghosh}},
  \bibinfo{author}{\bibfnamefont{K.}~\bibnamefont{Sinha}}, \bibnamefont{and}
  \bibinfo{author}{\bibfnamefont{J.-H.} \bibnamefont{Yu}}
  (\bibinfo{year}{2015}), \eprint{1506.07511}.

\bibitem[{\citenamefont{Thamm et~al.}(2015)\citenamefont{Thamm, Torre, and
  Wulzer}}]{Thamm:2015csa}
\bibinfo{author}{\bibfnamefont{A.}~\bibnamefont{Thamm}},
  \bibinfo{author}{\bibfnamefont{R.}~\bibnamefont{Torre}}, \bibnamefont{and}
  \bibinfo{author}{\bibfnamefont{A.}~\bibnamefont{Wulzer}}
  (\bibinfo{year}{2015}), \eprint{1506.08688}.

\bibitem[{\citenamefont{Alves et~al.}(2015)\citenamefont{Alves, Berlin,
  Profumo, and Queiroz}}]{Alves:2015mua}
\bibinfo{author}{\bibfnamefont{A.}~\bibnamefont{Alves}},
  \bibinfo{author}{\bibfnamefont{A.}~\bibnamefont{Berlin}},
  \bibinfo{author}{\bibfnamefont{S.}~\bibnamefont{Profumo}}, \bibnamefont{and}
  \bibinfo{author}{\bibfnamefont{F.~S.} \bibnamefont{Queiroz}}
  (\bibinfo{year}{2015}), \eprint{1506.06767}.

\bibitem[{\citenamefont{Hsieh et~al.}(2010)\citenamefont{Hsieh, Schmitz, Yu,
  and Yuan}}]{Hsieh:2010zr}
\bibinfo{author}{\bibfnamefont{K.}~\bibnamefont{Hsieh}},
  \bibinfo{author}{\bibfnamefont{K.}~\bibnamefont{Schmitz}},
  \bibinfo{author}{\bibfnamefont{J.-H.} \bibnamefont{Yu}}, \bibnamefont{and}
  \bibinfo{author}{\bibfnamefont{C.-P.} \bibnamefont{Yuan}},
  \bibinfo{journal}{Phys.Rev.} \textbf{\bibinfo{volume}{D82}},
  \bibinfo{pages}{035011} (\bibinfo{year}{2010}), \eprint{1003.3482}.

\bibitem[{\citenamefont{Berger et~al.}(2011{\natexlab{a}})\citenamefont{Berger,
  Cao, Chen, and Zhang}}]{Berger:2011hn}
\bibinfo{author}{\bibfnamefont{E.~L.} \bibnamefont{Berger}},
  \bibinfo{author}{\bibfnamefont{Q.-H.} \bibnamefont{Cao}},
  \bibinfo{author}{\bibfnamefont{C.-R.} \bibnamefont{Chen}}, \bibnamefont{and}
  \bibinfo{author}{\bibfnamefont{H.}~\bibnamefont{Zhang}},
  \bibinfo{journal}{Phys.Rev.} \textbf{\bibinfo{volume}{D83}},
  \bibinfo{pages}{114026} (\bibinfo{year}{2011}{\natexlab{a}}),
  \eprint{1103.3274}.

\bibitem[{\citenamefont{Cao et~al.}(2012)\citenamefont{Cao, Li, Yu, and
  Yuan}}]{Cao:2012ng}
\bibinfo{author}{\bibfnamefont{Q.-H.} \bibnamefont{Cao}},
  \bibinfo{author}{\bibfnamefont{Z.}~\bibnamefont{Li}},
  \bibinfo{author}{\bibfnamefont{J.-H.} \bibnamefont{Yu}}, \bibnamefont{and}
  \bibinfo{author}{\bibfnamefont{C.-P.}~\bibnamefont{Yuan}},
  \bibinfo{journal}{Phys.Rev.} \textbf{\bibinfo{volume}{D86}},
  \bibinfo{pages}{095010} (\bibinfo{year}{2012}), \eprint{1205.3769}.

\bibitem[{\citenamefont{Frampton}(1992)}]{Frampton:1992wt}
\bibinfo{author}{\bibfnamefont{P.}~\bibnamefont{Frampton}},
  \bibinfo{journal}{Phys.Rev.Lett.} \textbf{\bibinfo{volume}{69}},
  \bibinfo{pages}{2889} (\bibinfo{year}{1992}).

\bibitem[{\citenamefont{Pisano and Pleitez}(1992)}]{Pisano:1991ee}
\bibinfo{author}{\bibfnamefont{F.}~\bibnamefont{Pisano}} \bibnamefont{and}
  \bibinfo{author}{\bibfnamefont{V.}~\bibnamefont{Pleitez}},
  \bibinfo{journal}{Phys.Rev.} \textbf{\bibinfo{volume}{D46}},
  \bibinfo{pages}{410} (\bibinfo{year}{1992}), \eprint{hep-ph/9206242}.

\bibitem[{\citenamefont{Aad et~al.}(2015{\natexlab{b}})}]{Aad:2014aqa}
\bibinfo{author}{\bibfnamefont{G.}~\bibnamefont{Aad}} \bibnamefont{et~al.}
  (\bibinfo{collaboration}{ATLAS}), \bibinfo{journal}{Phys.Rev.}
  \textbf{\bibinfo{volume}{D91}}, \bibinfo{pages}{052007}
  (\bibinfo{year}{2015}{\natexlab{b}}), \eprint{1407.1376}.

\bibitem[{\citenamefont{Khachatryan
  et~al.}(2015{\natexlab{b}})}]{Khachatryan:2015sja}
\bibinfo{author}{\bibfnamefont{V.}~\bibnamefont{Khachatryan}}
  \bibnamefont{et~al.} (\bibinfo{collaboration}{CMS}),
  \bibinfo{journal}{Phys.Rev.} \textbf{\bibinfo{volume}{D91}},
  \bibinfo{pages}{052009} (\bibinfo{year}{2015}{\natexlab{b}}),
  \eprint{1501.04198}.

\bibitem[{\citenamefont{Aad et~al.}(2015{\natexlab{c}})}]{Aad:2014xea}
\bibinfo{author}{\bibfnamefont{G.}~\bibnamefont{Aad}} \bibnamefont{et~al.}
  (\bibinfo{collaboration}{ATLAS}), \bibinfo{journal}{Phys.Lett.}
  \textbf{\bibinfo{volume}{B743}}, \bibinfo{pages}{235}
  (\bibinfo{year}{2015}{\natexlab{c}}), \eprint{1410.4103}.

\bibitem[{\citenamefont{Khachatryan
  et~al.}(2015{\natexlab{c}})}]{Khachatryan:2015sma}
\bibinfo{author}{\bibfnamefont{V.}~\bibnamefont{Khachatryan}}
  \bibnamefont{et~al.} (\bibinfo{collaboration}{CMS})
  (\bibinfo{year}{2015}{\natexlab{c}}), \eprint{1506.03062}.

\bibitem[{\citenamefont{Aad et~al.}(2014{\natexlab{a}})}]{Aad:2014cka}
\bibinfo{author}{\bibfnamefont{G.}~\bibnamefont{Aad}} \bibnamefont{et~al.}
  (\bibinfo{collaboration}{ATLAS}), \bibinfo{journal}{Phys.Rev.}
  \textbf{\bibinfo{volume}{D90}}, \bibinfo{pages}{052005}
  (\bibinfo{year}{2014}{\natexlab{a}}), \eprint{1405.4123}.

\bibitem[{\citenamefont{Khachatryan
  et~al.}(2015{\natexlab{d}})}]{Khachatryan:2014fba}
\bibinfo{author}{\bibfnamefont{V.}~\bibnamefont{Khachatryan}}
  \bibnamefont{et~al.} (\bibinfo{collaboration}{CMS}), \bibinfo{journal}{JHEP}
  \textbf{\bibinfo{volume}{1504}}, \bibinfo{pages}{025}
  (\bibinfo{year}{2015}{\natexlab{d}}), \eprint{1412.6302}.

\bibitem[{\citenamefont{Aad et~al.}(2014{\natexlab{b}})}]{ATLAS:2014wra}
\bibinfo{author}{\bibfnamefont{G.}~\bibnamefont{Aad}} \bibnamefont{et~al.}
  (\bibinfo{collaboration}{ATLAS}), \bibinfo{journal}{JHEP}
  \textbf{\bibinfo{volume}{1409}}, \bibinfo{pages}{037}
  (\bibinfo{year}{2014}{\natexlab{b}}), \eprint{1407.7494}.

\bibitem[{\citenamefont{Khachatryan
  et~al.}(2015{\natexlab{e}})}]{Khachatryan:2014tva}
\bibinfo{author}{\bibfnamefont{V.}~\bibnamefont{Khachatryan}}
  \bibnamefont{et~al.} (\bibinfo{collaboration}{CMS}),
  \bibinfo{journal}{Phys.Rev.} \textbf{\bibinfo{volume}{D91}},
  \bibinfo{pages}{092005} (\bibinfo{year}{2015}{\natexlab{e}}),
  \eprint{1408.2745}.

\bibitem[{\citenamefont{Mohapatra and
  Pati}(1975{\natexlab{a}})}]{Mohapatra:1974gc}
\bibinfo{author}{\bibfnamefont{R.}~\bibnamefont{Mohapatra}} \bibnamefont{and}
  \bibinfo{author}{\bibfnamefont{J.~C.} \bibnamefont{Pati}},
  \bibinfo{journal}{Phys.Rev.} \textbf{\bibinfo{volume}{D11}},
  \bibinfo{pages}{2558} (\bibinfo{year}{1975}{\natexlab{a}}).

\bibitem[{\citenamefont{Mohapatra and
  Pati}(1975{\natexlab{b}})}]{Mohapatra:1974hk}
\bibinfo{author}{\bibfnamefont{R.~N.} \bibnamefont{Mohapatra}}
  \bibnamefont{and} \bibinfo{author}{\bibfnamefont{J.~C.} \bibnamefont{Pati}},
  \bibinfo{journal}{Phys.Rev.} \textbf{\bibinfo{volume}{D11}},
  \bibinfo{pages}{566} (\bibinfo{year}{1975}{\natexlab{b}}).

\bibitem[{\citenamefont{Mohapatra and Senjanovic}(1981)}]{Mohapatra:1980yp}
\bibinfo{author}{\bibfnamefont{R.~N.} \bibnamefont{Mohapatra}}
  \bibnamefont{and}
  \bibinfo{author}{\bibfnamefont{G.}~\bibnamefont{Senjanovic}},
  \bibinfo{journal}{Phys.Rev.} \textbf{\bibinfo{volume}{D23}},
  \bibinfo{pages}{165} (\bibinfo{year}{1981}).

\bibitem[{\citenamefont{Barger et~al.}(1980{\natexlab{a}})\citenamefont{Barger,
  Keung, and Ma}}]{Barger:1980ix}
\bibinfo{author}{\bibfnamefont{V.~D.} \bibnamefont{Barger}},
  \bibinfo{author}{\bibfnamefont{W.-Y.} \bibnamefont{Keung}}, \bibnamefont{and}
  \bibinfo{author}{\bibfnamefont{E.}~\bibnamefont{Ma}},
  \bibinfo{journal}{Phys.Rev.} \textbf{\bibinfo{volume}{D22}},
  \bibinfo{pages}{727} (\bibinfo{year}{1980}{\natexlab{a}}).

\bibitem[{\citenamefont{Barger et~al.}(1980{\natexlab{b}})\citenamefont{Barger,
  Keung, and Ma}}]{Barger:1980ti}
\bibinfo{author}{\bibfnamefont{V.~D.} \bibnamefont{Barger}},
  \bibinfo{author}{\bibfnamefont{W.-Y.} \bibnamefont{Keung}}, \bibnamefont{and}
  \bibinfo{author}{\bibfnamefont{E.}~\bibnamefont{Ma}},
  \bibinfo{journal}{Phys.Rev.Lett.} \textbf{\bibinfo{volume}{44}},
  \bibinfo{pages}{1169} (\bibinfo{year}{1980}{\natexlab{b}}).

\bibitem[{\citenamefont{Georgi et~al.}(1989)\citenamefont{Georgi, Jenkins, and
  Simmons}}]{Georgi:1989ic}
\bibinfo{author}{\bibfnamefont{H.}~\bibnamefont{Georgi}},
  \bibinfo{author}{\bibfnamefont{E.~E.} \bibnamefont{Jenkins}},
  \bibnamefont{and} \bibinfo{author}{\bibfnamefont{E.~H.}
  \bibnamefont{Simmons}}, \bibinfo{journal}{Phys.Rev.Lett.}
  \textbf{\bibinfo{volume}{62}}, \bibinfo{pages}{2789} (\bibinfo{year}{1989}).

\bibitem[{\citenamefont{Georgi et~al.}(1990)\citenamefont{Georgi, Jenkins, and
  Simmons}}]{Georgi:1989xz}
\bibinfo{author}{\bibfnamefont{H.}~\bibnamefont{Georgi}},
  \bibinfo{author}{\bibfnamefont{E.~E.} \bibnamefont{Jenkins}},
  \bibnamefont{and} \bibinfo{author}{\bibfnamefont{E.~H.}
  \bibnamefont{Simmons}}, \bibinfo{journal}{Nucl.Phys.}
  \textbf{\bibinfo{volume}{B331}}, \bibinfo{pages}{541} (\bibinfo{year}{1990}).

\bibitem[{\citenamefont{Li and Ma}(1981)}]{Li:1981nk}
\bibinfo{author}{\bibfnamefont{X.}~\bibnamefont{Li}} \bibnamefont{and}
  \bibinfo{author}{\bibfnamefont{E.}~\bibnamefont{Ma}},
  \bibinfo{journal}{Phys.Rev.Lett.} \textbf{\bibinfo{volume}{47}},
  \bibinfo{pages}{1788} (\bibinfo{year}{1981}).

\bibitem[{\citenamefont{Malkawi et~al.}(1996)\citenamefont{Malkawi, Tait, and
  Yuan}}]{Malkawi:1996fs}
\bibinfo{author}{\bibfnamefont{E.}~\bibnamefont{Malkawi}},
  \bibinfo{author}{\bibfnamefont{T.~M.} \bibnamefont{Tait}}, \bibnamefont{and}
  \bibinfo{author}{\bibfnamefont{C.-P.}~\bibnamefont{Yuan}},
  \bibinfo{journal}{Phys.Lett.} \textbf{\bibinfo{volume}{B385}},
  \bibinfo{pages}{304} (\bibinfo{year}{1996}), \eprint{hep-ph/9603349}.

\bibitem[{\citenamefont{He et~al.}(2000)\citenamefont{He, Tait, and
  Yuan}}]{He:1999vp}
\bibinfo{author}{\bibfnamefont{H.-J.} \bibnamefont{He}},
  \bibinfo{author}{\bibfnamefont{T.~M.} \bibnamefont{Tait}}, \bibnamefont{and}
  \bibinfo{author}{\bibfnamefont{C.-P.}~\bibnamefont{Yuan}},
  \bibinfo{journal}{Phys.Rev.} \textbf{\bibinfo{volume}{D62}},
  \bibinfo{pages}{011702} (\bibinfo{year}{2000}), \eprint{hep-ph/9911266}.

\bibitem[{\citenamefont{Chivukula et~al.}(2004)\citenamefont{Chivukula, He,
  Howard, and Simmons}}]{Chivukula:2003wj}
\bibinfo{author}{\bibfnamefont{R.~S.} \bibnamefont{Chivukula}},
  \bibinfo{author}{\bibfnamefont{H.-J.} \bibnamefont{He}},
  \bibinfo{author}{\bibfnamefont{J.}~\bibnamefont{Howard}}, \bibnamefont{and}
  \bibinfo{author}{\bibfnamefont{E.~H.} \bibnamefont{Simmons}},
  \bibinfo{journal}{Phys.Rev.} \textbf{\bibinfo{volume}{D69}},
  \bibinfo{pages}{015009} (\bibinfo{year}{2004}), \eprint{hep-ph/0307209}.

\bibitem[{\citenamefont{Chivukula et~al.}(2006)\citenamefont{Chivukula,
  Coleppa, Di~Chiara, Simmons, He et~al.}}]{Chivukula:2006cg}
\bibinfo{author}{\bibfnamefont{R.~S.} \bibnamefont{Chivukula}},
  \bibinfo{author}{\bibfnamefont{B.}~\bibnamefont{Coleppa}},
  \bibinfo{author}{\bibfnamefont{S.}~\bibnamefont{Di~Chiara}},
  \bibinfo{author}{\bibfnamefont{E.~H.} \bibnamefont{Simmons}},
  \bibinfo{author}{\bibfnamefont{H.-J.} \bibnamefont{He}},
  \bibnamefont{et~al.}, \bibinfo{journal}{Phys.Rev.}
  \textbf{\bibinfo{volume}{D74}}, \bibinfo{pages}{075011}
  (\bibinfo{year}{2006}), \eprint{hep-ph/0607124}.

\bibitem[{\citenamefont{Berger et~al.}(2011{\natexlab{b}})\citenamefont{Berger,
  Cao, Yu, and Yuan}}]{Berger:2011xk}
\bibinfo{author}{\bibfnamefont{E.~L.} \bibnamefont{Berger}},
  \bibinfo{author}{\bibfnamefont{Q.-H.} \bibnamefont{Cao}},
  \bibinfo{author}{\bibfnamefont{J.-H.} \bibnamefont{Yu}}, \bibnamefont{and}
  \bibinfo{author}{\bibfnamefont{C.-P.} \bibnamefont{Yuan}},
  \bibinfo{journal}{Phys.Rev.} \textbf{\bibinfo{volume}{D84}},
  \bibinfo{pages}{095026} (\bibinfo{year}{2011}{\natexlab{b}}),
  \eprint{1108.3613}.

\bibitem[{\citenamefont{Du et~al.}(2012)\citenamefont{Du, He, Kuang, Zhang,
  Christensen et~al.}}]{Du:2012vh}
\bibinfo{author}{\bibfnamefont{C.}~\bibnamefont{Du}},
  \bibinfo{author}{\bibfnamefont{H.-J.} \bibnamefont{He}},
  \bibinfo{author}{\bibfnamefont{Y.-P.} \bibnamefont{Kuang}},
  \bibinfo{author}{\bibfnamefont{B.}~\bibnamefont{Zhang}},
  \bibinfo{author}{\bibfnamefont{N.~D.} \bibnamefont{Christensen}},
  \bibnamefont{et~al.}, \bibinfo{journal}{Phys.Rev.}
  \textbf{\bibinfo{volume}{D86}}, \bibinfo{pages}{095011}
  (\bibinfo{year}{2012}), \eprint{1206.6022}.

\bibitem[{\citenamefont{Abe et~al.}(2013)\citenamefont{Abe, Chen, and
  He}}]{Abe:2012fb}
\bibinfo{author}{\bibfnamefont{T.}~\bibnamefont{Abe}},
  \bibinfo{author}{\bibfnamefont{N.}~\bibnamefont{Chen}}, \bibnamefont{and}
  \bibinfo{author}{\bibfnamefont{H.-J.} \bibnamefont{He}},
  \bibinfo{journal}{JHEP} \textbf{\bibinfo{volume}{1301}}, \bibinfo{pages}{082}
  (\bibinfo{year}{2013}), \eprint{1207.4103}.

\bibitem[{\citenamefont{Wang et~al.}(2013)\citenamefont{Wang, Du, and
  He}}]{Wang:2013jwa}
\bibinfo{author}{\bibfnamefont{X.-F.} \bibnamefont{Wang}},
  \bibinfo{author}{\bibfnamefont{C.}~\bibnamefont{Du}}, \bibnamefont{and}
  \bibinfo{author}{\bibfnamefont{H.-J.} \bibnamefont{He}},
  \bibinfo{journal}{Phys.Lett.} \textbf{\bibinfo{volume}{B723}},
  \bibinfo{pages}{314} (\bibinfo{year}{2013}), \eprint{1304.2257}.

\bibitem[{\citenamefont{Cao et~al.}(2015)\citenamefont{Cao, Yan, Yu, and
  Zhang}}]{Cao:2015doa}
\bibinfo{author}{\bibfnamefont{Q.-H.} \bibnamefont{Cao}},
  \bibinfo{author}{\bibfnamefont{B.}~\bibnamefont{Yan}},
  \bibinfo{author}{\bibfnamefont{J.-H.} \bibnamefont{Yu}}, \bibnamefont{and}
  \bibinfo{author}{\bibfnamefont{C.}~\bibnamefont{Zhang}}
  (\bibinfo{year}{2015}), \eprint{1504.03785}.

\bibitem[{\citenamefont{Patra et~al.}(2015)\citenamefont{Patra, Queiroz, and
  Rodejohann}}]{Patra:2015bga}
\bibinfo{author}{\bibfnamefont{S.}~\bibnamefont{Patra}},
  \bibinfo{author}{\bibfnamefont{F.~S.} \bibnamefont{Queiroz}},
  \bibnamefont{and}
  \bibinfo{author}{\bibfnamefont{W.}~\bibnamefont{Rodejohann}}
  (\bibinfo{year}{2015}), \eprint{1506.03456}.

\bibitem[{\citenamefont{Dulat et~al.}(2015)\citenamefont{Dulat, Hou, Gao,
  Guzzi, Huston et~al.}}]{Dulat:2015mca}
\bibinfo{author}{\bibfnamefont{S.}~\bibnamefont{Dulat}},
  \bibinfo{author}{\bibfnamefont{T.~J.} \bibnamefont{Hou}},
  \bibinfo{author}{\bibfnamefont{J.}~\bibnamefont{Gao}},
  \bibinfo{author}{\bibfnamefont{M.}~\bibnamefont{Guzzi}},
  \bibinfo{author}{\bibfnamefont{J.}~\bibnamefont{Huston}},
  \bibnamefont{et~al.} (\bibinfo{year}{2015}), \eprint{1506.07443}.

\bibitem[{\citenamefont{Berger and Cao}(2010)}]{Berger:2009qy}
\bibinfo{author}{\bibfnamefont{E.~L.} \bibnamefont{Berger}} \bibnamefont{and}
  \bibinfo{author}{\bibfnamefont{Q.-H.} \bibnamefont{Cao}},
  \bibinfo{journal}{Phys.Rev.} \textbf{\bibinfo{volume}{D81}},
  \bibinfo{pages}{035006} (\bibinfo{year}{2010}), \eprint{0909.3555}.

\bibitem[{\citenamefont{Lai et~al.}(2010)\citenamefont{Lai, Guzzi, Huston, Li,
  Nadolsky et~al.}}]{Lai:2010vv}
\bibinfo{author}{\bibfnamefont{H.-L.} \bibnamefont{Lai}},
  \bibinfo{author}{\bibfnamefont{M.}~\bibnamefont{Guzzi}},
  \bibinfo{author}{\bibfnamefont{J.}~\bibnamefont{Huston}},
  \bibinfo{author}{\bibfnamefont{Z.}~\bibnamefont{Li}},
  \bibinfo{author}{\bibfnamefont{P.~M.} \bibnamefont{Nadolsky}},
  \bibnamefont{et~al.}, \bibinfo{journal}{Phys.Rev.}
  \textbf{\bibinfo{volume}{D82}}, \bibinfo{pages}{074024}
  (\bibinfo{year}{2010}), \eprint{1007.2241}.

\bibitem[{\citenamefont{Buras et~al.}(2014)\citenamefont{Buras, De~Fazio, and
  Girrbach-Noe}}]{Buras:2014yna}
\bibinfo{author}{\bibfnamefont{A.~J.} \bibnamefont{Buras}},
  \bibinfo{author}{\bibfnamefont{F.}~\bibnamefont{De~Fazio}}, \bibnamefont{and}
  \bibinfo{author}{\bibfnamefont{J.}~\bibnamefont{Girrbach-Noe}},
  \bibinfo{journal}{JHEP} \textbf{\bibinfo{volume}{1408}}, \bibinfo{pages}{039}
  (\bibinfo{year}{2014}), \eprint{1405.3850}.

\bibitem[{\citenamefont{Buras et~al.}(2013)\citenamefont{Buras, De~Fazio,
  Girrbach, and Carlucci}}]{Buras:2012dp}
\bibinfo{author}{\bibfnamefont{A.~J.} \bibnamefont{Buras}},
  \bibinfo{author}{\bibfnamefont{F.}~\bibnamefont{De~Fazio}},
  \bibinfo{author}{\bibfnamefont{J.}~\bibnamefont{Girrbach}}, \bibnamefont{and}
  \bibinfo{author}{\bibfnamefont{M.~V.} \bibnamefont{Carlucci}},
  \bibinfo{journal}{JHEP} \textbf{\bibinfo{volume}{1302}}, \bibinfo{pages}{023}
  (\bibinfo{year}{2013}), \eprint{1211.1237}.

\bibitem[{\citenamefont{Ninh and Long}(2005)}]{Ninh:2005su}
\bibinfo{author}{\bibfnamefont{L.~D.} \bibnamefont{Ninh}} \bibnamefont{and}
  \bibinfo{author}{\bibfnamefont{H.~N.} \bibnamefont{Long}},
  \bibinfo{journal}{Phys.Rev.} \textbf{\bibinfo{volume}{D72}},
  \bibinfo{pages}{075004} (\bibinfo{year}{2005}), \eprint{hep-ph/0507069}.

\bibitem[{\citenamefont{Martinez and Ochoa}(2012)}]{Martinez:2012ni}
\bibinfo{author}{\bibfnamefont{R.}~\bibnamefont{Martinez}} \bibnamefont{and}
  \bibinfo{author}{\bibfnamefont{F.}~\bibnamefont{Ochoa}},
  \bibinfo{journal}{Phys.Rev.} \textbf{\bibinfo{volume}{D86}},
  \bibinfo{pages}{065030} (\bibinfo{year}{2012}), \eprint{1208.4085}.

\bibitem[{\citenamefont{Montalvo et~al.}(2012)\citenamefont{Montalvo, Ulloa,
  and Tonasse}}]{Montalvo:2012qg}
\bibinfo{author}{\bibfnamefont{J.~C.} \bibnamefont{Montalvo}},
  \bibinfo{author}{\bibfnamefont{G.~R.} \bibnamefont{Ulloa}}, \bibnamefont{and}
  \bibinfo{author}{\bibfnamefont{M.}~\bibnamefont{Tonasse}},
  \bibinfo{journal}{Eur.Phys.J.} \textbf{\bibinfo{volume}{C72}},
  \bibinfo{pages}{2210} (\bibinfo{year}{2012}), \eprint{1205.3822}.

\bibitem[{\citenamefont{Alves et~al.}(2011)\citenamefont{Alves, Barreto, and
  Dias}}]{Alves:2011mz}
\bibinfo{author}{\bibfnamefont{A.}~\bibnamefont{Alves}},
  \bibinfo{author}{\bibfnamefont{E.~R.} \bibnamefont{Barreto}},
  \bibnamefont{and} \bibinfo{author}{\bibfnamefont{A.}~\bibnamefont{Dias}},
  \bibinfo{journal}{Phys.Rev.} \textbf{\bibinfo{volume}{D84}},
  \bibinfo{pages}{075013} (\bibinfo{year}{2011}), \eprint{1105.4849}.

\bibitem[{\citenamefont{Cieza~Montalvo
  et~al.}(2008{\natexlab{a}})\citenamefont{Cieza~Montalvo, Cortez, and
  Tonasse}}]{Montalvo:2008cx}
\bibinfo{author}{\bibfnamefont{J.}~\bibnamefont{Cieza~Montalvo}},
  \bibinfo{author}{\bibfnamefont{N.~V.} \bibnamefont{Cortez}},
  \bibnamefont{and} \bibinfo{author}{\bibfnamefont{M.}~\bibnamefont{Tonasse}}
  (\bibinfo{year}{2008}{\natexlab{a}}), \eprint{0812.4000}.

\bibitem[{\citenamefont{Cieza~Montalvo
  et~al.}(2008{\natexlab{b}})\citenamefont{Cieza~Montalvo, Cortez, and
  Tonasse}}]{CiezaMontalvo:2008sa}
\bibinfo{author}{\bibfnamefont{J.}~\bibnamefont{Cieza~Montalvo}},
  \bibinfo{author}{\bibfnamefont{N.~V.} \bibnamefont{Cortez}},
  \bibnamefont{and} \bibinfo{author}{\bibfnamefont{M.}~\bibnamefont{Tonasse}},
  \bibinfo{journal}{Phys.Rev.} \textbf{\bibinfo{volume}{D78}},
  \bibinfo{pages}{116003} (\bibinfo{year}{2008}{\natexlab{b}}),
  \eprint{0804.0618}.

\bibitem[{\citenamefont{Cieza~Montalvo
  et~al.}(2008{\natexlab{c}})\citenamefont{Cieza~Montalvo, Cortez, and
  Tonasse}}]{CiezaMontalvo:2008ew}
\bibinfo{author}{\bibfnamefont{J.}~\bibnamefont{Cieza~Montalvo}},
  \bibinfo{author}{\bibfnamefont{N.~V.} \bibnamefont{Cortez}},
  \bibnamefont{and} \bibinfo{author}{\bibfnamefont{M.}~\bibnamefont{Tonasse}},
  \bibinfo{journal}{Phys.Rev.} \textbf{\bibinfo{volume}{D77}},
  \bibinfo{pages}{095015} (\bibinfo{year}{2008}{\natexlab{c}}),
  \eprint{0804.0033}.

\bibitem[{\citenamefont{Soa et~al.}(2007)\citenamefont{Soa, Thuy, Thuc, and
  Huong}}]{Soa:2007zz}
\bibinfo{author}{\bibfnamefont{D.}~\bibnamefont{Soa}},
  \bibinfo{author}{\bibfnamefont{D.}~\bibnamefont{Thuy}},
  \bibinfo{author}{\bibfnamefont{L.}~\bibnamefont{Thuc}}, \bibnamefont{and}
  \bibinfo{author}{\bibfnamefont{T.}~\bibnamefont{Huong}},
  \bibinfo{journal}{J.Exp.Theor.Phys.} \textbf{\bibinfo{volume}{105}},
  \bibinfo{pages}{1107} (\bibinfo{year}{2007}).

\bibitem[{\citenamefont{Van~Soa and Le~Thuy}(2006)}]{VanSoa:2006ea}
\bibinfo{author}{\bibfnamefont{D.}~\bibnamefont{Van~Soa}} \bibnamefont{and}
  \bibinfo{author}{\bibfnamefont{D.}~\bibnamefont{Le~Thuy}}
  (\bibinfo{year}{2006}), \eprint{hep-ph/0610297}.

\bibitem[{\citenamefont{Cieza~Montalvo
  et~al.}(2006)\citenamefont{Cieza~Montalvo, Cortez, Sa~Borges, and
  Tonasse}}]{CiezaMontalvo:2006zt}
\bibinfo{author}{\bibfnamefont{J.}~\bibnamefont{Cieza~Montalvo}},
  \bibinfo{author}{\bibfnamefont{N.~V.} \bibnamefont{Cortez}},
  \bibinfo{author}{\bibfnamefont{J.}~\bibnamefont{Sa~Borges}},
  \bibnamefont{and} \bibinfo{author}{\bibfnamefont{M.~D.}
  \bibnamefont{Tonasse}}, \bibinfo{journal}{Nucl.Phys.}
  \textbf{\bibinfo{volume}{B756}}, \bibinfo{pages}{1} (\bibinfo{year}{2006}),
  \eprint{hep-ph/0606243}.

\bibitem[{\citenamefont{Van~Soa et~al.}(2009)\citenamefont{Van~Soa, Dong,
  Huong, and Long}}]{VanSoa:2008bm}
\bibinfo{author}{\bibfnamefont{D.}~\bibnamefont{Van~Soa}},
  \bibinfo{author}{\bibfnamefont{P.~V.} \bibnamefont{Dong}},
  \bibinfo{author}{\bibfnamefont{T.~T.} \bibnamefont{Huong}}, \bibnamefont{and}
  \bibinfo{author}{\bibfnamefont{H.~N.} \bibnamefont{Long}},
  \bibinfo{journal}{J.Exp.Theor.Phys.} \textbf{\bibinfo{volume}{108}},
  \bibinfo{pages}{757} (\bibinfo{year}{2009}), \eprint{0805.4456}.

\bibitem[{\citenamefont{Coutinho et~al.}(2013)\citenamefont{Coutinho,
  Salustino~Guimarães, and Nepomuceno}}]{Coutinho:2013lta}
\bibinfo{author}{\bibfnamefont{Y.}~\bibnamefont{Coutinho}},
  \bibinfo{author}{\bibfnamefont{V.}~\bibnamefont{Salustino~Guimarães}},
  \bibnamefont{and}
  \bibinfo{author}{\bibfnamefont{A.}~\bibnamefont{Nepomuceno}},
  \bibinfo{journal}{Phys.Rev.} \textbf{\bibinfo{volume}{D87}},
  \bibinfo{pages}{115014} (\bibinfo{year}{2013}), \eprint{1304.7907}.

\bibitem[{\citenamefont{Martinez and Ochoa}(2009)}]{Martinez:2009ik}
\bibinfo{author}{\bibfnamefont{R.}~\bibnamefont{Martinez}} \bibnamefont{and}
  \bibinfo{author}{\bibfnamefont{F.}~\bibnamefont{Ochoa}},
  \bibinfo{journal}{Phys.Rev.} \textbf{\bibinfo{volume}{D80}},
  \bibinfo{pages}{075020} (\bibinfo{year}{2009}), \eprint{0909.1121}.

\bibitem[{\citenamefont{Ramirez~Barreto
  et~al.}(2007)\citenamefont{Ramirez~Barreto, Coutinho, and
  Sa~Borges}}]{RamirezBarreto:2007mt}
\bibinfo{author}{\bibfnamefont{E.}~\bibnamefont{Ramirez~Barreto}},
  \bibinfo{author}{\bibfnamefont{Y.~D.~A.} \bibnamefont{Coutinho}},
  \bibnamefont{and}
  \bibinfo{author}{\bibfnamefont{J.}~\bibnamefont{Sa~Borges}},
  \bibinfo{journal}{Eur.Phys.J.} \textbf{\bibinfo{volume}{C50}},
  \bibinfo{pages}{909} (\bibinfo{year}{2007}), \eprint{hep-ph/0703099}.

\bibitem[{\citenamefont{Ramirez~Barreto
  et~al.}(2006)\citenamefont{Ramirez~Barreto, Coutinho, and
  Sa~Borges}}]{RamirezBarreto:2006tn}
\bibinfo{author}{\bibfnamefont{E.}~\bibnamefont{Ramirez~Barreto}},
  \bibinfo{author}{\bibfnamefont{Y.~D.~A.} \bibnamefont{Coutinho}},
  \bibnamefont{and} \bibinfo{author}{\bibfnamefont{J.}~\bibnamefont{Sa~Borges}}
  (\bibinfo{year}{2006}), \eprint{hep-ph/0605098}.

\bibitem[{\citenamefont{Coutinho et~al.}(1999)\citenamefont{Coutinho,
  Queiroz~Filho, and Tonasse}}]{Coutinho:1999hf}
\bibinfo{author}{\bibfnamefont{Y.~D.~A.} \bibnamefont{Coutinho}},
  \bibinfo{author}{\bibfnamefont{P.}~\bibnamefont{Queiroz~Filho}},
  \bibnamefont{and} \bibinfo{author}{\bibfnamefont{M.}~\bibnamefont{Tonasse}},
  \bibinfo{journal}{Phys.Rev.} \textbf{\bibinfo{volume}{D60}},
  \bibinfo{pages}{115001} (\bibinfo{year}{1999}), \eprint{hep-ph/9907553}.

\bibitem[{\citenamefont{Alves et~al.}(2013)\citenamefont{Alves,
  Ramirez~Barreto, Dias, de~S.~Pires, Queiroz et~al.}}]{Alves:2012yp}
\bibinfo{author}{\bibfnamefont{A.}~\bibnamefont{Alves}},
  \bibinfo{author}{\bibfnamefont{E.}~\bibnamefont{Ramirez~Barreto}},
  \bibinfo{author}{\bibfnamefont{A.}~\bibnamefont{Dias}},
  \bibinfo{author}{\bibfnamefont{C.}~\bibnamefont{de~S.~Pires}},
  \bibinfo{author}{\bibfnamefont{F.~S.} \bibnamefont{Queiroz}},
  \bibnamefont{et~al.}, \bibinfo{journal}{Eur.Phys.J.}
  \textbf{\bibinfo{volume}{C73}}, \bibinfo{pages}{2288} (\bibinfo{year}{2013}),
  \eprint{1207.3699}.

\bibitem[{\citenamefont{Ruiz-Alvarez et~al.}(2012)\citenamefont{Ruiz-Alvarez,
  de~S.~Pires, Queiroz, Restrepo, and Rodrigues~da Silva}}]{Alvares:2012qv}
\bibinfo{author}{\bibfnamefont{J.}~\bibnamefont{Ruiz-Alvarez}},
  \bibinfo{author}{\bibfnamefont{C.}~\bibnamefont{de~S.~Pires}},
  \bibinfo{author}{\bibfnamefont{F.~S.} \bibnamefont{Queiroz}},
  \bibinfo{author}{\bibfnamefont{D.}~\bibnamefont{Restrepo}}, \bibnamefont{and}
  \bibinfo{author}{\bibfnamefont{P.}~\bibnamefont{Rodrigues~da Silva}},
  \bibinfo{journal}{Phys.Rev.} \textbf{\bibinfo{volume}{D86}},
  \bibinfo{pages}{075011} (\bibinfo{year}{2012}), \eprint{1206.5779}.

\bibitem[{\citenamefont{Kelso et~al.}(2014)\citenamefont{Kelso, de~S.~Pires,
  Profumo, Queiroz, and Rodrigues~da Silva}}]{Kelso:2013nwa}
\bibinfo{author}{\bibfnamefont{C.}~\bibnamefont{Kelso}},
  \bibinfo{author}{\bibfnamefont{C.~A.} \bibnamefont{de~S.~Pires}},
  \bibinfo{author}{\bibfnamefont{S.}~\bibnamefont{Profumo}},
  \bibinfo{author}{\bibfnamefont{F.~S.} \bibnamefont{Queiroz}},
  \bibnamefont{and} \bibinfo{author}{\bibfnamefont{P.~S.}
  \bibnamefont{Rodrigues~da Silva}}, \bibinfo{journal}{Eur.Phys.J.}
  \textbf{\bibinfo{volume}{C74}}, \bibinfo{pages}{2797} (\bibinfo{year}{2014}),
  \eprint{1308.6630}.

\bibitem[{\citenamefont{Barreto et~al.}(2013)\citenamefont{Barreto, Coutinho,
  and Borges}}]{Barreto:2013paa}
\bibinfo{author}{\bibfnamefont{E.~R.} \bibnamefont{Barreto}},
  \bibinfo{author}{\bibfnamefont{Y.}~\bibnamefont{Coutinho}}, \bibnamefont{and}
  \bibinfo{author}{\bibfnamefont{J.~S.} \bibnamefont{Borges}},
  \bibinfo{journal}{Phys.Rev.} \textbf{\bibinfo{volume}{D88}},
  \bibinfo{pages}{035016} (\bibinfo{year}{2013}), \eprint{1307.4683}.

\end{thebibliography}

\end{document}